\documentclass[11pt]{article}
\usepackage{amsmath}
\usepackage{amssymb}
\usepackage{hhline}
\usepackage[scale={.75,.75}]{geometry}
\usepackage{url}
\usepackage{lscape}
\usepackage{caption}
\usepackage[numbers, sort&compress]{natbib}
\usepackage{epsfig}
\usepackage{multirow} 
\usepackage{color}
\usepackage{verbatim}
\usepackage{nicefrac}
\usepackage{upgreek}
\usepackage{bbm}
\usepackage{setspace}
\usepackage{pstricks}
\usepackage{color}

\newcommand{\slashed}[1]{\displaystyle{\not}{#1}}

\usepackage{hyperref}

\definecolor{green}{rgb}{0,0.5,0}

\begin{document}
\date{}

\title{\vspace{-2.5cm} 
\begin{flushright}
\vspace{-0.4cm}
{\scriptsize \tt TUM-HEP-881/13}  
\end{flushright}
\vspace{-0.3cm}
{\bf The Phenomenology of Right Handed Neutrinos}}

\author{Marco Drewes\\
\footnotesize{Physik Department T70, Technische Universit\"at M\"unchen, }\\
\footnotesize{James Franck Stra\ss e 1, D-85748 Garching, Germany}}

\maketitle

\vspace{-0.5cm}
\begin{abstract}
Neutrinos are the only matter particles in the Standard Model of particle physics that have only been observed with left handed chirality to date. 
If right handed neutrinos exist, they could be responsible for several phenomena that have no explanation within the Standard Model, including neutrino oscillations, the baryon asymmetry of the universe, dark matter and dark radiation. 
After a pedagogical introduction, we review recent progress in the phenomenology of right handed neutrinos. We in particular discuss the mass ranges suggested by hints for neutrino oscillation anomalies and dark radiation (eV),
sterile neutrino dark matter scenarios (keV) and experimentally testable theories of baryogenesis (GeV to TeV). We summarize constraints from theoretical considerations, laboratory experiments, astrophysics and cosmology for each of these.  
\end{abstract}


\footnotesize
\tableofcontents
\normalsize
\newpage
\section{Introduction}
This review is intended to give a comprehensive overview of what we know (more precisely: what I know) about the phenomenology of right handed neutrinos. 
Faced with the difficulty of writing a text that is 
readable within reasonable time, and is at the same time is as precise as possible, 
I provide extra details, explanations and comments for the interested reader in an extensive set of footnotes.
For a quick read you may essentially ignore all of them. 
Furthermore, each section has been written to be as self-contained as possible,
so that they can also be read individually as ``mini reviews''.
The basic notations used throughout the review are introduced in sections \ref{missingpiece} and \ref{NeutrinoMasses}. A very quick summary is given in appendix \ref{overview}. \\ 

\noindent I will maintain a list of corrections
online at \href{http://res-publica.eu/RightHandedNeutrinos.html}{\it http://res-publica.eu/RightHandedNeutrinos.html}. If you find any errors or typos I would be very grateful if you could inform me; you can find  my most up-to-date contact information on the above website.
This preprint is identical to the published article \cite{RHNeutrinosPublished} up to a few rephrasings and
a number of additional references that appeared after the article had been accepted for publication.

\subsection{Physics beyond the Standard Model}
The Standard Model of particle physics (SM) 
and theory
of General Relativity (GR) form the basic pillars of modern physics. 
Together they can describe almost all phenomena
observed in nature\footnote{The basic laws of other areas of natural science and technology can be understood as effective theories, which in principle can be derived from the SM and GR.
Though there exist many complex phenomena that we do not understand or cannot predict in detail, this lack of predictivity is almost certainly related to the complexity of the system rather than a lack of understanding of its basic components, the elementary quantum fields.} in terms of a small number of underlying principles - general covariance, gauge invariance and quantum mechanics - and
a handful of numbers\footnote{There are $19$ free parameters in the SM; these
are usually chosen as six quark masses, three mixing angles and one CP violating phase for the quarks, three charged lepton masses,  three gauge couplings, two parameters in the Higgs potential and a QCD vacuum angle. The neutrinos are massless in the SM.
GR adds two additional parameters to this barcode of nature, the Planck mass and the
cosmological constant.} \cite{Beringer:1900zz}.
All elementary particles we have observed to date can be understood as fundamental excitations of a few quantum fields, the properties of which are constrained by the local structure of space and time\footnote{All known elementary particles transform under irreducible representations of the Poincar\'e group
.}. 
Interactions between them are the result of (gauge) symmetries of the Lagrangian.

In spite of its enormous success, this cannot be a complete theory of nature for two reasons.
On one hand, it treats gravitational fields as a classical background, 
while matter and other interactions are described by quantum field theory in the SM. 
This approximation certainly becomes invalid and has to be extended to a theory of quantum gravity at energies near the Planck scale\footnote{We use natural units $c=\hbar=1$.} $M_P=1.22\times10^{19}$ GeV.
We do not address this problem in the following because it is of little relevance for experiments in foreseeable time. 
On the other hand, there are four experimental and observational facts which cannot be understood in the framework SM+GR.
Three of them are widely believed to be related to particle physics,
\begin{itemize}
\item[(I)] flavour violation in neutrino experiments, section \ref{NeutrinoMasses},
\item[(II)] the cosmological origin of the baryonic matter in the universe, section \ref{Leptogenesis},
\item[(III)] the composition and origin of the observed dark matter (DM), section \ref{DMsection}.
\end{itemize}
In addition to the above {\it evidence} for the existence of ``new physics'', there are a number of {\it hints} in experimental data that may point towards the existence of physics beyond the SM; these have not (yet?) led to a claim of discovery and may also be explained by systematics.
Of these, we will only discuss two in detail in this review,\footnote{Others include the long standing issues of the muon magnetic moment (see e.g. \cite{Jegerlehner:2009ry} for a review), the annual modulation in the DAMA data \cite{Bernabei:2010mq}, the excess of positrons in the cosmic radiation \cite{Adriani:2008zr,Aguilar:2013qda}, the "forward-backward asymmetry" observed at the Tevatron \cite{Aaltonen:2011kc}, the disputed evidence for neutrinoless double $\beta$-decay  claimed by the Heidelberg-Moscow experiment \cite{KlapdorKleingrothaus:2001ke}, the cosmological lithium problem \cite{Fields:2011zzb}, unexplained features in the galactic $\gamma$-ray spectrum that may be related to DM \cite{Weniger:2012tx} and shifts in quasar absorption lines that have been interpreted as signatures of a varying fine structure constant \cite{King:2012id}.}
\begin{itemize} 
\item[(i)] the statistical preference for additional relativistic particles dubbed ``dark radiation'' (DR)\footnote{The term ``dark radiation'' refers to relativistic particles in the early universe 
with no or tiny interaction with the SM  
at temperatures $T< 2$ MeV.}  in some fits to cosmological data, section \ref{ThermalHistory}, 
\item[(ii)] the anomalies seen in some short baseline and reactor neutrino experiments, section \ref{LowEnergyExp}.
\end{itemize} 
All of the above phenomena may be related to {\it right handed (RH) neutrinos} with different masses. It is the purpose of this article to summarize how they can be connected to these hypothetical particles and review bounds from theoretical considerations, laboratory experiments and cosmology on 
RH neutrino properties.

To complete the list, let us add the fourth piece of evidence for physics beyond SM+GR, which is related to gravitation of cosmology,
\begin{itemize}
\item[(IV)]  the overall geometry of the universe (isotropy, homogeneity and spatial flatness), as e.g. seen in the cosmic microwave background (CMB).
\end{itemize}
We do not address (IV) here, as it is not related to RH neutrinos (or particle physics in general) in an obvious way\footnote{It has been speculated that this point may be related to the RH neutrinos' superpartners \cite{Murayama:1992ua}, see also \cite{Mazumdar:2012qk}.}. An intuitive explanation is given by cosmic inflation, see section \ref{briefhistory}.
Finally, the observed acceleration of the universe's expansion is often included in this list. However, all observations can currently be explained in terms of a cosmological constant $\Lambda$, which is simply a free parameter in GR. Hence, the accelerated expansion can be accommodated in the framework of SM+GR. The question of the microphysical "origin" of $\Lambda$ (and its smallness) only arises when the SM and GR are interpreted as low energy limits of a more general theory, including a complete description of quantum gravity. 
To date, (I)-(III) and (IV) are the only confirmed empirical proofs of physics that cannot be explained by SM+GR.\footnote{There are various aspects of the SM that may be considered ``problems'' from an aesthetic viewpoint or physical intuition, such as the hierarchy between the electroweak and Planck scale, the strong CP problem, the factorization of the gauge group and the flavour structure. We do not discuss these here. We also do not discuss the issue of vacuum stability, which seems inconclusive at this stage due to uncertainties in the top mass \cite{Shaposhnikov:2009pv,Bezrukov:2012sa,Degrassi:2012ry,Alekhin:2012py,EliasMiro:2011aa}.}.\\

In the remainder of this section we introduce the concept of RH neutrinos and define our notation. In section \ref{NeutrinoMasses} we review how they can generate masses for the known neutrinos. In section \ref{LowEnergyExp} we summarize bounds from past laboratory experiments on RH neutrino properties, discuss the interpretation of the observed neutrino oscillation anomalies (ii) in terms of RH neutrinos and comment on possible future searches. 
In sections \ref{ThermalHistory}-\ref{DMsection} we discuss various cosmological constraints, starting with a general summary in section \ref{briefhistory}. 
The perspectives to interpret the hints for ``dark radiation'' (i) in terms of RH neutrinos and reconcile them with the oscillation anomalies (ii) are addressed in section \ref{RHasDR}.
Section \ref{Leptogenesis} is devoted to the idea that RH neutrinos are the origin of the baryonic matter in the universe (leptogenesis) and possible implications for their properties. 
Section \ref{DMsection} discusses RH neutrinos as DM candidates.
In section \ref{nuMSMsec} we address the question how many of these phenomena can be explained {\it simultaneously} by RH neutrinos {\it alone}. 
We conclude in section \ref{conclusions} and give a tabular summary of possible RH neutrino mass scales and their implications for known and future observations in appendix \ref{overview}.

\subsection{The missing piece?}\label{missingpiece} 
\begin{figure}
  \centering
    \includegraphics[width=9cm]{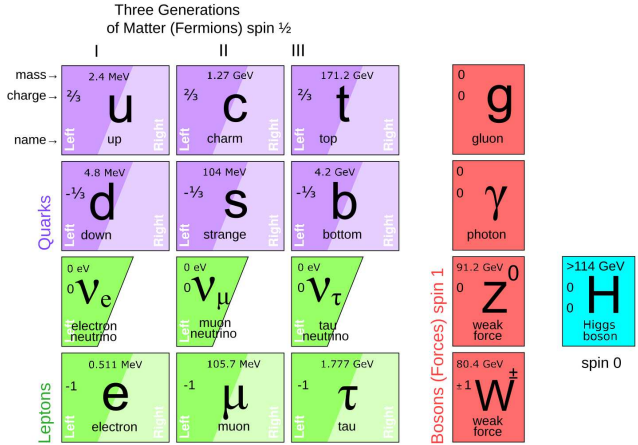}
    \caption{The particle content of the SM. Are we missing the right handed partner of the neutrinos? Picture taken from \cite{Shaposhnikov:2013dra}.\label{SM1}}
\end{figure}

All matter we know is composed of elementary fermions with spin $\frac{1}{2}$. These can be described by Weyl spinors, which transform under irreducible representations of the Poncair\'e group, and combinations thereof. There are two such representations, known as ``left chiral'' and ``right chiral'' spinors.
Remarkably, all known elementary fermions except neutrinos come in pairs of opposite chirality, i.e. have been observed as ``left handed'' (LH) and ``right handed'' (RH) particles\footnote{In this article ``RH'' and ``LH'' always refer to the chirality of the fields and {\it not} to helicity eigenstates.}, see figure \ref{SM1}. 
For unknown reasons the interactions of the SM are such that both can be combined into a Dirac spinor, see appendix \ref{appendixDiracMajorana}. 
Neutrinos, however, so far have only been observed as LH particles.
One conclusion that could be derived from this is that no right chiral ``partner'' for the observed LH neutrinos exists in nature.
Another possible conclusion is that we have not seen RH neutrinos just because their interaction with other matter is too weak. 
Indeed LH neutrinos are electrically and colour neutral; in the SM they only participate in the weak interaction, which does not couple to RH fields. This suggests that their RH partners are singlet under all gauge interactions\footnote{Here we refer to the $SU(3)\times SU(2) \times U(1)$ gauge group of the SM. They can of course be charged under some extended gauge group, which either acts only in a ``hidden sector'' or is broken at energies above the electroweak scale.}. Such particles are referred to as ``sterile neutrinos''. 

Let us add $n$ RH fermions $\nu_{R,i}$ to the SM that are singlet under all gauge interactions and couple to LH neutrinos in same way as RH charged leptons couple to LH charged leptons, i.e. via Yukawa interactions. We will refer to these fields as {\it RH neutrinos} and to the index $_i$ that labels them as {\it flavour index}.
Then the most general renormalizable Lagrangian in Minkowski space that only contains SM fields and $\nu_R$ reads
\begin{eqnarray}
	\label{L}
	\mathcal{L} =\mathcal{L}_{SM}+ 
	i \overline{\nu_{R}}\slashed{\partial}\nu_{R}-
	\overline{l_{L}}F\nu_{R}\tilde{\Phi} -
	\overline{\nu_{R}}F^{\dagger}l_L\tilde{\Phi}^{\dagger} 
	-{\rm \frac{1}{2}}(\overline{\nu_R^c}M_{M}\nu_{R} 
	+\overline{\nu_{R}}M_{M}^{\dagger}\nu^c_{R}). 
	\end{eqnarray}
Here we have suppressed flavour and isospin indices.
$\mathcal{L}_{SM}$ is the Lagrangian of the SM. $F$ is a matrix of
Yukawa couplings and $M_{M}$ a Majorana mass term for the right handed
neutrinos $\nu_{R}$. $l_{L}=(\nu_{L},e_{L})^{T}$ are the left handed
lepton doublets in the SM and $\Phi$ is the Higgs doublet. 
$\tilde{\Phi}=(\epsilon\Phi)^\dagger$, where $\epsilon$ is the $SU(2)$ antisymmetric tensor, and $\nu_R^c=C\bar{\nu_R}^T$, where the charge conjugation matrix is $C=i\gamma_2\gamma_0$ in the Weyl representation.
We choose a flavour basis where the charged lepton Yukawa couplings and $M_{M}$ are
diagonal and the neutrino coupling to weak currents has the form (\ref{WeakWW}).\footnote{This corresponds to a ``mass basis'' for charged leptons and ``flavour basis'' for active neutrinos. For vanishing Higgs field value it would coincide with the mass basis for all fields.} Throughout this article we assume that there are only three ``active'' neutrinos $\nu_L$ that are charged under the weak interaction. If there are more families, then these must be heavier than $m_Z/2$, otherwise they would 
contribute to the width of the Z-boson \cite{ALEPH:2005ab}, see \cite{Delepine:2013sca} for some recent discussion and references.

In the Lagrangian (\ref{L}) the fields $\nu_R$ only interact via the Yukawa couplings $F$. In the early universe, when the temperature was high enough that Higgs particles were present in the primordial plasma ($T>T_{EW}\sim 140$ GeV for a Higgs mass $m_H\sim 125$ GeV \cite{:2012gk,:2012gu}), this 
interaction allowed $\nu_R$-particles to participate in various different scattering processes. 
At energies much below the mass of the W-boson one can in good
approximation replace the Higgs field $\Phi$ by its vacuum expectation
value $v=174$ GeV. Then (\ref{L}) can be written as
\begin{eqnarray}
	\label{Lagrangian_flavour_base}
	\mathcal{L} = \mathcal{L}_{SM}+i\overline{\nu_{R}}_{,I}
	\slashed{\partial}\nu_{R,I} - (m_D)_{\alpha I} \overline{\nu_{L}}_{,\alpha} \nu_{R, I} - (m_D)^{*}_{\alpha I}
	\overline{\nu_{R}}_{,I} \nu_{L, \alpha} \nonumber\\- {\small
	\frac{1}{2}}\big[(M_{M})_{IJ} \overline{\nu_{R}^c}_{,I} \nu_{R,J}
	+  (M_{M})^{*}_{IJ} \overline{\nu_{R}}_{,I} \nu_{R,J}^c\big]
	\end{eqnarray}
where we defined the {\it Dirac mass matrix} $m_{D}=Fv$. 
Thus, at $T\ll T_{EW}$ the only effect of the Yukawa interaction is the generation of the Dirac mass term $m_D$, and the only way how the fields $\nu_R$ interact with the SM is via their mixing with $\nu_L$ due to $m_D$.

\subsection{The range of right handed neutrino masses}
While in the SM the Higgs mass $m_H$ is the only dimensionful parameter (apart from the Planck mass), the Lagrangian (\ref{L}) introduces $n$ new dimensional parameters in $M_M$. The scale(s) associated with these provide a convenient way to classify different RH neutrino scenarios. 
Various embeddings of (\ref{L}) into a bigger framework make different predictions for $M_M$ (see e.g. \cite{Mohapatra:1998rq,Fukugita:2003en,Mohapatra:2005wg} for a general overview and \cite{Abazajian:2012ys,Antusch:2013ti,Merle:2013gea} for recent developments in model building), but empirically there are only few constraints. The following scenarios are particularly motivated, a summary in table form is given in appendix \ref{overview}. Of course, $M_M$ can have eigenvalues in several different mass ranges, so that several of these scenarios may be combined in nature.

\textbf{$\pmb{M_M\gtrsim 10^9}$ GeV} - This range is motivated by embeddings of (\ref{L}) into GUT scenarios \cite{Georgi:1974sy}, such as  SO(10) unification \cite{Georgi:1974my,Fritzsch:1974nn}. SO(10) models necessarily require the existence of $\nu_R$.\footnote{Any model that contains a $U(1)_{B-L}$ gauge symmetry requires this for anomaly freedom. This can be used as an argument for the existence of $\nu_R$: The conservation of $B-L$ in the SM is not related to a gauge symmetry. If there is such symmetry, then $\nu_R$ must exist.} For Yukawa couplings $F$ of order one, RH neutrinos with masses favoured by GUT models reproduce the scale of observed neutrino oscillations (I) via (\ref{SeesawNeutrinomass}). In addition, typical parameter values allow to generate the observed baryon density in the universe (II) in CP-violating decays of RH neutrinos, see section \ref{DecayLeptogenesis}. The regime $M_M<10^{15}$ GeV is favoured \cite{Maltoni:2000iq}.

\textbf{$\pmb{M_M\sim}$ TeV} - Theoretically this mass range is interesting because it follows from a {\it no new scale} principle of minimality: If $M_M$ is near the electroweak scale, the origin of both scales may be related. 
The origin of matter (II) may be explained by leptogenesis from CP-violating $\nu_R$-oscillations (see section \ref{LeptogenfromOsc}) or, if two $\nu_R$ masses are degenerate, decays (see section \ref{DecayLeptogenesis}). Neutrino masses (I) are explained by the seesaw mechanism.
From an experimental viewpoint this mass range is favourable because it is accessible by high energy experiments, such as LHC. 

\textbf{$\pmb{M_M\sim}$ GeV} - If $M_M$ has at least two eigenvalues $\gtrsim2$ GeV and another eigenvalue in the keV range, then the observations (I)-(III) can be described by (\ref{L}) alone,  and no other physics between the electroweak and Planck scales is required, see section \ref{nuMSMsec}. Experimentally $\nu_R$ with GeV masses may be found using high intensity experiments, see section \ref{collidersec}.

\textbf{$\pmb{M_M\sim}$ keV} - RH neutrinos with keV masses are promising candidates for the DM (III), see section \ref{DMsection}.

\textbf{$\pmb{M_M\sim}$ eV} - RH neutrinos with eV masses can provide an explanation for the anomalies (i) and/or (ii), which are observed in some neutrino experiments (see section \ref{anomalies}) and cosmological data (see section \ref{ThermalHistory}, in particular \ref{RHasDR}).

$\pmb{M_M=0}$ -  
For $n=3$ the leptonic sector exactly resembles the quark sector without strong interactions.
In this case neutrinos are Dirac particles.\footnote{For $n\neq 3$ it is in general not possible to combine all $\nu_L$ and $\nu_R$ into Dirac spinors, see appendix \ref{appendixDiracMajorana}.}
Then neutrino masses are generated by the Higgs mechanism in precisely the same way as other fermion masses, and their smallness can only be assigned to very tiny Yukawa couplings.
Though in principle possible, this may appear ``unnatural'' unless there is a deeper reason for it\footnote{Some speculations on such reasons can e.g. be found in \cite{Dienes:1998sb,ArkaniHamed:1998vp,Grossman:1999ra,ArkaniHamed:1998pf}.}. 
Furthermore, there is no known principle that forbids $M_M$ for the gauge singlet fields $\nu_R$\footnote{A small value of $M_M$ is, however, ``natural'' in the technical sense because the symmetry of $\mathcal{L}$ increases in the limit $M_M\rightarrow 0$ (there is a global $U(1)_{B-L}$).}.
This is in contrast to quarks and charged leptons, for which an explicit mass term is forbidden by gauge symmetry. 

\section{Neutrino oscillations}\label{NeutrinoMasses}
The probably strongest motivation for the existence of $\nu_R$ are neutrino oscillations, the only processes amongst (I)-(IV) that have been observed in the laboratory.
The neutrinos $\nu_L$ are massless in the SM\footnote{We do not consider neutrino masses as a part of the SM because we do not know what the nature (Dirac or Majorana) or absolute scale of the mass term is.}.
In the past two decades an increasing number of neutrino experiments has observed neutrino flavour changes, which indicate that neutrinos are massive and oscillate, see e.g. \cite{Strumia:2006db,Abazajian:2012ys} for reviews with many references. 
The experimental results can be divided into two categories: the {\it standard $3$-scenario} (SS) of three massive neutrinos, which we discuss here, and deviations from it, which we discuss in section \ref{LowEnergyExp}.

The interactions of neutrinos in the SM are described by the Lagrangian term
\begin{equation}\label{WeakWW}
-\frac{g}{\sqrt{2}}\overline{\nu_L}\gamma^\mu e_L W^+_\mu
-\frac{g}{\sqrt{2}}\overline{e_L}\gamma^\mu \nu_L W^-_\mu  
- \frac{g}{2\cos\theta_W}\overline{\nu_L}\gamma^\mu\nu_L Z_\mu ,
\end{equation}
where $g$ is the $SU(2)$ gauge coupling constant and $\theta_W$ the weak mixing angle. 
This defines the basis of weak interaction eigenstates (electron, muon and tau neutrino).\footnote{More precisely, if one considers the general form of the interaction $\frac{g}{\sqrt{2}}\overline{\nu_{L,\alpha}}\gamma^\mu U_{\alpha\beta}e_{L,\beta} W_\mu^+$ in the basis where charged Yukawa couplings are diagonal, then the basis of weak interaction eigenstates for $\nu_L$ is the one where $U_{\alpha\beta}=\delta_{\alpha\beta}$.} 
If neutrinos have a mass, then the mass term need not be diagonal in this basis in flavour space.
In the SS, the weak interaction eigenstates $\nu_{L,e}$, $\nu_{L,\mu}$ and $\nu_{L,\tau}$ are superpositions of three mass eigenstates $\upsilon_{L,i}$ with masses $m_i$. For a given momentum, these have different energies if their masses are different, and their wave functions oscillate with different frequencies. Thus, the flavour decomposition of a neutrino state changes in time.
This can explain the observed neutrino oscillations. 
\subsection{The standard scenario of massive neutrinos}
There are two different ways to effectively realize the SS of three massive neutrinos with no extra particles.
\paragraph{Majorana neutrinos} -
A Majorana mass term of the form 
\begin{equation}\label{majoranamassterm}
\frac{1}{2}\overline{\nu_{L}}m_\nu \nu_{L}^{c} + h.c.
\end{equation} 
can be constructed without adding any new degrees of freedom to the SM. 
This term, however, breaks gauge invariance unless it is generated by spontaneous symmetry breaking from a gauge invariant term like \cite{Weinberg:1979sa}
\begin{equation}\label{weinbergoperator}
\frac{1}{2}\overline{l_{L}}\tilde{\Phi}f\tilde{\Phi}^{T}l_{L}^{c} + h.c.
,\end{equation}
where $f$ is some flavour matrix of dimension $1/{\rm mass}$. 
The dimension-5 operator (\ref{weinbergoperator}) is not renormalizable; 
in an effective field theory approach it can be understood as the low energy limit 
of renormalizable operators that is obtained after "integrating out" heavier degrees of freedom.
These can, for example (but not necessarily!), be right handed neutrinos, cf. (\ref{intout}).
Therefore a Majorana mass term (\ref{majoranamassterm}) clearly hints towards the existence of new physics although it can be constructed from SM fields only. 
At low energies one effectively observes the SS with only three massive neutrinos if the energy scale related to the new physics (in case of RH neutrinos the mass $M_M$) is sufficiently high that all new particles are too heavy to be seen in neutrino experiments. The Majorana mass term (\ref{majoranamassterm}) can be diagonalized by a transformation 
\begin{equation}\label{MajoranaDiagonalization}
m_{\nu}=U_\nu{\rm diag}(m_1,m_2,m_3)U_\nu^{T}.
\end{equation}
Such a transformation is always possible because the most general $m_\nu$ is a symmetric matrix. Any antisymmetric part is unphysical due to the simple spinor relation $2\overline{\nu_{L}}m_\nu \nu_{L}^{c}=\overline{\nu_{L}}(m_\nu+m_\nu^T) \nu_{L}^{c}$.\footnote{For the same reason the Majorana matrix $M_M$ in (\ref{L}) is symmetric in any flavour basis.} The matrix $U_\nu$ is constructed from the eigenvectors of $m_\nu m_\nu^\dagger$.
In the mass base, the neutrino mixing matrix  $U_\nu$  appears in the coupling to $W_\mu$ in (\ref{WeakWW}). 
\paragraph{Dirac neutrinos} -
If neutrinos are Dirac particles, the existence of $\nu_R$ is directly required to construct the mass term 
\begin{equation}\label{diracmassterm}
\overline{\nu_L} m_D\nu_R + h.c.
.\end{equation}
 Though this means adding new degrees of freedom to the SM at low energies, it is still a realization of the SS, i.e. only three massive neutrinos are observed. The  $\nu_R$ can be combined with the $\nu_L$ into Dirac spinors and there are only three different masses. One could say that there are no new particles in the strict sense, but just additional spin states for neutrinos. We refer to both of these scenarios as the SS.

A bi-unitary transformation $m_D=U_\nu {\rm diag}(m_1,m_2,m_3)V_\nu^\dagger$ can diagonalize the mass term (\ref{diracmassterm}), with real and positive $m_i$.  One can define a Dirac spinor $\Psi_\nu\equiv V_\nu^\dagger\nu_R+U_\nu^\dagger\nu_L$ with a diagonal mass term $m_\nu^{\rm diag}={\rm diag}(m_1,m_2,m_3)$, i.e. $\overline{\Psi_\nu}(i\slashed{\partial}-m_\nu^{\rm diag})\Psi_\nu$.  
The matrix $V_\nu$ is not physical and can be absorbed into a redefinition of the flavour vector $\nu_R$.
The neutrino mixing matrix  $U_\nu$ then appears in the coupling of that Dirac spinor to $W_\mu$ in (\ref{WeakWW}). A phenomenological prediction of Dirac neutrinos is that there is no neutrinoless double $\beta$-decay.
In order to be consistent with observations, the Yukawa couplings $F$ have to be very small ($F\sim 10^{-12}$) compared to those of the charged leptons and quarks.

\paragraph{The neutrino mixing matrix} - In the basis where charged Yukawa couplings are diagonal, the mixing matrix $U_\nu$ is identical to the Pontecorvo-Maki-Nakagawa-Sakata matrix \cite{Maki:1962mu,Pontecorvo:1967fh} and can be parametrized as
\begin{equation}
U_\nu=V^{(23)}U_{\delta}V^{(13)}U_{-\delta}V^{(12)}{\rm diag}(e^{i\alpha_1 /2},e^{i\alpha_2 /2},1)
\end{equation}
with $U_{\pm\delta}={\rm diag}(e^{\mp i\delta/2},1,e^{\pm i\delta/2})$.  
The 
matrices $V^{(ij)}$ are given by
\begin{eqnarray}
V^{(23)}=\left(\begin{tabular}{ccc}
$1$ & $0$ & $0$ \\ $0$ & $c_{23}$ & $s_{23}$ \\ $0$ & $-s_{23}$ & $c_{23}$
\end{tabular},\
\right), \
V^{(13)}=\left(
\begin{tabular}{ccc}
$c_{13}$ & 0 & $s_{13}$ \\ 0 & 1 & 0 \\ $-s_{13}$ & 0 & $c_{13}$
\end{tabular}
\right),\
V^{(12)}=\left(
\begin{tabular}{ccc}
$c_{12}$ & $s_{12}$ &  0\\ $-s_{12}$ & $c_{12}$ & 0 \\ 0 & 0 & 1
\end{tabular}
\right), 
\end{eqnarray}
where $c_{ij}$ and $s_{ij}$ denote $\cos(\uptheta_{ij})$ and $\sin(\uptheta_{ij})$, respectively. $\uptheta_{ij}$ are the neutrino mixing angles, $\alpha_1$, $\alpha_2$ and $\delta$ are $CP$-violating phases. 
Many parameters of the mixing matrix $U_\nu$ have been measured in recent years. 
In particular, two mass square differences have been determined as 
$\Delta m_{\rm sol}^2\equiv m_2^2-m_1^2\simeq 7.5\times10^{-5}{\rm eV}^2$ and $\Delta m_{\rm atm}^2\equiv|m_3^2-m_1^2|\simeq 2.4\times10^{-3}{\rm eV}^2$,
the mixing angles are $\uptheta_{12}\simeq 34^{\circ}$, $\uptheta_{23}\simeq39^{\circ}$ and $\uptheta_{13}\simeq 9^{\circ}$; the precise best fit values differ  for normal and inverted hierarchy (but the difference is smaller than the 1$\sigma$ ranges) and are given in \cite{Fogli:2012ua,GonzalezGarcia:2012sz,Tortola:2012te,Beringer:1900zz}, see also \cite{natureofmassiveneutrinos} and \cite{Antonelli:2012qu,AntonelliIJMPE} for recent reviews. 
What remains unknown are 
\begin{itemize}
\item{\it the hierarchy of neutrino masses} - One can distinguish between two non-equivalent setups. The {\it normal hierarchy} corresponds to $m_1<m_2<m_3$,  
with $\Delta m_{\rm sol}^2=m_2^2-m_1^2 $ and $\Delta m_{\rm atm}^2=m_3^2-m_1^2\simeq m_3^2-m_2^2\gg  \Delta m_{\rm sol}^2 $. The  {\it inverted hierarchy} corresponds to $m_2^2>m_1^2>m_3^2$, with $\Delta m_{\rm sol}^2=m_2^2-m_1^2 $ and $\Delta m_{\rm atm}^2=m_1^2-m_3^2\simeq m_2^2-m_3^2\gg  \Delta m_{\rm sol}^2 $. 
\item{\it the CP-violating phases} - The Dirac phase $\delta$ is the analogue to the CKM phase and remains present even for $M_M=0$. Global fits to neutrino data tend to prefer $\delta\sim\pi$ \cite{Fogli:2012ua}, but are not conclusive. The Majorana phases $\alpha_1$ and $\alpha_2$ become unphysical in the limit $M_M\rightarrow 0$ because they can be absorbed into redefinitions of the fields.
\item{\it the absolute mass scale} - The mass of the lightest neutrino is unknown, but the sum of masses is bound from above as $\sum_i m_i< 0.23$ eV \cite{Ade:2013lta} by cosmology\footnote{The bound and preferred value for $\sum_i m_i$ slightly change, depending on the dataset and analysis. It becomes more stringent when different data sets are combined, see e.g. \cite{Riemer-Sorensen:2013jsa}.}. It is also bound from below by the measured mass squares, $\sum_i m_i>0.06$ eV for normal and $\sum_i m_i>0.1$ eV for inverted hierarchy.  
\end{itemize}

\subsection{Neutrino masses from right handed neutrinos} 
In this section we discuss how sterile neutrinos described by the Lagrangian (\ref{L}) can give Dirac or Majorana mass terms to the active neutrinos.
It is obvious that the Lagrangian (\ref{Lagrangian_flavour_base}) with $M_M=0$ represents Dirac neutrinos with a mass term (\ref{diracmassterm}).
For $M_M\neq 0$ they generate a Majorana mass term (\ref{majoranamassterm}). 
If the mass $M_M$ is sufficiently large, RH neutrinos are heavy and neutrino oscillation experiments can be described by an effective Lagrangian that is obtained by integrating them out of (\ref{Lagrangian_flavour_base}), 
\begin{equation}\label{SeesawNeutrinomass}
\mathcal{L}=\mathcal{L}_{SM} - \frac{1}{2}\overline{\nu_{L}}m_\nu \nu_{L}^{c} \ {\rm with} \ m_\nu=-m_DM_M^{-1}m_D^T
.\end{equation}
If the Majorana mass is above the electroweak scale ($M_M\gg v$) one can already integrate out the $\nu_R$ in (\ref{L}) to obtain (\ref{weinbergoperator}) with $f=F M_M^{-1} F^T$, 
\begin{equation}\label{intout}
\mathcal{L}=\mathcal{L}_{SM}+\frac{1}{2}\overline{l_{L}}\tilde{\Phi} F M_M^{-1} F^T \tilde{\Phi}^{T}l_{L}^{c}
.\end{equation}

Hence, the SS with three massive neutrinos is effectively realized if $M_M$ is either zero or so much bigger than the observed neutrino masses that (\ref{SeesawNeutrinomass}) can be used to describe neutrino oscillations.
If one or more eigenvalues of $M_M$ are not that large, then the light sterile neutrinos appear as new particles in neutrino experiments. 
To explore the full range of masses, we write the mass term as
\begin{eqnarray}\label{neutrinomassfull}
\frac{1}{2}
(\overline{\nu_L} \  \overline{\nu_R^c})
\mathfrak{M}
\left(
\begin{tabular}{c}
$\nu_L^c$\\
$\nu_R$
\end{tabular}
\right) + h.c.
\equiv
\frac{1}{2}
(\overline{\nu_L} \ \overline{\nu_R^c})
\left(
\begin{tabular}{c c}
$0$ & $m_D$\\
$m_D^T$ & $M_M$
\end{tabular}
\right)
\left(
\begin{tabular}{c}
$\nu_L^c$\\
$\nu_R$
\end{tabular}
\right) + h.c. 
\end{eqnarray}
Here we have used the identity $\overline{\nu_R^c}M_M\nu_R^c=\overline{\nu_R}M_M^T\nu_R$ and the symmetry of $M_M$.
If $\nu_R$ exist, they generate a mass term for $\nu_L$ except if $F$ either vanishes\footnote{In the case $F=0$ it is not clear from the viewpoint of (\ref{L}) why the fields $\nu_R$ should be called ``neutrinos'', as they have nothing in common with the known neutrinos except being neutral. Of course, such {\it pariah neutrinos} [which only interact gravitationally in (\ref{L})] can be charged in a way that justifies this classification under an extended gauge group that it broken at high energies.}, $F$ and $M_M$ have a particular flavour structure that leads to cancellations in $m_\nu$\footnote{Such cancellation occurs in models where $M_M$ and $F$ are chosen in a way that there is no total lepton number violation.} or there are not enough RH neutrino flavours $\nu_R$ to give mass to all $\upsilon_L$\footnote{For the Lagrangian (\ref{L}) the number of massive active neutrinos $\upsilon_{L,\alpha}$ cannot exceed the number $n$ of RH fields $\nu_{R,I}$; each $\nu_{R,I}$ can only generate one neutrino mass. Hence, at least two RH neutrino flavours are required to explain the observed mass splittings $\Delta m_{\rm atm}$ and $\Delta m_{\rm sol}$ (for $n=2$ the lightest active neutrino is massless).}
; this is seen as a strong motivation to postulate their existence\footnote{Of course there are alternative ways to generate a neutrino mass term, see e.g. \cite{Froggatt:1978nt,Zee:1980ai,Schechter:1980gr,Gelmini:1980re,Magg:1980ut,Cheng:1980qt,Lazarides:1980nt,Mohapatra:1980yp,Mohapatra:1986bd,Mohapatra:1986aw,Foot:1988aq,Babu:1988ki,Roncadelli:1983ty,Ma:1998dn,Gherghetta:2003hf,Gu:2006dc} or \cite{Mohapatra:1998rq,Fukugita:2003en,Mohapatra:2005wg} for a general overview with many references.}.
The relative size of $M_M$ and $m_D$ (i.e. their eigenvalues) allows to distinguish different scenarios. Note that this does not say anything about the absolute scale of $m_D$, which may lie anywhere between $0$ and the electroweak scale (assuming perturbative couplings $F$).

$\pmb{M_M\ll m_D}$ - This is a pseudo-Dirac scenario; for $n=3$ the neutrinos in this case effectively behave like Dirac particles\footnote{This is not only way how Dirac fermions can arise. For instance, if the splitting between two eigenvalues of $M_M$ is sufficiently small, then one can form Dirac spinor by combining different flavours, see appendix \ref{appendixDiracMajorana}.}. This case is excluded unless $M_M\lesssim 10^{-9}$ eV, as otherwise solar neutrino oscillations into $\nu_R$ should have been observed \cite{deGouvea:2009fp}.

$\pmb{M_M\sim m_D}$ - In this case $M_M$ and $m_D$ should both be of the order of the observed neutrino mass differences. The full mass term can be diagonalized by a $(3+n)\times(3+n)$ matrix $\mathcal{U}$ as $\mathfrak{M}=\mathcal{U}{\rm diag}(m_1,m_2,m_3,M_1,\ldots,M_n)\mathcal{U}^T$. This can lead to rather large mixing between $\nu_L$ and $\nu_R$.

$\pmb{M_M\gg m_D}$ - This is the seesaw limit, in which the neutrino mass term is described by (\ref{SeesawNeutrinomass}). 
 This scenario is discussed in detail in the following section \ref{SeeSawSec}. The seesaw limit roughly applies if $M_M>1$ eV.

\subsection{The seesaw mechanism}\label{SeeSawSec}
\begin{figure}
  \centering
    \includegraphics[width=9cm]{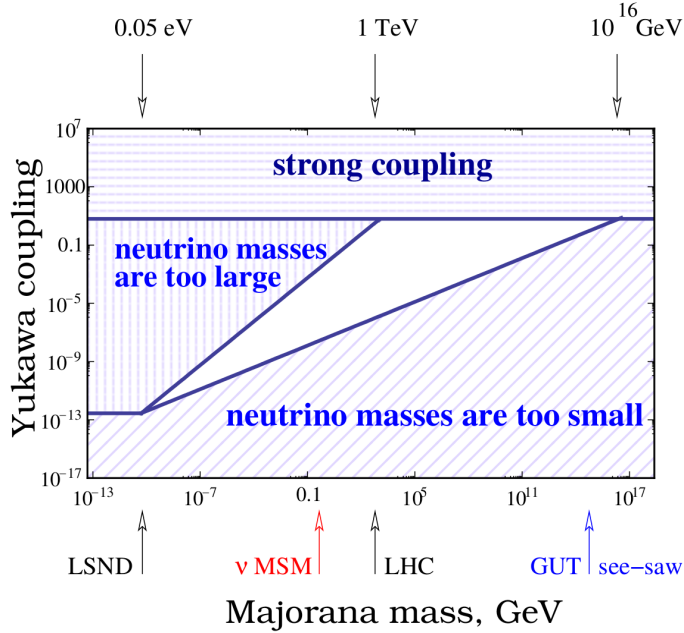}
    \caption{A schematic illustration of the relation between $F$ and $M_M$ in the seesaw limit $m_D\ll M_M$. 
Individual elements of the matrices $F$ and $M_M$ can deviate considerably from this if there are cancellations in (\ref{activeneutrinomasses}). Plot taken from \cite{Abazajian:2012ys}.\label{seesawfig}}
\end{figure}

In the limit $m_D\ll M_M$ (in terms of eigenvalues), the full $(3+n)\times(3+n)$ mass matrix $\mathfrak{M}$  for $\nu_L$ and $\nu_R$ has two distinct sets of eigenvalues. 
$n$ of them ($M_I$) are of the order of the eigenvalues of $M_M$, 
the remaining three ($m_i$) are suppressed by two powers of the {\it active - sterile mixing matrix} $\theta$,
\begin{equation}
\theta=m_D M_M^{-1}
. 
\end{equation}
The seesaw hierarchy $\theta\ll 1$ separates the two distinct sets of mass eigenstates. 
One can block-diagonalize (\ref{neutrinomassfull}) by expanding in $\theta$;
on one hand one obtains the $3\times3$ mass matrix 
\begin{eqnarray}
		 m_{\nu} &=& -\theta M_M \theta^T~,
\label{activeneutrinomasses}
\end{eqnarray}
with eigenvalues $m_i$; it corresponds to the operator (\ref{SeesawNeutrinomass}). The known bounds on neutrino masses impose constraints 
between the values of $F$ and $M_M$, see figure \ref{seesawfig}.
On the other hand there is the $n\times n$ matrix
\begin{eqnarray}
 M_N &=& M_M + \frac{1}{2}
\big(\theta^{\dagger} \theta M_M + M_M^T \theta^T \theta^{*}
\big)~
\label{sterileneutrinomassesAll}
\end{eqnarray}
with eigenvalues $M_I$.
The  matrices $m_{\nu}$ and $M_N$ are not diagonal and lead to flavour oscillations. 
Diagonalizing them yields the mass term 
\begin{equation}\label{massterm}
\frac{1}{2}\left(\overline{\upsilon_L}m_\nu^{\rm diag} \upsilon_L^c + \overline{\upsilon_R^c}M_N^{\rm diag}\upsilon_R\right) + h.c.
\end{equation}
with
\begin{equation}
m_\nu^{\rm diag}={\rm diag}(m_1,m_2,m_3) \ , \ M_N^{\rm diag}=(M_1,\ldots,M_n)
\end{equation}
In summary, the full matrix $\mathcal{U}$ that diagonalizes $\mathfrak{M}$ reads 
\begin{equation}
\mathcal{U}=
\left[
\left(
\begin{tabular}{c c}
$\mathbbm{1}-\frac{1}{2}\theta\theta^\dagger$ & $\theta$\\
$-\theta^\dagger$ & $\mathbbm{1}-\frac{1}{2}\theta^\dagger\theta$
\end{tabular}
\right)  \ + \ \mathcal{O}[\theta^3]
\right]
\left(
\begin{tabular}{c c}
$U_\nu$ & $ $\\
$ $ & $U_N^*$
\end{tabular}
\right) 
.\end{equation}
Three mass states have light masses $m_i\sim \mathcal{O}[\theta^2M_M]$ and are mainly mixings of the SU(2) charged fields $\nu_L$, 
\begin{equation}
\upsilon_L
\equiv
U_\nu^{\dagger}\left((\mathbbm{1}-
\frac{1}{2}\theta\theta^{\dagger})\nu_L-
\theta\nu_{R}^c\right)
\end{equation}
The remaining $n$ mass states have masses $M_I\sim\mathcal{O}[M_M]$ and are mainly
mixings of the singlet fields $\nu_R$,
\begin{equation}
\label{NIdef}
\upsilon_R\equiv
U_N^{\dagger}\left((\mathbbm{1}-
\frac{1}{2}\theta^{T}\theta^{*})\nu_{R}+
\theta^{T}\nu_{L}^{c}\right)
.
\end{equation}
We refer to $\upsilon_L$ as {\it active neutrinos} because they take part in the unsuppressed weak interactions (\ref{WeakWW}). The $\upsilon_R$ also participate in the weak interaction (\ref{WeakWW}), but only with an amplitude suppressed by $\theta$; hence they are {\it sterile neutrinos}.
The matrix $U_N$ diagonalises the sterile neutrino mass matrix $M_N$,\footnote{We choose the phase convention in $U_N$ such that $M_N=U_N^*{\rm diag}(M_1,\ldots,M_n)U_N^\dagger$ with $M_I$ real and positive.
In this convention we observe the same relations $\nu_L\simeq U_\nu\upsilon_L$ and $\nu_R\simeq U_N\upsilon_R$ for left and right handed fields (rather than $U_N^*$ for $\nu_R$).}
it can be seen as analogue to $U_\nu$.  
The diagonal elements of $M_N$ are much bigger than the off-diagonals and very close to the entries of $M_M$. 
Therefore one can in most cases neglect all terms of second order in $\theta$ and approximate $M_N=M_M$, $U_N=\mathbbm{1}$. However, if two eigenvalues of $M_M$ are degenerate, then the $U_N$ can contain large sterile-sterile mixing angles.
The active-sterile mixing angles are determined by the entries of the matrix $\theta$ and always small.
More precisely, the experimentally relevant mixing between active and sterile species is given by the matrix $\Theta$ with\footnote{In (\ref{NIdef}) the matrix
$U_N^\dagger\theta^T=\Theta^T$ appears (rather than
$\Theta$) because the
$N_I$ couple to $\nu_{L,\alpha}$, but overlap with
$\nu_{L,\alpha}^c$.} 
\begin{equation}
\label{ThetaDef}
\Theta_{\alpha I}\equiv(\theta U_N^*)_{\alpha I} 
\end{equation}
Practically, experiments to date constrain the quantities
\begin{equation}
U_\alpha^2\equiv\sum_I \Theta_{\alpha I}\Theta_{\alpha I}^*=\sum_I
\theta_{\alpha I}\theta_{\alpha I}^*\label{UalphaDef}
\end{equation} 
and combinations thereof.

This is known as the seesaw mechanism \cite{Minkowski:1977sc,GellMann:seesaw,Mohapatra:1979ia,Yanagida:1980xy} because increasing the eigenvalues of $M_M$ pushes the masses of the sterile neutrinos up and those of active neutrinos down, just as if they sat on a seesaw\footnote{
More precisely, this is the type-I seesaw mechanism. The term ``seesaw'' is also used for several modifications of this idea \cite{McDonald:2013kca}, such as the ``type-II seesaw'' \cite{Schechter:1980gr,Magg:1980ut,Cheng:1980qt,Lazarides:1980nt,Mohapatra:1980yp}, ``type-III'' seesaw \cite{Foot:1988aq}, ``split seesaw'' \cite{Kusenko:2010ik}, ``radiative seesaw'' \cite{Ma:2006km,Krauss:2002px}  or ``inverse seesaw'' \cite{Mohapatra:1986bd,Mohapatra:1986aw}.}.
If the couplings $F$ are of order one and $M_M$ in the range suggested by GUT models, then $m_\nu$ roughly reproduces the observed neutrino mass splittings. Hence, the mechanism provides a natural explanation for the smallness of neutrino masses. However, the scale $M_M$ of the seesaw is phenomenologically almost unconstrained and may be as low as $1$ eV \cite{deGouvea:2005er}.

To make the Majorana nature of the fields explicit (and get rid of the charge conjugation matrix in the mass term) one can describe them in terms of Majorana spinors; we define the flavour vectors\footnote{Obviously $\upsilon_R=P_RN$ and $\upsilon_L=P_L\upnu$, where $P_{R,L}$ are chiral projectors.}
\begin{equation}\label{Nandupnudef}
N \equiv \upsilon_R+\upsilon_R^c \ , \ \upnu \equiv \upsilon_L+\upsilon_L^c.
\end{equation}
Obviously, the elements $\upnu_i$ are active neutrinos and the $N_I$ sterile neutrinos.
The mass and kinetic terms then can be combined as 
\begin{equation}
\frac{1}{2}\bar{N}(i\slashed{\partial}-M_N^{\rm diag})N + 
\frac{1}{2}\bar{\upnu}(i\slashed{\partial}-m_\nu^{\rm diag})\upnu
\end{equation}
Up to the normalization, this looks like the Lagrangian for a Dirac field, but one has to keep in mind the Majorana conditions $N_I=N_I^c$, $\upnu_i=\upnu_i^c$.

\section{Other laboratory experiments}\label{LowEnergyExp}
The production and study of $N_I$ particles in the laboratory is
in principle possible if $M_I$ is below the electroweak scale. 
At energies $\ll M_I$, the $N_I$ only leave indirect traces in the laboratory. They manifest as higher dimensional operators \cite{Abada:2007ux}, such as the mass term (\ref{SeesawNeutrinomass}). These can lead to deviations from SM predictions in different observables, such as lepton number violation or $\beta$-decays. These signatures provide valuable information, but are usually not specific to RH neutrinos.
Here we list a number of experimental setups that can constrain the properties of $N_I$. So far almost all but those in section \ref{anomalies} have reported negative results, i.e. only allow to exclude certain parameter regions.

\subsection{Neutrino oscillation anomalies}\label{anomalies}
\begin{figure}
  \centering
\begin{tabular}{c c}
    \includegraphics[width=8cm]{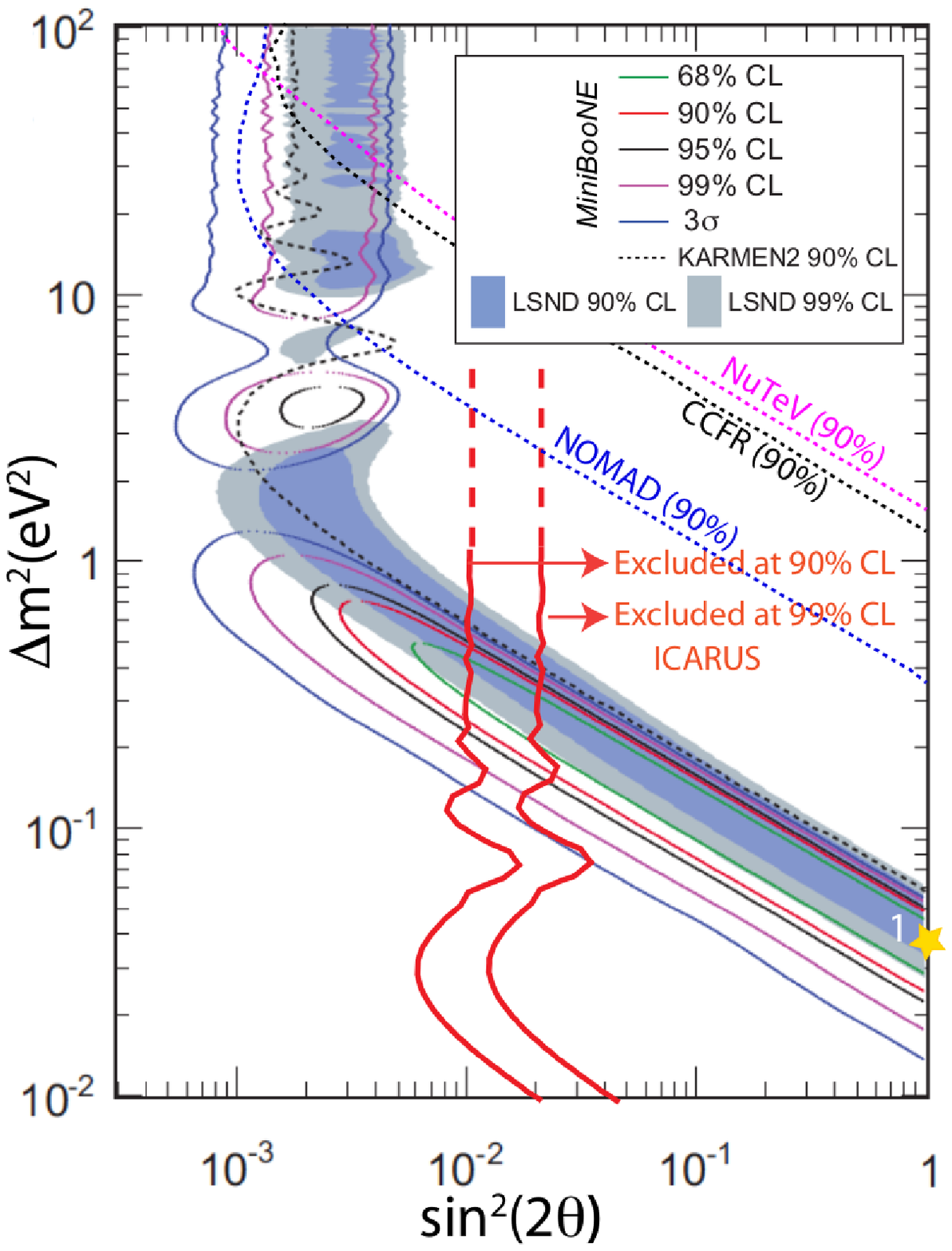} &
    \includegraphics[width=8cm
]{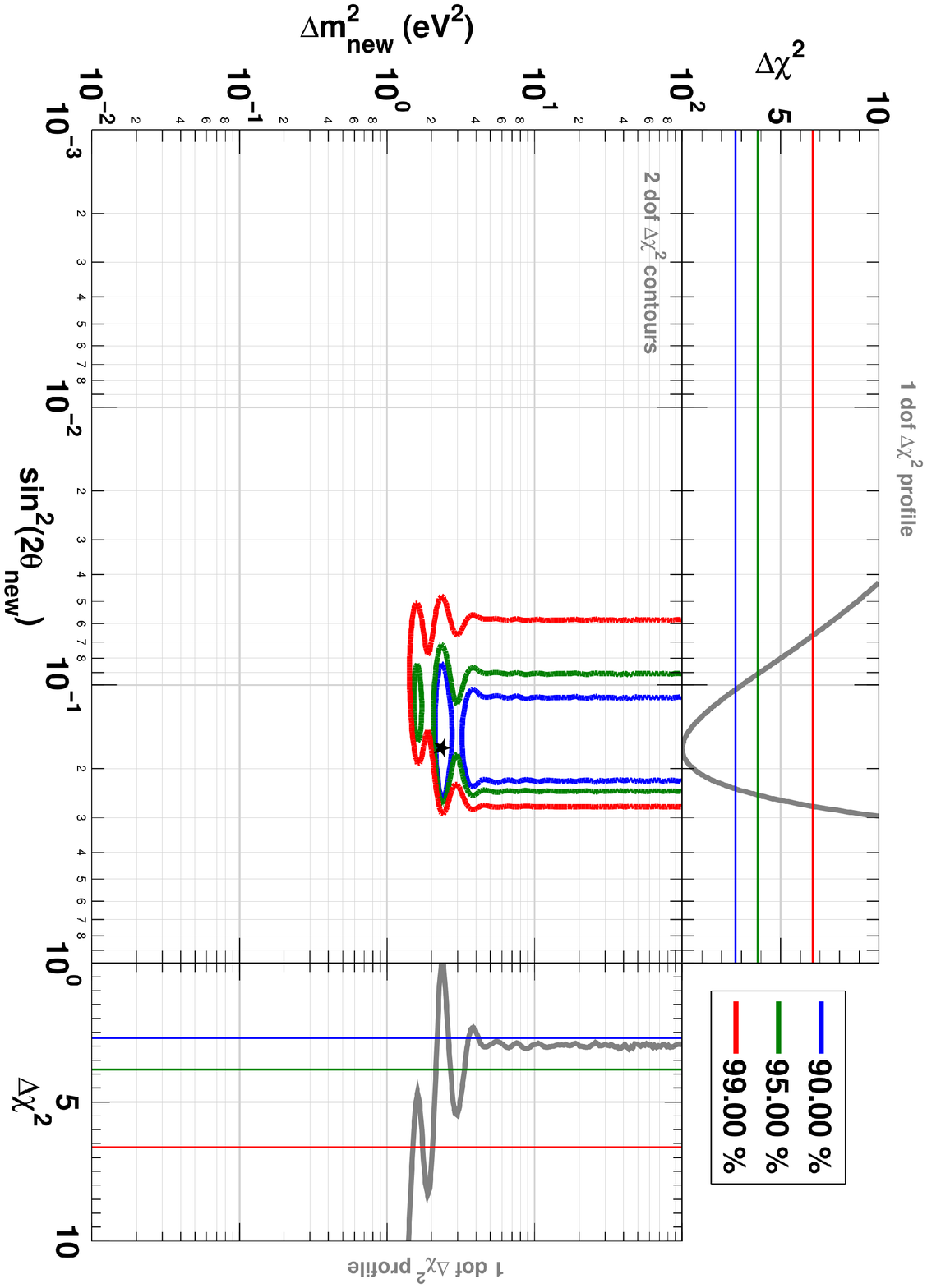}
\end{tabular}
    \caption{{\it Left plot}: The region in the mass-mixing plane preferred by the LSND anomaly (coloured bands) is compared to the region allowed by MiniBooNe (coloured lines) and exclusion plots by other experiments as indicated in the plot; plot taken from \cite{Antonello:2012pq}. {\it Right plot}: region allowed by the combined reactor and gallium anomalies, taken from \cite{Abazajian:2012ys}. Data from both is e.g. combined in figure 4 in \cite{Kopp2013}.\label{anomaliesfig}}
\end{figure}

\paragraph{Accelerator experiments} - Some short baseline and reactor neutrino experiments have reported deviations (ii) from the SS. A more detailed review of these results can found in \cite{Palazzo:2013me}, which we follow closely here. 
The most prominent findings come from the LNSD experiment \cite{Athanassopoulos:1996jb,Aguilar:2001ty}, which studied transitions $\bar{\nu}_\mu\rightarrow\bar{\nu}_e$ and saw an excess of $\bar{\nu}_e$-events. The similar KARMEN experiment did not see such excess \cite{Armbruster:2002mp}, but could not exclude the entire parameter space of the LSND anomaly \cite{Church:2002tc}, mostly due to its shorter baselength. The most recent results from the MiniBooNE experiment \cite{AguilarArevalo:2012va} from $\nu_\mu\rightarrow\nu_e$ and $\bar{\nu}_\mu\rightarrow\bar{\nu}_e$ studies seem compatible with the LSND anomaly, which is in contrast to previous results. However, it has been pointed out that much of the signal comes from energies below $475$ MeV, where the background evaluation is problematic \cite{Palazzo:2013me}.
The ICARUS experiment \cite{Antonello:2012pq} is also sensitive to sterile neutrinos; active-sterile oscillations get averaged over, but lead to an energy independent enhancement of event rates. 
Together these experiments roughly restrict the parameter space for one sterile neutrino to a mass splitting $\Delta m^2\sim 0.5 {\rm eV}^2$ and mixing $\sin^22\theta\sim 5\times10^{-3}$ \cite{Palazzo:2013me}, see figure \ref{anomaliesfig}.
\paragraph{Reactor and gallium anomalies} - 
There are two more anomalies that can be interpreted as a sign for sterile neutrinos with eV masses. 
One is the reactor anomaly \cite{Mention:2011rk}. 
The neutrino fluxes from nuclear reactors appeared to be in agreement with theory until recently. It was not new experimental data, but more refined theoretical calculations that lead to tension with the SS \cite{Mueller:2011nm,Huber:2011wv}. However, it should be pointed out that even in these results, about $10\%$ of the $\beta$-decay branches are not known and can only be estimated.\footnote{If the uncertainty in $\beta$-transitions is bigger than usually assumed the anomaly could be statistically insignificant \cite{Hayes:2013wra}.} 
This involves using the ILL-experiment's measurement for the electron spectrum 
\cite{VonFeilitzsch:1982jw,Hahn:1989zr} as a reference point, hence the anomaly could be due to a systematic in that measurement. 
The other anomaly arises in the calibration of the GALLEX and SAGE experiments \cite{Abdurashitov:2005tb,Giunti:2012tn}. \\

These anomalies have resulted in a great interest in searches for eV-mass sterile neutrinos. Global fits to all data have e.g. been performed in \cite{Conrad:2012qt,Archidiacono:2013xxa,Kopp2013}. As discussed in detail in \cite{Kopp2013}, the situation at this stage is not clear. The accelerator anomalies come from appearance measurements (muon to electron). If they are caused by sterile neutrino oscillations, there should also be a $\nu_\mu$ disappearance, which is not observed.
This tension is present in the $3+1$ (three active and one eV mass sterile neutrino) model and remains in $3+2$ scenarios (three active plus two eV mass sterile neutrinos). It reduces a bit in a $1+3+1$ model with two sterile neutrinos (where one sterile neutrino is lighter than an active one). The reactor and gallium anomalies, which are due to disappearance, do not show such tension to other experiments.
They come, however, from $\nu_e$ disappearance, which is controlled by other parameters than $\nu_\mu$ disappearance.
Hence, the situation remains puzzling. A detailed list of various proposals for future experiments \cite{Agarwalla:2010zu,deGouvea:2011zz,Rubbia:2013ywa,Kose:2013zsa} can be found in \cite{Abazajian:2012ys}.

\subsection{Lepton flavour violation}
The most studied consequence of the lepton number violation due to $M_M$ is the neutrinoless double $\beta$-decay discussed in the following subsection \ref{0nubbsec}.
$M_M$ and $F$ also mediate flavour violation in the charged lepton sector \cite{Bilenky:1977du,Cheng:1980qt,Cheng:1980tp,Bilenky:1987ty,Casas:2001sr,Lavignac:2001vp,Pascoli:2003rq,Ibarra:2003up,Ibarra:2011xn,Lello:2012in,Abada:2012mc,Alonso:2012ji}, leading e.g. to muon decays $\mu\rightarrow e\gamma$ and unitarity violation of the PMNS matrix \cite{Antusch:2006vwa,Forero:2011pc}.
Searches for these processes in proposed experiments, such as COMET \cite{Hungerford:2009zz,Cui:2009zz} or Mu2e \cite{Abrams:2012er}, can help to constrain RH neutrino properties \cite{Alonso:2012ji}.

In the framework of (\ref{L}), with no other physics added to the SM, it is hard to observe these processes. 
For a generic choice of parameters the seesaw relation (\ref{activeneutrinomasses}) implies that either the suppressing scale $M_I$ is too heavy or the Yukawa couplings $F$ are too small. A class of models that offers relatively good chances to observe lepton flavour violation are those where the structure of $F$ and $M_M$ is such that an approximately conserved generalized lepton number can be defined \cite{Wyler:1982dd,Mohapatra:1986bd,GonzalezGarcia:1988rw,Branco:1988ex,Kersten:2007vk,Abada:2007ux,Shaposhnikov:2006nn,Gavela:2009cd,Blanchet:2009kk}. In this case the scales where lepton flavour violation and total lepton number violation occur can be rather different. In such models even RH neutrinos responsible for the baryon asymmetry of the universe may give a measurable contribution to $\mu\rightarrow e \gamma$ \cite{Blanchet:2009kk}.\footnote{Right handed neutrinos can also mediate lepton number violation in more exotic models. For instance, $\mu\rightarrow e\gamma$ may be measurable \cite{Duerr:2011ks} in models where the $\nu_R$ are composite objects that are hold together by strong couplings in a hidden sector \cite{ArkaniHamed:1998pf,McDonald:2010jm}.}

Models that embed (\ref{L}) into an extended framework, such as supersymmetric and grand unified theories, often contain interactions that violate lepton flavour or even the total lepton number. 
This includes left-right symmetric models, see e.g. \cite{Barry:2013xxa} and references therein.
It also applies to various ``bottom up'' models that e.g. include an extended scalar sector which couples to neutrinos.
Hence, while non-observation of $\mu\rightarrow e\gamma$ and other processes can constrain $N_I$-properties, the observation would not necessarily be a clear sign for their existence.

\subsection{Neutrinoless double $\beta$-decay}\label{0nubbsec}
If neutrinos are Majorana particles, neutrinoless double-$\beta$ decays ($0\nu\beta\beta$) are possible \cite{Mohapatra:1980yp} (see \cite{Rodejohann:2011mu,Bilenky:2012qi} for recent reviews).
\begin{figure}[!h]
\centering
\includegraphics[width=8cm]{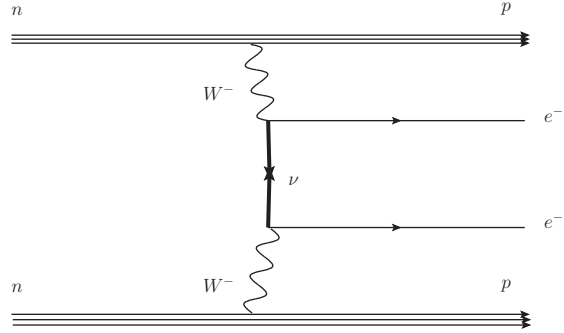}
\caption{Diagram for neutrinoless double beta decay. Here we have made the lepton number flow explicit by assigning arrows to fermion lines. The ``clashing arrows'' in the center of the diagram are allowed because neutrinos and antineutrinos are indistinguishable if they are Majorana particles. If some $N_I$ are light enough, they may also be exchanged instead of $\upnu_i$.
The amplitude for this process vanishes in the limit $M_M\rightarrow 0$.\label{0nubb}}
\end{figure}
Whether $0\nu\beta\beta$-decay occurs at observable rates depends on the Majorana mass matrix $M_M$. The lepton number violation is sufficient if at least one eigenvalue of $M_M$ is larger than the exchanged momentum $\sim 100$ MeV \cite{Blennow:2010th,Barry:2011fp,Mitra:2011qr}, see also \cite{Ibarra:2011xn,Mitra:2012qz}.
The $0\nu\beta\beta$-decay can be pictured as exchange of an electron neutrino that acts as its own antiparticle between the nuclei, see figure \ref{0nubb}.
The amplitude is given by the convolution of nuclear matrix elements with \begin{displaymath}
(U_\nu)_{e i}^2\frac{m_i^2}{p^2-m_i^2}+\Theta_{e I}^2\frac{m_I^2}{p^2-m_I^2}
.\end{displaymath}
Here $p^2=p_0^2-\textbf{p}^2$ is the exchanged momentum, with $|\textbf{p}|\sim 100$ MeV.
In the regimes $M_I\ll 100$ MeV and $M_I\gg 100$ MeV one can make analytical approximations and approximate the inverse half life time for $0\nu\beta\beta$-decays
\begin{equation}
\tau^{-1}\simeq G_{0\nu}\left|
\sum_i \mathcal{M}_i (U_\nu)_{e i}^2\frac{m_i}{m_e}
+ \sum_I^{M_I\ll 100 {\rm MeV}} \mathcal{M}_I \Theta_{e I}^2\frac{m_i}{m_e}
+ \sum_I^{M_I\gg 100 {\rm MeV}} \tilde{\mathcal{M}}_I \Theta_{e I}^2\frac{m_p}{m_i}\right|.
\end{equation}
Here $G_{0\nu}$ is the ``phase space factor'' (for instance $G_{0\nu} = 7.93 \times 10^{−15} {\rm yr}^{-1}$ for $^{76}$Ge \cite{Mitra:2011qr}). The dependence in the intermediate regime is more complicated \cite{Mitra:2011qr}.
The $\mathcal{M}_i$ and $\mathcal{M}_I$ are nuclear matrix elements corresponding to light active or sterile neutrino exchange. The $\tilde{\mathcal{M}}_I$ are matrix elements for the contribution from heavy sterile neutrinos, sometimes called ``direct contribution'' or ``contact term'' (the $\tilde{ }$ indicates the different normalization conventions for both regimes). Such heavy neutrinos are not really ``exchanged'' as propagating particles; because their mass is larger than the exchanged momentum, they can be integrated out and their contribution can be understood in terms of an effective operator as (\ref{majoranamassterm}).   
The nuclear matrix elements are a source of considerable uncertainty. 

Usually light neutrinos (active or sterile) with masses $\ll 100 $ MeV strongly dominate in the exchange \cite{LopezPavon:2012zg}. However, without any extra sources of $L$-violation beyond (\ref{L}), the existence of a mass state with $M_I>100$ MeV remains a necessary condition for observable rates of $0\nu\beta\beta$-decay 
even if the direct exchange of these heavy particles does not give a significant contribution to $\tau^{-1}$. 
When only light neutrinos are exchanged one can approximately factorize the dependencies on nuclear and neutrino physics and parametrize the latter in terms of the effective Majorana mass
\footnote{The effective Majorana mass is different from the kinetic mass $(\sum_i |\mathcal{U}_{e i}|^2 m_i^2 )^{1/2}$.}
\begin{equation}\label{meff}
m_{ee}=\left|
\sum_i (U_\nu)_{e i}^2m_i + \sum_{I}^{M_I\ll 100 {\rm MeV}}\Theta_{e I}^2M_I
\right|
.\end{equation}
For very light $N_I$ one has to go beyond the seesaw approximation (\ref{ThetaDef}) to calculate the active-sterile mixing matrix $\Theta$.

\begin{figure}
\begin{center}
\includegraphics*[width=10cm
]{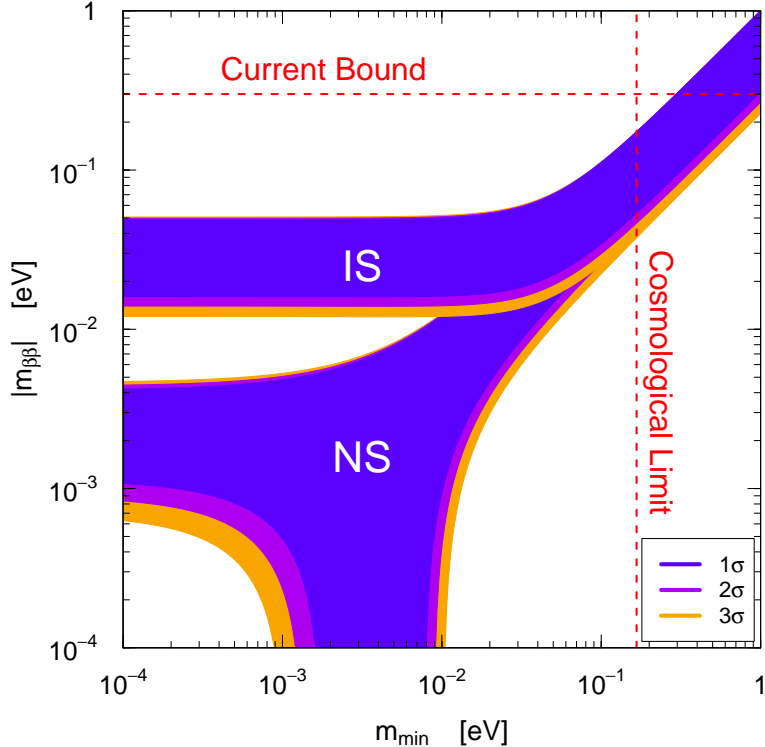}
\end{center}
\caption{Current bounds on $m_{ee}$ (here called $m_{\beta\beta}$) as a function of the smallest neutrino mass, as summarized in \cite{Bilenky:2012qi}. ``NS'' refers to normal hierarchy, ``IS'' to inverted hierarchy. The cosmological limit has tightened with the publication of the Planck results \cite{Ade:2013lta}.\label{0nubbfigure}
}
\end{figure}

Leaving aside some tuned special cases \cite{LopezPavon:2012zg}, the contribution from active neutrinos dominates in (type I) seesaw models \cite{Bezrukov:2005mx,Mitra:2011qr,Asaka:2011pb,Asaka:2013jfa}.
If all sterile neutrino masses are below $100$ MeV, they usually give a negative contribution, i.e. reduce the rate of $0\nu\beta\beta$-decays compared to the case where only active neutrinos are exchanged (which in any case is undetectably small if there is no $M_I$ above $100$ MeV), see \cite{Girardi:2013zra} for a recent discussion.  
The existence of light sterile neutrinos is strongly constrained by cosmology. 
In particular, if sterile neutrinos compose the observed DM (see section \ref{DMsection}), then the bounds on their mixing imply that they do not contribute \cite{Bezrukov:2005mx,Merle:2013ibc}. 
If only the active neutrino mass states $\upnu_i$  contribute to (\ref{meff}) and have an inverted mass hierarchy, then there exists a lower bound $m_{ee}>20$ meV, see figure \ref{0nubbfigure}. For normal hierarchy there is no such bound.
If also light sterile states $N_I$ contribute \cite{Mohapatra:1980yp,Bamert:1994qh,Benes:2005hn,Bezrukov:2005mx,Mitra:2011qr,Asaka:2011pb,Merle:2013ibc}  there is no lower bound for any hierarchy.

So far there is no clear observation of neutrinoless double-$\beta$ decays.
The only claim of a detection \cite{KlapdorKleingrothaus:2001ke,KlapdorKleingrothaus:2004wj} suggests $m_{ee}=0.32\pm0.03$ eV 
\cite{KlapdorKleingrothaus:2006ff}, but is disputed, as it appears to be in conflict with other observations \cite{Fogli:2008ig,Bergstrom:2012nx,Dev:2013vxa}. For instance, the EXO-200
experiment finds $m_{ee}<0.14 - 0.38$ eV \cite{deGouvea:2013zba}. 
Other experiments  so far only put upper bounds on the rate for this process, see figure \ref{0nubbfigure} (an overview is also given in \cite{Schwingenheuer:2012zs,deGouvea:2013zba}).

\subsection{Collider searches}\label{collidersec}
Depending on the masses $M_I$, there are different ways to look for $N_I$-signatures in collider experiments, see \cite{Atre:2009rg} for a review.

\paragraph{Intensity frontier} -
\begin{figure}[!h]
\centering
\includegraphics[width=12cm]{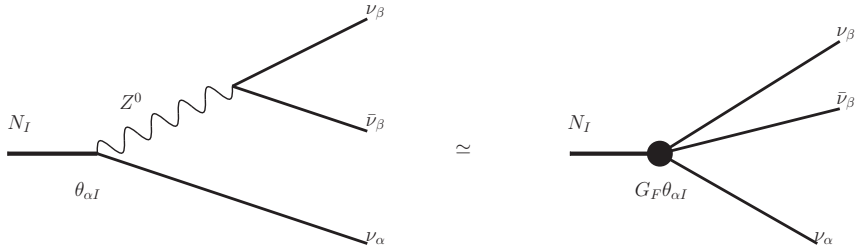}
\caption{Example for a contribution to $N_I$-decay at low energies. The $N_I$ itself can be produced in meson decays. Instead of the $\bar{\nu}_\beta$-$\nu_\beta$ pair there can also be charged lepton-antilepton pairs or quark-antiquark pairs (which hardronize) in the final state if this in kinematically allowed.\label{colliderdecay}}
\end{figure}
If kinematically possible, the sterile neutrinos $N_I$ participate in all processes that involve active neutrinos, 
but with a probability that is suppressed by the small mixings $U_\alpha^2$. 
This makes it possible to produce them in meson decays for $M_I\lesssim$ 
a few GeV \cite{Gorbunov:2007ak,Ruchayskiy:2011aa,ESbau,Lello:2012gi,Asaka:2012bb,Gninenko:2013tk}, see also \cite{Quintero:2011yh,Helo:2011yg,Helo:2010cw}\footnote{Sterile neutrinos may also be produced in the decay of tauons \cite{Castro:2012gi,Helo:2011yg,Helo:2010cw}.}. 
One can distinguish two ways to look for $N_I$.
First, they could be seen as missing energy in the meson decays that produce them; in two body decays this allows to determine their mass and (via the branching ratio) their mixing. 
Second, the subsequent decay of the $N_I$ shown in figure \ref{colliderdecay} may also be observed (e.g. as ``${\rm nothing}\rightarrow {\rm leptons}$'' process) if one places a detector along the beamline. For some parameter choices, it can even be possible to observe both events in the same detector \cite{Achard:2001qw}.
Several experiments of this type have been set up in the past \cite{Yamazaki:1984sj,Daum:2000ac,Bernardi:1985ny,Bernardi:1987ek,Vaitaitis:1999wq,Bergsma:1985is,Astier:2001ck,CooperSarkar:1985nh}, 
in particular CERN PS191 \cite{Bernardi:1985ny,Bernardi:1987ek},
NuTeV \cite{Vaitaitis:1999wq}, CHARM \cite{Bergsma:1985is}, NOMAD
\cite{Astier:2001ck} and WA66 \cite{CooperSarkar:1985nh}. The bounds derived from these for $n=2$ are shown in figure \ref{BAUnuMSM}. 
They are valid under the assumption that $N_I$ have no interactions other than those in (\ref{L}) below the electroweak scale. If the mass $M_I$ is larger than a few GeV, the $N_I$ are too heavy to be produced efficiently (D or B meson decays are not possible) and it is unlikely that direct searches in the near future can find them. If they are, on the other hand, lighter than $\sim$MeV, then their Yukawa coupling $F$ must be very small due to (\ref{activeneutrinomasses}) and the branching ration is very small\footnote{It has been suggested that, if $M_M$ is generated by spontaneous symmetry breaking below the electroweak scale, it may be possible to detect even lighter (keV) sterile neutrinos \cite{Shoemaker:2010fg}.}.

\paragraph{High energy frontier} -
Sterile neutrinos with masses up to a few TeV in principle are within reach of the LHC or other future high energy experiments \cite{Datta:1993nm,Almeida:1997em,Almeida:2000pz,Panella:2001wq,Cheung:2004xm,Han:2006ip,Shaposhnikov:2006nn,Kersten:2007vk,delAguila:2007em,Huitu:2008gf,Atre:2009rg,Kovalenko:2009td,Matsumoto:2010zg,Aoki:2010tf,Das:2012ze,Baldes:2013eva,Shoemaker:2010fg,Dinh:2013vya,Dev:2013wba}.
A promising signal in different scenarios are $N_I$-decays that involve same sign dileptons (two leptons of the same charge) in the final state. 
However, the Yukawa coupling in (\ref{L}) have to be tiny due to the seesaw relation (\ref{activeneutrinomasses}) and $\mu\rightarrow e\gamma$ data \cite{Ibarra:2011xn}.
If the $\nu_R$ only interact via these Yukawa couplings $F$, the branching ratio is usually too small \cite{Kovalenko:2009td};
a detection is only possible if $F$ has a particular structure that leads to cancellations in the contributions of different elements to $m_\nu$ \cite{Shaposhnikov:2006nn,Kersten:2007vk}, hence allows for larger individual entries of $F$. 
Such cancellation can occur if the total lepton number $L$ is approximately conserved \cite{Kersten:2007vk}.If the $\nu_R$ only interact via these Yukawa couplings $F$, the branching ratio is usually too small \cite{Kovalenko:2009td};
a detection is only possible if $F$ has a particular structure that leads to cancellations in the contributions of different elements to $m_\nu$ \cite{Shaposhnikov:2006nn,Kersten:2007vk}, hence allows for larger individual entries of $F$. 
Such cancellation can occur if the total lepton number $L$ (summed over active and sterile flavours) is approximately conserved \cite{Kersten:2007vk}.

The perspectives are much better if the $\nu_R$ have interactions in addition to those in (\ref{L}). 
For instance, it may be possible to study RH neutrinos in high energy collisions if the scalar sector is extended in comparison to the SM \cite{Cheung:2004xm,Shoemaker:2010fg,Huitu:2008gf,Aoki:2010tf}, in models where neutrino masses are generated via the inverse seesaw mechanism \cite{Dias:2011sq,Das:2012ze,Baldes:2013eva} or if spacetime has more than four dimensions \cite{Haba:2009sd,Matsumoto:2010zg}.
In  left - right symmetric models \cite{Pati:1974yy,Mohapatra:1974hk,Senjanovic:1975rk}
the $\nu_R$ are charged under a right chiral $SU(2)$ gauge symmetry, which can be broken near the TeV sale.
One promising possibility to test such models is to search for the associated gauge bosons \cite{Keung:1983uu},
but also the properties of the $N_I$ themselves can be probed \cite{Maiezza:2010ic,AguilarSaavedra:2012fu,Nemevsek:2012iq,Chen:2013foz}. Current bounds are given in \cite{CMS:2012zv,ATLAS:2012ak}, the right $W$-boson should be heavier than about $2.5$ TeV \cite{Barry:2013xxa}.
In supersymmetric (SUSY) theories the seesaw mechanism may also be studied indirectly at colliders \cite{Deppisch:2003wt}; if their SUSY-partner (sneutrino) is the lightest sparticle, then observations of the decays of heavier SUSY-particles into sterile sneutrinos can constrain their properties \cite{Cheung:2011ph}. 
It has been suggested that $N_I$ can give a contribution to the Higgs mass, see e.g. \cite{Heinemeyer:2010eg,Gogoladze:2012jp,Wang:2013jya}, though this is disputed \cite{Draper:2013ava}.
Finally, LHC searches can also be promising for models that couple $N_I$ to $Z'$ gauge bosons \cite{King:2004cx,Anchordoqui:2012qu}, including those where $N_I$ are responsible for leptogenesis \cite{Blanchet:2009bu}.

\subsection{Direct dark matter searches }
Sterile neutrinos are a promising DM candidate, which we discuss in detail in  section \ref{DMsection}.
If they compose all DM, then bounds from X-ray and structure formation imply that their Yukawa couplings are so small that they cannot contribute to the observed neutrino oscillations \cite{Boyarsky:2009ix}\footnote{It has been shown that the bounds on DM sterile neutrinos can constrain the active neutrino mass spectrum if additional assumptions are made \cite{Merle:2012xq}.}. This means that, if one at the same time requires RH neutrinos to explain the two observed neutrino mass differences, there must be at least three of them. 
The small coupling also makes it practically impossible to directly observe these particles in collider experiments due to the tiny branching ratio for their production. 
They could at least in principle be found in direct detection experiments that look for interactions of DM sterile neutrinos from the interstellar medium  with atomic nuclei in the laboratory, and some experiments have been suggested \cite{Li:2010vy,Liao:2010yx}. However, such detection would be extremely challenging (most likely impossible) due to the small mixing angle and the background from solar and stellar active neutrinos \cite{Ando:2010ye}.

\subsection{Other constraints}
Several other ways to constrain $N_I$ properties have been suggested. 
If they have eV masses as suggested by the oscillation anomalies (ii), they should affect $\beta$-decays \cite{deGouvea:2005er,deGouvea:2006gz,Formaggio:2011jg,Kraus:2012he,Esmaili:2012vg}, which has not been observed.
They may also leave traces in neutrino telescopes \cite{Asaka:2012hc} like IceCube \cite{Nunokawa:2003ep,Choubey:2007ji,Esmaili:2012nz,Esmaili:2013cja,Esmaili:2013vza} or detectors for direct DM searches \cite{Gutlein:2010tq}.
The effect of keV sterile neutrinos on nuclear decays has been studied in \cite{Bezrukov:2006cy,deVega:2011xh}, a detection at the current stage seems very unlikely due to the background.
In \cite{Akhmedov:2013hec} it was found that sterile neutrinos with TeV masses can improve the fits to electroweak precision data.\footnote{Electroweak precision data also allows to constrain models of right handed neutrinos with electroweak scale masses beyond (\ref{L}), see e.g. \cite{Hoang:2013jfa}.}

\section{Thermal history of the universe}\label{ThermalHistory}
If RH neutrinos $\nu_R$ with $F\neq 0$ exist, they are necessarily produced thermally in the early universe. At temperatures above the electroweak scale $T_{EW}\sim 140$ GeV [assuming a Higgs mass of $\sim 125$ GeV \cite{:2012gk,:2012gu}], Higgs particles are present in the plasma. This allows for $N_I$ production
 as long as the temperature is high enough [$T\gtrsim M_I(T)$, where $M_I(T)$ is an effective mass in the plasma]. 
The relevant processes are decays and inverse decays $N_I\leftrightarrow \Phi l_\alpha$ (or $\Phi\leftrightarrow N_I l_\alpha$, depending on the effective masses of the (quasi)particles in the plasma \cite{Giudice:2003jh,Garbrecht:2010sz,Anisimov:2010gy}) and scatterings (such as $\bar{t}t\leftrightarrow N_I l_\alpha$).
If the masses $M_N$ are below the electroweak scale, $N_I$ are produced at $T<T_{EW}$ via active-sterile mixing [The same process can be viewed as a $\nu_L$-$\nu_R$ oscillation in the flavour basis used in (\ref{L})]. 
In addition to that, there may be other production mechanisms if $\nu_R$ have additional interactions with a hidden sector or extended Higgs sector, couple directly to the inflaton or are charged under a gauge symmetry that is broken above the electroweak scale.
The presence of $N_I$ in the plasma can have different effects in the early universe, which we will summarize in the following.

All astronomical observations to date are in rather good agreement with the $\Lambda$CDM model of cosmology\footnote{A brief review of the $\Lambda$CDM model is given in \cite{Beringer:1900zz}, see e.g. \cite{Mukhanov:2005sc,Weinberg:2008zzc} for detailed introductions.}, sometimes dubbed the ``Concordance Model'' or ``Standard Model of Cosmology'' in analogy to the SM.  
The most important cosmological parameters  in the context of RH neutrinos are the fractions of the total energy budget of the observable universe from baryons ($\Omega_B\sim 0.049$) and dark matter ($\Omega_{DM}\sim 0.265$ without active neutrinos) \cite{Ade:2013lta}, as well as the {\it effective number of neutrino species} $N_{\rm eff}$ in the radiation dominated epoch\footnote{``Radiation dominated epoch'' refers to the period in the universe's history when more energy was stored in relativistic than in nonrelativistic degrees of freedom. }. The latter can be understood as a measure for the expansion history of the universe.
The (Hubble) rate of the universe's expansion is given by
\begin{equation}\label{hubble}
H^2=\frac{8\pi}{3}G\rho
,\end{equation}
where $G=M_P^{-2}$ is Newton's constant and $\rho$ the energy density of the universe. 
The contribution of ``known neutrinos plus unknown physics'' to $\rho$ is usually parametrized as $N_{\rm eff}\times\rho_\nu$, where $\rho_\nu$ is the contribution from one ultrarelativistic species and $N_{\rm eff}$ is the {\it effective number of neutrino species}. 
It can be identified with the actual number of neutrino species if all neutrinos are effectively massless and there is no other ``new physics''. 
It is common to parametrize
\begin{equation}\label{NeffParam}
\rho_\gamma + \rho_{\rm neutrinos}+[{\rm new \ physics \ effects}] \equiv\rho_\gamma+N_{\rm eff}\rho_\nu=\frac{\pi^2}{15}T_\gamma^4\left[1+N_{\rm eff}\frac{7}{8}\left(\frac{4}{11}\right)^{4/3}\right],
\end{equation}
where $T_\gamma$ and $\rho_\gamma$ are temperature and energy density of photons and $\rho_{\rm neutrinos}$ is the energy density of the SM-neutrinos.
The second equality holds only for $T< 0.2$ MeV, i.e. after electrons and positrons annihilated. This is the temperature regime where $N_{\rm eff}$ is practically tested by observations.
In the standard scenario $N_{\rm eff}=3$ during BBN and $N_{\rm eff}=3.046$ at the time of photon decoupling.\footnote{The deviation from 3 parametrises a deviation from the equilibrium distribution of neutrinos caused by $e^\pm$ annihilation \cite{Mangano:2005cc}.}
This assumes that each neutrino flavour has two internal degrees of freedom. In the SM (without neutrino masses), these are particles and antiparticles. In contrast to charged leptons and quarks, $\nu_L$-states do not have an independent spin degree of freedom in the SM: For massless particles, the helicity must be equal to the chirality; hence neutrinos (i.e. one-particle states $|\nu_L\rangle$) must have left helicity and antineutrinos have right helicity.

If additional particles were present in the early universe, this would lead to a larger value of $N_{\rm eff}$ \cite{Shvartsman:1969mm,Steigman:1977kc}.
Each $\nu_{R,I}$ adds two neutrino degrees of freedom. If $n=3$ and all of these were thermalized, $N_{\rm eff}$ would be $6$. 
However, if neutrinos are Dirac fields, then the Yukawa couplings $F$ must be tiny to explain the smallness of the neutrino masses. Then the $\nu_R$ degrees of freedom do not get populated significantly by thermal production. In other words, the helicity changing processes are so suppressed at $T\gg m_i$ that practically all neutrinos, which are produced by the weak interaction, have left helicity. This argument of course assumes that $\nu_R$ have no other couplings than $F$ that could contribute to thermal production. Hence, the fact that we observe $N_{\rm eff}\simeq 3-4$ strongly constrains models of Dirac neutrinos in which $\nu_R$ are charged under some gauge group at high energies or otherwise produced in the early universe.   
If neutrinos are Majorana fields and $M_M\gtrsim 100$ MeV, then the heavy particles $N_I$ have decoupled and decayed long before big bang nucleosynthesis (BBN) and do not affect light element abundances or the CMB. Lighter sterile neutrinos can be long lived enough to contribute to $N_{\rm eff}$ during BBN and afterwards. If they are in thermal equilibrium and ultrarelativistic, then each $N_I$ increases $N_{\rm eff}$ by one. If their abundance is below equilibrium or their mass not negligible, they contribute less.
 
Since (\ref{NeffParam}) assumes that all neutrinos are massless and in thermal and chemical equilibrium, any deviation of the momentum distribution from $(e^{\textbf{p}/T}+1)^{-1}$ [such as a chemical potential, a non-negligible mass\footnote{
See \cite{Dolgov:2002wy,Lesgourgues:2012uu} for a detailed discussion of the effects of neutrino masses in cosmology.
For eV masses the deviation is small in the early universe.
An early discussion in Russian can be found in \cite{Zeldovich:1981wf}.
} or a nonequilibrium distribution] leads to non-integer contribution to $N_{\rm eff}$; this applies to both, active and sterile neutrinos.
Furthermore, in the SM the ratio of the temperatures of the neutrino and photon backgrounds at the time of photon decoupling is $(4/11)^{1/3}$. If this ratio is different (e.g. because some decaying particle injects energy into either of them), then this would lead to deviations from $N_{\rm eff}=3.046$ even if there are only three neutrinos \cite{Boehm:2012gr,Steigman:2013yua,Boehm:2013jpa}.

\subsection{A brief history of the universe}\label{briefhistory}
Observations of the CMB show that the universe was homogeneous and isotropic to one part in $\sim 100000$ at redshift $z\sim1100$, when photons decoupled from the primordial plasma \cite{Hinshaw:2012fq}. This is puzzling, as the radiation we receive from different directions originates from regions that were causally disconnected at that time if the universe only contained radiation and matter (``horizon problem''). Furthermore, the inferred overall spatial curvature is zero or very small \cite{Hinshaw:2012fq}, which means that it was extremely close to zero at earlier times (``flatness problem''). Both problems can be understood as the result of cosmic inflation \cite{Starobinsky:1980te,Guth:1980zm}, a phase of accelerated expansion in the universe's very early history.
Inflation can also explain the small density perturbations that served as seeds for structure formation in the universe as quantum fluctuations that were ``streched out'' to macroscopic scales by the rapid expansion, and predict the correct properties of their spectrum. However, while inflation is an excellent model for cosmology, we do not know much about the fundamental physics mechanism that drove it.
 
If inflation was driven by the potential energy of an inflaton field, then the quantum fluctuations of this field lead to small perturbations in the (otherwise homogeneous) gravitational potential. After the inflaton dissipates its energy into relativistic particles during {\it cosmic reheating} \cite{Dolgov:1982th,Kofman:1994rk,Kofman:1997yn},\footnote{See \cite{Drewes:2013iaa} for a recent discussion.} these lead to density fluctuations in the primordial plasma, which manifest in temperature fluctuations in the CMB and form the seeds for the formation of structures in the universe \cite{Mukhanov:1981xt}. 
The power spectrum of CMB fluctuations is in very good agreement with the above hypothesis \cite{Hinshaw:2012fq}. 
After reheating, $N_I$ can affect the thermal history of the expanding universe in several ways. 
\paragraph{Baryogenesis via leptogenesis} ($T>T_{EW}$) - There is good evidence \cite{Canetti:2012zc} that the observed $\Omega_B$ is the thermal relic of a small matter-antimatter asymmetry of order $\sim 10^{-10}$ in the early universe (BAU), which survived after all other particles and antiparticles annihilated into CMB photons and neutrinos and is reflected in today's baryon to photon ratio\footnote{The relation between these parameters is given by $\eta_B\simeq 2.739\cdot10^{-8}h^2\Omega_B$, where h parametrises the the Hubble rate $H_0 = 100h (km/s)/Mpc$.} $\eta_B\sim 10^{-10}$. If this asymmetry was produced by {\it leptogenesis}, then the observed $\Omega_B$ allows to constrain $F$ and $M_M$, cf. section \ref{Leptogenesis}.
\paragraph{Dark Matter production} ($T\sim 100$ MeV if produced by mixing) - If DM consists of thermally produced $N_I$, then there are various different bounds on their properties, which we discuss in section \ref{DMsection}.
\paragraph{Late time leptogenesis} ($T_{EW}>T>$ few MeV) - Sphaleron processes \cite{Kuzmin:1985mm} are the only source of baryon number violation\cite{Adler:1969gk,Bell:1969ts,'tHooft:1976up}  in the early universe in (\ref{L}). They freeze out\footnote{A reaction ``freezes out'' when the temperature dependent rate at which it occurs falls below the Hubble rate $H$ of cosmic expansion. This essentially means that the density of the primordial plasma has become so low that the rate at which the participating particles meet is negligible. 
A particle freezes out when the last process (other than decay) that changes its comoving number density freezes out.} at $T\sim T_{EW}$, below which baryon number is conserved. The production of a lepton asymmetry can, however, continue afterwards if some $N_I$ are out of equilibrium (e.g. during their freezeout and decay) because $F$ and $M_M$ violate flavoured and total lepton number, respectively. The generated asymmetries can be much bigger than $\eta_B$ and differ in each flavour \cite{Canetti:2012vf,Canetti:2012kh}. 
A weak constraint on the asymmetry  may be derived from its effect on hadronisation at $T\sim 200$ MeV \cite{Schwarz:2009ii}.
Stronger constraints can be derived from BBN, see below.
Late time asymmetries can be very important if there are long lived sterile neutrinos because they can amplify or suppress their production rate.
For keV-mass sterile neutrinos an amplification of their production rate \cite{Laine:2008pg,Kishimoto:2008ic} can be so efficient that they are abundant enough to constitute all DM \cite{Canetti:2012vf}. 
A suppression of the thermal production rate, on the other hand, can help to ease the tension between cosmological and laboratory hints for eV-mass sterile neutrinos, see section \ref{RHasDR}.
In return, light sterile neutrinos can also amplify an active neutrino asymmetry \cite{Foot:1995qk}.

\paragraph{Neutrino freezeout} ($T\sim 1.1$ MeV for active neutrinos) - The freezeout of active neutrinos leads to a cosmic background of relativistic neutrinos, analogue to the CMB. They affect the expansion by their energy density $\rho_{\rm neutrinos}$.
The neutrino background may carry a lepton asymmetry that is orders of magnitude larger than the baryon asymmetry $\sim \eta_B\sim 10^{-10}$.
The main constrains on such asymmetry come from BBN \cite{Kang:1991xa}.
In the SS, active neutrino oscillations tend to make the asymmetries in individual flavours equal\footnote{This is
often referred to as ``flavour equilibration'', though the process occurs close 
to the neutrino freezeout and neutrinos may not reach thermal equilibrium. 
This means that the lepton asymmetry cannot be translated into a chemical potential in the strict sense.}
\cite{Dolgov:2002ab,Wong:2002fa,Abazajian:2002qx}. 
How complete this equilibration is depends on the mixing angle $\uptheta_{13}$. The measured value for $\uptheta_{13}$ suggests a high degree of equilibration \cite{Mangano:2011ip}.
Lepton asymmetries in the neutrino background can be constrained due to their effect on the momentum distribution, which changes the relation between temperature and energy density (\ref{NeffParam}) and mimics an $N_{\rm eff}\neq 3$. The bounds on $N_{\rm eff}$ from BBN (see below) allow to constrain the asymmetry to roughly $|\eta_L| \lesssim 0.1$ \cite{Mangano:2011ip}, where $\eta_L$ is defined analogously to (\ref{etadef}); see also \cite{Serpico:2005bc,Pastor:2008ti,Mangano:2010ei,Castorina:2012md}.

If sterile neutrinos are relativistic, then they can form a similar background and contribute to $N_{\rm eff}$. 
For an active-sterile mass splitting $\Delta m^2<1.3\times 10^{-7} {\rm eV}^2$ active-sterile oscillations are effective only after the active neutrino freezeout, then they simply distort the momentum distributions \cite{Lesgourgues:2012uu}. For much larger splittings they can be produced efficiently via their mixing at $T>1$ MeV.
The preference for $N_{\rm eff}>3$ in different cosmological data sets (see below) can be interpreted as a hint for eV mass sterile neutrinos as DR, see section \ref{RHasDR}.
However, the constraints on $N_{\rm eff}$ from BBN, see below, imply that such background must either have frozen out considerably earlier if it ever was in equilibrium or never thermalized (e.g. because the production was suppressed by some mechanism).

\paragraph{Big bang nucleosynthesis} ($T\lesssim 100$ keV) - There was a brief period in the early universe
during which the temperature was low enough for nuclei heavier than hydrogen (H) 
to exist and still high enough for thermonuclear reactions to occur.  
During this period most of the deuterium (D), helium ($^3$He,
$^4$He) and lithium (mainly $^7$Li) in the universe were formed \cite{Gamow:1946eb}, see e.g. chapter 22 in \cite{Beringer:1900zz} for a review. These light elements, in particular $^4$He, 
make up the vast majority of all nuclei other than H in the universe. 
Sterile neutrinos can affect this big bang nucleosynthesis (BBN) in different ways, depending on their mass. 
$N_I$ with masses far above the electroweak scale have no effect on BBN, as they have decayed long before.
If the masses are in the GeV to TeV range, these particles can be long lived enough that the entropy released during their decay affects BBN or the thermal history afterwards. The good agreement between BBN calculations and the observed H and He abundances implies that, if sterile neutrinos with GeV$\lesssim M_I \lesssim$TeV exist, they must have decayed sufficiently long before BBN. The resulting bounds \cite{Ruchayskiy:2012si,Canetti:2012vf,Canetti:2012kh} in the mass-mixing plane are plotted in figure \ref{BAUnuMSM}.
If DM is composed of keV-mass sterile neutrinos, these would have no visible effect on BBN because their number density is too low to affect the expansion history in that era.
Light sterile neutrinos (with eV masses) would, on the other hand, significantly contribute to $N_{\rm eff}$ \cite{Barbieri:1989ti,Kainulainen:1990ds,Dolgov:2002wy} as additional radiation and increase the rate of expansion of the universe via (\ref{hubble}) and (\ref{NeffParam}). 
This, for instance, determines the precise moment of neutron freeze-out ($T\sim 0.8$ MeV), which roughly occurs when the expansion rate (\ref{hubble})  equals the rate for the reaction $n+e^+\leftrightarrow p+\bar{\nu}_e$.
It also affects the amount of time that passes until the formation of elements ($T\sim 10-100$ keV), during which neutrons decay. Both determine the number of neutrons available for fusions \cite{Lesgourgues:2006nd,Steigman:2012ve}. 
However, a change in the expansion rate is not specific to light neutrinos, 
hence the resulting change in the value for $N_{\rm eff}$ extracted from fits to light element abundances does not actually "measure" the number of light neutrinos or other particles (see e.g. \cite{Brust:2013ova} for a summary of examples).
It simply parametrizes any deviation from the SM prediction.
Sterile neutrinos can also affect BBN if they cause deviations of the neutrino momentum distributions from a Fermi-Dirac spectrum, e.g. by active-sterile oscillations or by inducing chemical potentials \cite{Khlopov:1981nq}. On one hand, this slightly modifies the relation (\ref{NeffParam}) between $T$ and $\rho_\nu$, which again affects the rate of expansion. More importantly, the He abundance is directly affected by a distortion of the $\nu_e$ spectrum \cite{Dolgov:2002wy}. D is less affected, hence provides a more direct probe of the expansion rate \cite{DiBari:2001ua}.

BBN depends on $\Omega_B$, $N_{\rm eff}$ and the lepton asymmetries $Y_\alpha$.
In the SM, where $N_{\rm eff}=3$ is fixed and $Y_\alpha=0$, the only free parameter during BBN is $\Omega_B$ (or $\eta_B$), 
and there is a rather impressive agreement between theoretical predictions and the value obtained 
from the light elements, see e.g. \cite{Steigman:2010zz}.
If one treats $N_{\rm eff}$ as a free parameter using (\ref{NeffParam}), a deviation from $3$ mainly reflects in the $^4$He abundance.
The values for $N_{\rm eff}$ obtained from BBN alone (see e.g. \cite{Steigman:2010zz,Mangano:2011ar,Pettini:2012ph}, cf. also \cite{Nollett:2011aa,Hamann:2011ge}) show a slight preference for $N_{\rm eff}>3$, but are consistent with $N_{\rm eff}=3$. 
Different interpretation of the data are discussed in \cite{Steigman:2012ve}: 
In absence of significant lepton asymmetry, BBN alone yields $N_{\rm eff}=3.71^{+0.47}_{-0.45}$. 
If one fixes the He abundance to the value inferred from the CMB. this tightens to $N_{\rm eff}=3.53^{+0.66}_{-0.63}$ ($N_{\rm eff}=3.22\pm0.55$ when using the D abundance measured in \cite{Pettini:2012ph} alone). The likelihood functions given in \cite{Mangano:2011ar} show a preference for $N_{\rm eff}>0$, but clearly constrain $N_{\rm eff}<4$.
Hence, sterile neutrinos with eV masses, as motivated by the oscillation anomalies (ii), can only be made consistent with BBN if they are not thermalized.
This strongly constrains models of light sterile neutrinos unless some mechanism suppresses their production. 
If DM is composed of sterile neutrinos with keV masses (see section \ref{DMsection}), these do not affect $N_{\rm eff}$ because their number density is well below the equilibrium value (their number density can simply be obtained by putting today's $\Omega_B$ in relation to their mass). Hence, they are unaffected by these bounds. 
\paragraph{Matter - radiation equality} ($T\sim 0.8$ eV) -
As the universe expands, the matter density is diluted as $\propto a^3$, where $a$ is the scale factor, due to the increasing physical volume. The radiation energy density is diluted faster ($\propto a^4$) due to the stretching of the wavelengths, hence an initially radiation dominated universe becomes matter dominated at the point of {\it matter-radiation equality}.  
In cosmology, one refers to all relativistic degrees of freedom (particle energy dominated by momentum) as ``radiation'', while nonrelativistic degrees of freedom (particle energy dominated by mass)  are ``matter''.
A first principles derivation of the kinetic equation in the expanding universe that covers both regimes is given in \cite{Drewes:2012qw}.
Due to Hubble expansion, constituents of the primordial plasma change their identity from ``radiation''  to ``matter'' when $T$ 
(more precisely: their average momentum) falls below their mass, which has to be taken into account when determining the point of equality.  
If sterile neutrinos are relativistic near $T\sim 1$ eV, they change the temperature of matter radiation equality. 
This is crucial for the growth of density perturbations in the primordial plasma, see following paragraph.
If they become nonrelativistic just around this time, their equation of state is neither that of ``radiation'' nor ``matter'' in the intermediate regime. The effect of the time dependence of the equation of state has e.g. been studied in \cite{Wei:2013et}.

\paragraph{Photon decoupling} ($T\sim 0.25$ eV) - The universe becomes transparent when the temperature is so low that photons cannot dissociate H atoms any more\footnote{The temperature at which this happens is much lower than the H binding energy because of the small baryon to photon ratio $\eta_B$, which implies that there are enough photons in the high energy tail of the Bose-Einstein distribution to dissociate the atoms.}. While photons previously scattered frequently with free electrons, the cross section with neutral atoms is so small that the average photon has not interacted with matter ever since. The cosmic microwave background (CMB) formed by these primordial photons allows us to observe the universe at a very early stage (redshift $z \sim 1100$). 
The CMB contains an enormous amount of information and is, altogether, one of the most impressive confirmations of the $\Lambda$CDM model. The vast amount of data also makes it an excellent tool to look for hints of physics beyond $\Lambda$CDM and the SM. 
The small temperature fluctuations (of relative order $\delta T/T\sim 10^{-5}$) in the CMB yield the earliest probe of structure in the universe. 
They were generated by acoustic oscillations of the coupled baryon-photon plasma 
in the primordial gravitational potential wells (DM also falls into these wells, but does not feel the radiation pressure and does not oscillate).

The fluctuations can be decomposed into modes with wave numbers $k$, their spatial extension is characterized by the inverse of $k$. When $1/k$ is larger than the causal horizon, the mode remains ``frozen''. 
After the end of inflation, density fluctuations successively ``(re)enter the horizon''; that is, the causal horizon becomes bigger than their spatial extension. Then their amplitude starts to grow due to gravitational infall.
The evolution in the radiation dominated era is governed by the competition between gravity and the radiation pressure (which depends on $N_{\rm eff}$); this competition leads to {\it baryon acoustic oscillations} (BAO) of the plasma.   
The modes oscillate under the action of both forces until they decouple (when the baryons stop feeling the pressure) and start to collapse under the action of gravity, eventually forming the the structures that we observe in the universe. The observed peaks in the CMB power spectrum 
can be identified with multipoles $l$ in the multipole expansion of the CMB fluctuations that
correspond to the modes $k$ which reached the maximal elongation at the moment of decoupling. The first peak corresponds to the smallest $k$ that reached maximal compression by the time of decoupling and so on.  
The evolution of perturbations can be studied quantitatively by a coupled set of Boltzmann and Einstein equations.
Before photon decoupling, the density perturbations are small and $\delta_k\rho/\rho\ll 1$\footnote{Here $\delta\rho_k/\rho$ collectively refers to the contrasts in the densities of photons, baryons, neutrinos and metric degrees of freedom, which of course have to be studied independently.} can be used as as an expansion parameter in calculations. 
 
The power spectrum of temperature fluctuations can be affected by RH neutrinos in different ways \cite{Bashinsky:2003tk,Lesgourgues:2012uu,Abazajian:2012ys}. 
If they are nonrelativistic, they act as DM. Then they only affect the growth of perturbations via their free streaming length (the DM sterile neutrinos discussed in section \ref{DMsection} are relativistic at freezeout, but become nonrelativistic between BBN and the decoupling of photons). 
If they are relativistic (as e.g. the eV-mass-$N_I$ motivated by oscillation anomalies), 
they act as radiation and modify the power spectrum  in several ways \cite{Silk:1967kq,Hu:1994uz,Hu:1995fqa,Hu:1995en,Hu:1998tk,Bowen:2001in,Bashinsky:2003tk}. 
Their contribution to $N_{\rm eff}>3$ increases the rate of expansion (\ref{hubble}), hence reduces the comoving sound horizon, which moves the peaks in the power spectrum to higher multipoles. It also enhances the hight of the first two peaks via the integrated Sachs-Wolfe effect \cite{Hu:1994uz}.
Light sterile neutrinos are essentially collisionless and can, in contrast to baryons and photons, not be treated as a perfect fluid. Their anisotropic stress affects the gravitational potential via Einstein's equations. Since this effect is more relevant in the radiation dominated era, it affects modes that enter the horizon before and after matter-radiation equality in a different way, which reflects in a change in the relative height of high and low $l$ peaks \cite{Hu:1995en}; it suppresses the modes $l\gtrsim 200$ and also shifts the positions of the peaks \cite{Bashinsky:2003tk}. Finally, sterile neutrinos contribute to the damping of small scales via their free streaming and affect the high-$l$ modes \cite{Silk:1967kq,Bashinsky:2003tk,Hou:2011ec}. 
All these effects can be parametrized in terms of $N_{\rm eff}$.\footnote{Note that individually they are degenerate with other effects, cf. e.g. the discussion in \cite{Abazajian:2012ys} and references therein, and only a global fit allows to reliably determine $N_{\rm eff}$.} 
\begin{figure}[!h]
\centering
\includegraphics[width=14cm]{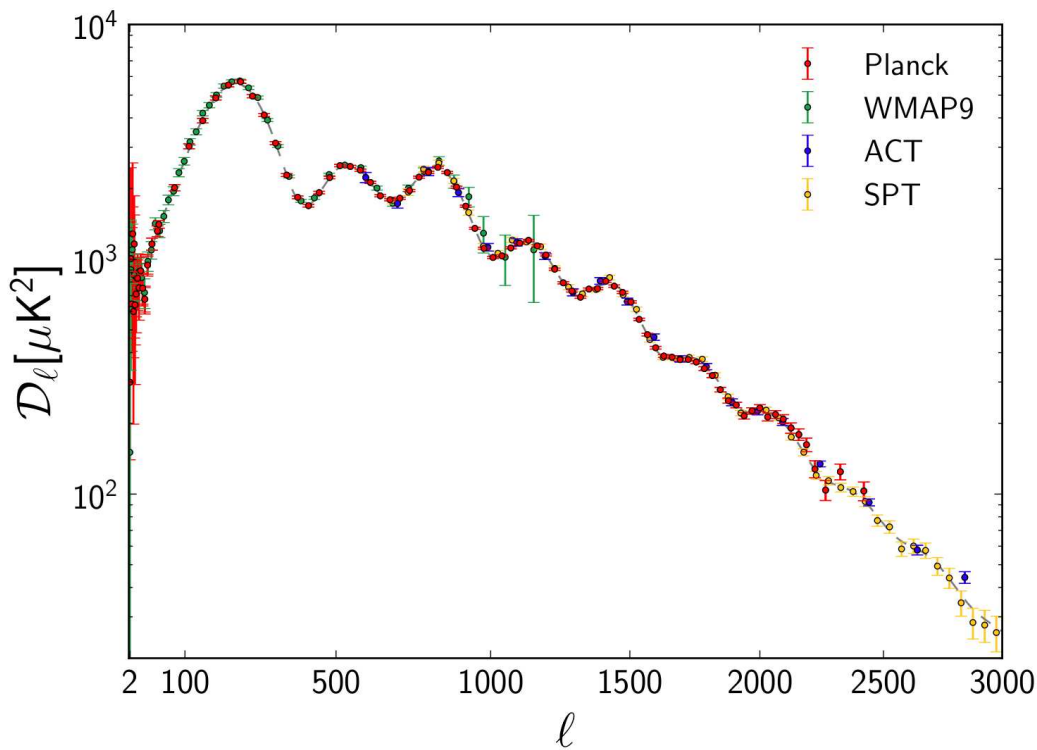} 
\caption{CMB power spectrum as measured by WMAP \cite{Hinshaw:2012fq}, SPT \cite{Keisler:2011aw}, ACT \cite{Dunkley:2010ge} and Planck \cite{Ade:2013xsa}; plot taken from \cite{Ade:2013xsa}. 
\label{cmbpowerspectra}}
\end{figure}

The CMB power spectrum has been studied with different instruments, including the Wilkinson Microwave Anisotropy Probe (WMAP) satellite, South Pole Telescope (SPT), Atacama Cosmology Telescope (ACT) and Planck satellite, see figure \ref{cmbpowerspectra}.
A few years ago these datasets consistently preferred $N_{\rm eff}>3$. 
In order to constrain $N_{\rm eff}$, it is crucial to have information from high and low multipoles $l$. While WMAP with its full sky coverage can measure the low $l$ at good precision, its angular resolution is not high enough to go beyond the third peak. ACT and SPT have a better resolution, but can only observe parts of the sky and cannot go below $l\sim 500$. 
Thus, these data sets need to be combined to reduce the degeneracies between $N_{\rm eff}$ and other parameters. 
Planck has, for the first time, measured low and high $l$ with a single instrument.
Combining WMAP 7 year data with ACT in 2011 preferred $N_{\rm eff}=5.3\pm1.3$ (which reduced to $4.6\pm0.8$ if data from BAO and measurements of the Hubble rate $H_0$ are added) \cite{Dunkley:2010ge}, combining SPT with WMAP7 gave $N_{\rm eff}=3.85\pm0.62$ ($3.86\pm0.42$ with BAO and $H_0$) \cite{Keisler:2011aw}.
More recent SPT results \cite{Hou:2012xq} continue to favour DR with $N_{\rm eff}=3.62\pm0.48$ ($3.71\pm0.43$ when combined with BAO and $H_0$ data).
In contrast to that, new ACT data \cite{Dunkley:2013vu,Sievers:2013wk} favours $N_{\rm eff}=2.79\pm0.56$, in accord with the SS.
This tension between SPT and ACT has e.g. been discussed in \cite{DiValentino:2013mt,Archidiacono:2013xxa}. 
The WMAP collaboration quotes $N_{\rm eff}=3.84\pm0.40$ using their 9 year data \cite{Hinshaw:2012fq}.
A combined analysis performed in \cite{Calabrese:2013jyk} yields $N_{\rm eff}=3.28\pm0.40$.
Very recently, Planck found $N_{\rm eff}=3.30 \pm 0.27$ \cite{Ade:2013lta}, which is perfectly consistent with the SS.
We discuss the implications of these measurements for sterile neutrinos as DR in section \ref{RHasDR}.\\

\begin{figure}[!h]
\centering
\includegraphics[width=13cm]{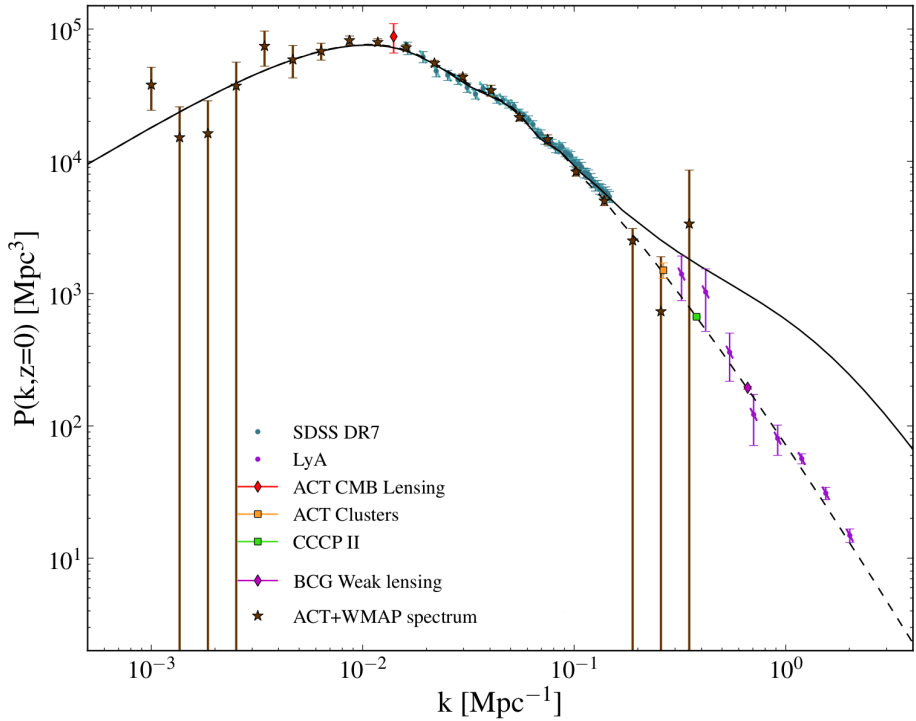}
\caption{Matter power spectrum from different observations as indicated in the plot
and combined in \cite{Hlozek:2011pc} (plot taken from there). 
\label{matterpowerspectra}}
\end{figure}
\paragraph{Large Scale Structure}
The formation of structures in the universe (see \cite{Bernardeau:2001qr} for a review) after photon decoupling can roughly be divided into two parts.
As long as $\delta\rho_k/\rho\ll1$ can be used as an expansion parameter (``linear regime''), semianalytic methods \cite{Zeldovich:1969sb} and fully automated codes (such as CAMB\footnote{http://camb.info} \cite{Lewis:1999bs,Lewis:2002ah} or CLASS\footnote{http://lesgourg.web.cern.ch/lesgourg/class.php} \cite{Blas:2011rf}) are applicable. These can test a large number of cosmological parameter sets in a short time. For $\delta\rho_k/\rho>1$ (``nonlinear regime'') the applicability of semianalytic methods (see e.g. \cite{Press:1973iz,Peacock:1990zz,Bond:1990iw}) is very limited; in general only expensive numerical n-body simulations \cite{Bullock:1999he,Springel:2005nw,Springel:2005mi,BoylanKolchin:2009nc,Prada:2011jf,Bolshoi,Lovell:2011rd} allow to make predictions. In the intermediate range, some sophisticated methods have been suggested (see e.g. \cite{Matarrese:2007wc,deVega:2010yk,Tassev:2011ac,Anselmi:2012cn,Rampf:2012xa,Rampf:2012pu,Tassev:2012cq}).
For perturbations on different scales, the evolution becomes nonlinear at different times; the behaviour on cosmic scales is still linear nowadays, while the matter distribution locally is extremely inhomogeneous.
Sterile neutrinos can affect the structure formation if they act as DM or DR. 

The implications of DM sterile neutrinos with keV masses are discussed in section \ref{DMsection}, the only big difference to CDM scenarios lies in their free streaming length $\lambda_{DM}$. This leads to a suppression of structures  on (comoving) scales smaller than $\lambda_{DM}$.
Lighter sterile neutrinos with $\sim$eV masses as suggested by the oscillation anomalies (ii) would act as DR and as {\it hot dark matter} (HDM) contribution to $\Omega_{DM}$. 
The effect that they have on the expansion rate (\ref{hubble}) would not only reflect in the abundances of light elements and the CMB, but also in the distribution of matter in the universe.

An important quantity that characterizes the distribution of large scale structures (LSS) in the universe is the {\it matter power spectrum}, cf. figure \ref{matterpowerspectra}, which is obtained from large galaxy surveys \cite{Maggiore:2009rv,Maggiore:2009rw,Maggiore:2009rx,Percival:2009xn,Papastergis:2011xe,Beutler:2011hx,Blake:2011en,Aihara:2011sj,Anderson:2012sa,Blake:2012pj,Padmanabhan:2012hf} and measures fluctuations on smaller scales than the CMB.    
Apart from small wiggles (related to baryonic acoustic oscillations), the matter power spectrum shows a ``turnover point'', the position of which is sensitive to $N_{\rm eff}$. This feature occurs because perturbations oscillated in the radiation dominated era after they enter the horizon and grew only slowly, while in the matter dominated era they grew more quickly.  The initial power spectrum is almost scale invariant, but those modes that entered the horizon during radiation domination grew only slowly until the moment of matter-radiation equality and are suppressed with respect to those that enter after that point. 
A larger $N_{\rm eff}$ leads to a later equality, which affects the position of the turnover point.
The damping of modes right of the turnover point 
is controlled by the ratio $\Omega_B/(\Omega_B+\Omega_{DM})$. It is also affected by the velocity dispersion of massive neutrinos \cite{Abazajian:2012ys}.
However, since the position of the turnover point is only known with limited accuracy \cite{Poole:2012ex} and LSS data cannot constrain all cosmological parameters, it is usually analysed in combination with CMB data (see below).

\subsection{Sterile neutrinos as dark radiation}\label{RHasDR}
For the past decade, cosmological data generally favoured $N_{\rm eff}>3.046$, though the significance was not strong enough to rule out the SS. 
It is tempting to think that sterile neutrinos may be the common origin of the oscillation anomalies (ii) and these observations (i).
For the masses and mixings indicated by LSND, the thermal production of the sterile neutrinos in the early universe would be efficient enough to thermalize them \cite{Dolgov:2003sg,Cirelli:2004cz,Mirizzi:2012we}, leading to a contribution to $N_{\rm eff}$ of order one per species.
\begin{figure}
  \centering
\begin{tabular}{r l}
    \includegraphics[width=9cm]{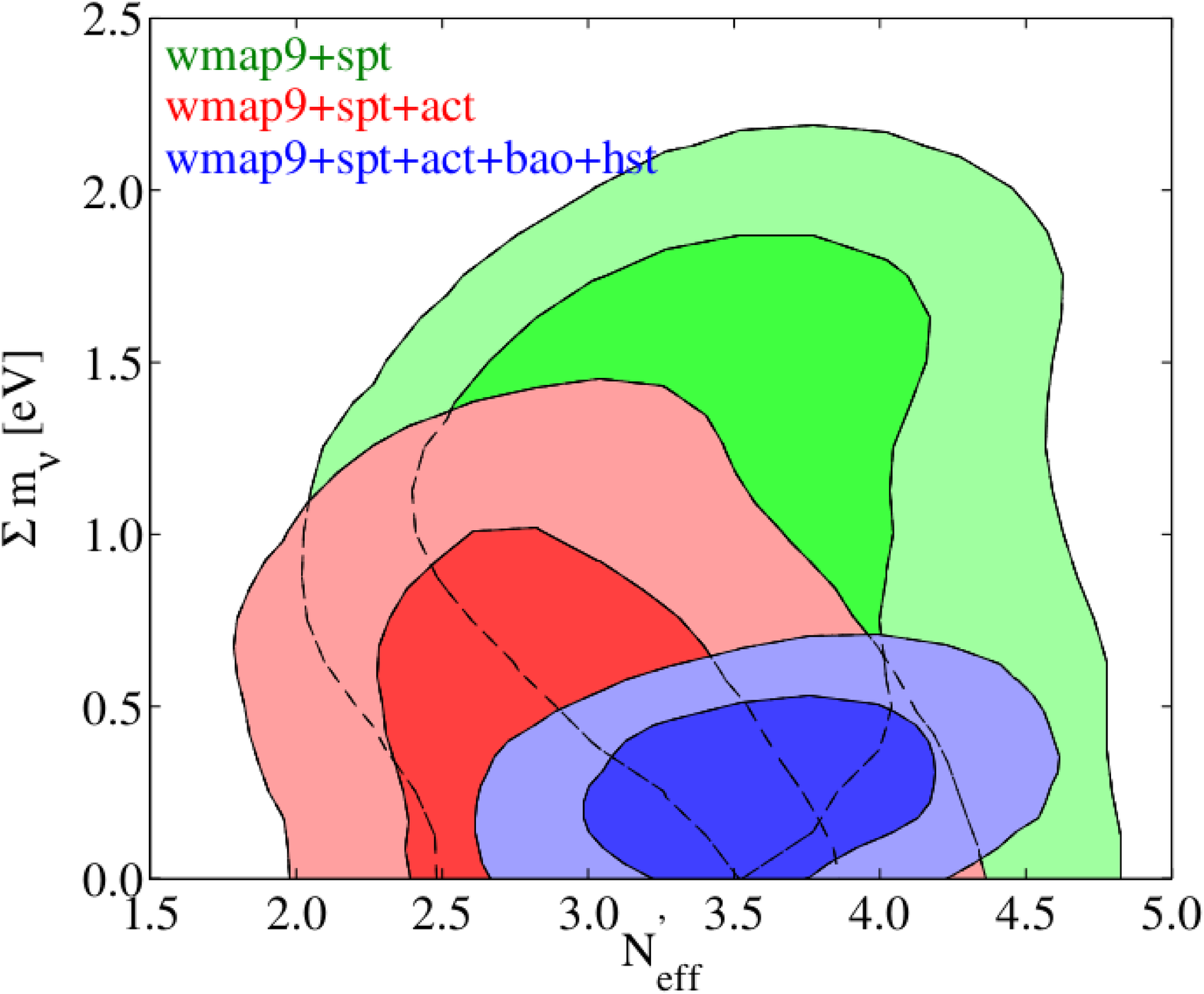} &
\includegraphics[width=7cm]{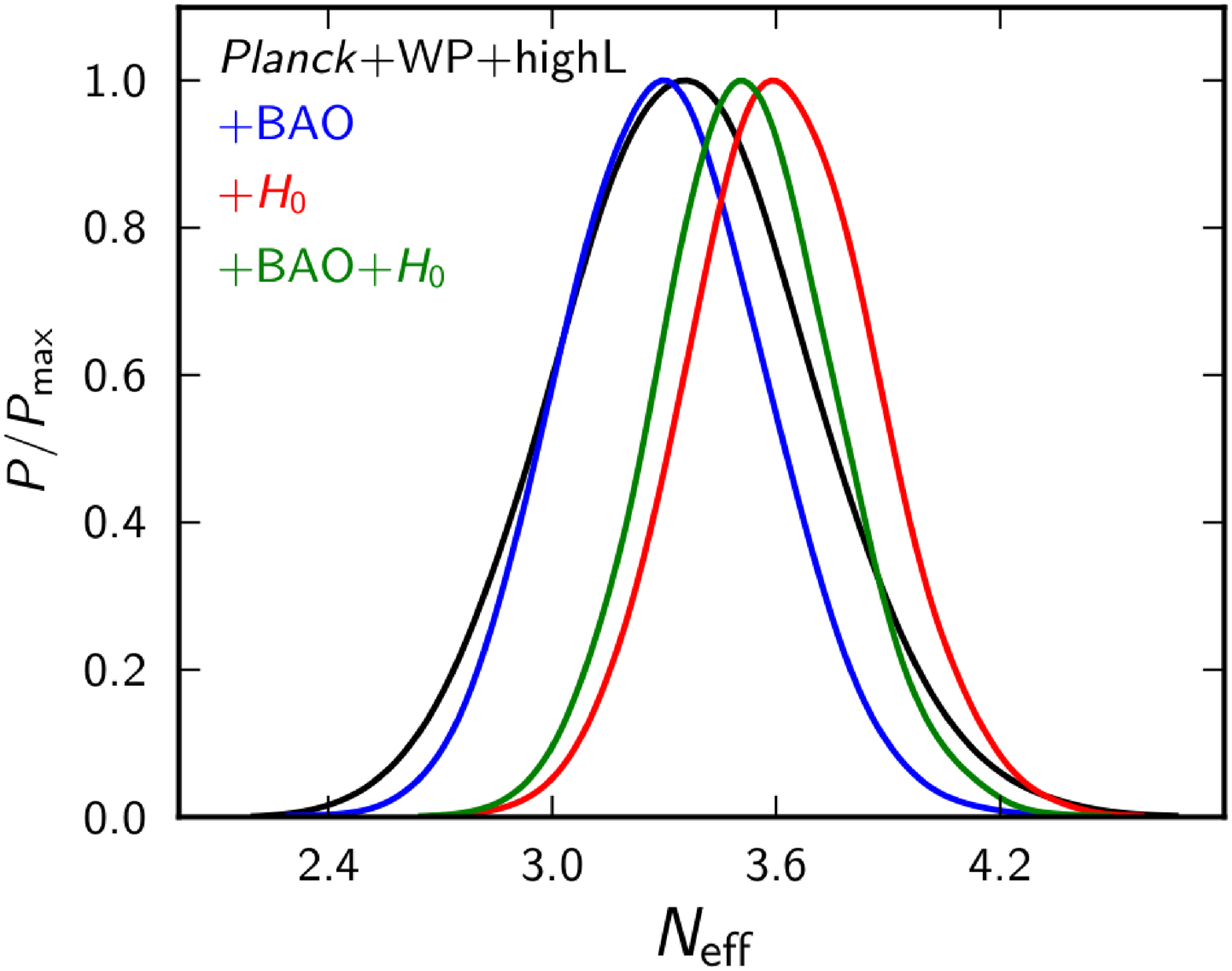}
\end{tabular}
    \caption{{\it Left plot}: Combinations of $N_{\rm eff}$ and the sum of neutrino masses favoured by the analysis in \cite{Archidiacono:2013xxa}. Note that the parameter $N_{\rm eff}$ on the horizontal axis is not exactly identical to the parameter in (\ref{NeffParam}) because the authors allowed for a common neutrino mass. 
{\it Right plot}: Marginalized posterior distribution of $N_{\rm eff}$ from
Planck (black) and Planck data supplemented by BAO data \cite{Percival:2009xn,Padmanabhan:2012hf,Blake:2011en,Beutler:2011hx,Anderson:2012sa} (blue),
a local $H_0$ measurement \cite{Riess:2011yx} (red), and both BAO and $H_0$ (green); plot from \cite{Ade:2013lta}.\label{cmbfig}
}
\end{figure} 
 
Unfortunately, the current situation is rather confusing. On one hand there is some tension amongst different oscillation experiments \cite{Kopp2013}; most of the region favoured by LNSD is excluded by ICARUS, KARMEN and MiniBooNE. 
On the other hand, there is also tension between the latest results from the CMB.
SPT+WMAP results \cite{Hou:2012xq} favour DR with $N_{\rm eff}=3.62\pm0.48$; this is consistent with the $N_{\rm eff}\simeq 4$ predicted by a thermalized sterile neutrino with masses and mixing suggested by the oscillation anomalies, but also does not rule out the SS.  
In contrast, ACT+WMAP  \cite{Dunkley:2013vu,Sievers:2013wk} favours $N_{\rm eff}=2.79\pm0.56$, which would clearly rule out thermal DR. This is in contrast to earlier ACT results \cite{Dunkley:2010ge}, which favoured DR. Planck found $3.30 \pm 0.27$ \cite{Ade:2013lta} (also using BAO data); though this shows a slight preference for $N_{\rm eff}>3$, it is perfectly consistent with the SS and clearly disfavours $N_{\rm eff}=4$.

The WMAP, ACT and SPT results have been combined with various other data sets, including measurements of the CMB, the LSS and BAO from galaxy surveys and measurements of the contemporary Hubble constant $H_0$.
For instance, in \cite{Archidiacono:2013xxa} it was recently argued that cosmology can be consistent with the neutrino oscillation anomalies. This conclusion was mainly possible because the authors excluded ACT and BBN data from the combined analysis.
This can be seen in figure \ref{cmbfig}. 
For small masses $< 0.5$ eV, WMAP+SPT+ACT data prefer extra neutrinos; but for masses around $1$eV the data favours $N_{\rm eff}\sim 3$, as predicted by the SS. Including BAO and $H_0$ data indicates $N_{\rm eff}>3$, but also disfavours the $\sim1$ eV region for the masses preferred by oscillation anomalies. Without ACT, $N=4$ and a mass of $1$ eV are perfectly consistent. 
Similar analyses, using pre-Planck data, have e.g. been performed in \cite{GonzalezGarcia:2010un,Krauss:2010xg,Hou:2011ec,Hamann:2011ge,Archidiacono:2012ri,Giusarma:2012ph,RiemerSorensen:2012ve,Joudaki:2012uk,DiValentino:2013mt,Feeney:2013wp}\footnote{A nice summary of various other ways to combine different data sets can be found in \cite{RiemerSorensen:2013ih}.}. The conclusions differ, depending on the selected data sets  and statistical method. There is a general preference for $N_{\rm eff}>3$, but no exclusion of the SS. 

However, while the authors of \cite{Archidiacono:2013xxa} presented arguments to exclude ACT from their analysis, the conclusions change dramatically when the Planck result $3.30 \pm 0.27$ \cite{Ade:2013lta} is included. Though a slight preference for $N_{\rm eff}$ so far seems to remain \cite{Archidiacono:2013fha,Wyman:2013lza,Hamann:2013iba,Battye:2013xqa,Gariazzo:2013gua}, the thermalized DR ($N_{\rm eff}\geq 4$) that the oscillation anomalies hint at is clearly disfavoured \cite{Mirizzi:2013kva}. 
The above result assumes massless neutrinos, but does not change significantly if the sum of neutrino masses is allowed to vary ($N_{\rm eff}=3.32^{+0.54}_{-0.52}$ and $\sum_I m_i<0.28$ eV when combined with BAO \cite{Ade:2013lta}).
There is some freedom to increase $N_{\rm eff}$ if local measurements are used to determine $H_0$ instead of the value obtained from the Planck data itself, see figure \ref{cmbfig}.
Local measurements of cosmological quantities such as the Hubble rate or the age of the universe (by studying old objects) depend less on the assumed model of cosmology and can be used complementary to high redshift data, see \cite{Verde:2013fva} for a recent discussion.  
The Planck collaboration concludes that the tension between direct $H_0$ measurements and the CMB
and BAO data in the $\Lambda$CDM model can be relieved by $N_{\rm eff}>3.046$, but there is no strong preference for this extension from the CMB damping tail \cite{Ade:2013lta}.
It remains to be seen whether this tendency continues when the full Planck data is analysed, which will provide the most stringent constraint on $N_{\rm eff}$ from a single instrument \cite{Planck:2006aa}.

The apparent tension between oscillation anomalies and the Planck and ACT results
could be eased if the thermal production of light sterile neutrinos in the early universe is somehow suppressed, e.g. by a chemical potential \cite{Foot:1995bm,Hamann:2011ge,Ho:2012br,Schwarz:2012yw}. 
Quantitative studies of the flavour evolution \cite{Foot:1995qk,Shi:1996ic,Bell:1998ds,Hannestad:1998zg,Dolgov:2002ab,Wong:2002fa,Abazajian:2002qx} are required to answer the question to which degree a lepton chemical potential can prevent the thermalization \cite{Abazajian:2001nj,Abazajian:2004aj,Chu:2006ua,Melchiorri:2008gq,Hamann:2011ge,Kirilova:2011an,Hannestad:2012ky,Mirizzi:2012we,Ho:2012br,Schwarz:2012yw,Jacques:2013xr,Saviano:2013ktj,Hannestad:2013pha}.

If one, on the other hand, interprets the Planck and ACT results as indicators that there is no DR, then it remains to be understood why so many previous analyses seemed to point towards its existence.
The possibility that the preference $N_{\rm eff}>3.046$ is an artefact of the prior choice in Bayesian analyses has been discussed and rejected in \cite{Hamann:2011hu}. On the other hand, the authors of \cite{Feeney:2013wp} found that Bayesian model selection prefers the SS, despite the fact that parameter estimates for $N_{\rm eff}$ are larger than $3.046$. In \cite{Audren:2012wb} it has been pointed out that there is no evidence for $N_{\rm eff}>3$ if one relaxes the assumptions on the late history of the universe suggested by $\Lambda$CDM when analysing the CMB. 
Generally, there is a lot more freedom if one allows for more physics beyond the $\Lambda$CDM scenario. 
In \cite{Steigman:2013yua} it has been discussed in detail that different modifications of the SM can predict values $N_{\rm eff}>3$ or $N_{\rm eff}<3$, irrespective of the number of neutrino degrees of freedom. However, it seems that at least the simplest scenario in which light sterile neutrinos with generic couplings are the common explanation for the oscillation anomalies and the preference for $N_{\rm eff}>3$ is disfavoured by Planck unless more new physics is added. At the same time, it seems unlikely that the SS can be ruled out by CMB observations in foreseeable time. 

\subsection{Sterile neutrinos in astrophysics}
In addition to possible signatures in high redshift observations, sterile neutrinos may also have an effect on astrophysical phenomena at present time. 
Their most studied role in astrophysics is that of a DM candidate, see section \ref{RHasDR}, but they may have effects on other phenomena.
They can, for instance, affect the transport in supernova explosions if they have eV \cite{Wu:2013gxa}, keV \cite{Hidaka:2006sg,Hidaka:2007se} or GeV \cite{Fuller:2009zz} masses. Sterile neutrinos with keV masses can also help to explain the high rotation velocities of pulsars \cite{Kusenko:1998bk,Fuller:2003gy}.

\section{Baryogenesis via leptogenesis}\label{Leptogenesis}
The observable universe does not contain any significant amounts of antibaryons \cite{Canetti:2012zc}, i.e. it is highly matter-antimatter asymmetric.
Given our knowledge about the thermal history of the universe, today's baryon density $\Omega_B$ is easily explained  as the remnant of a small matter-antimatter asymmetry at early times, when the temperature was high enough for pair creation processes to occur faster than the Hubble rate. 
The baryon asymmetry in the early universe (BAU) can be estimated  by the baryon-to-photon ratio at later times,
\begin{equation}\label{etadef}
\frac{n_B-n_{\bar{B}}}{n_B+n_{\bar{B}}}\Big|_{T\gg 1{\rm GeV}}\sim 
\frac{n_B-n_{\bar{B}}}{s}
\sim\frac{n_B}{n_\gamma}\Big|_{T\ll 1 {\rm GeV}}
\equiv \eta_B
.\end{equation} 
Here $n_B$ and $n_{\bar B}$ are comoving number densities for baryons and anti-baryons.
$\eta_B$ can be determined rather consistently from BBN \cite{Steigman:2010zz} or the CMB and LSS \cite{Hinshaw:2012fq},
 \begin{equation}
\eta_{B}^{\rm BBN}=5.80\pm 0.27 \ , \ \eta_{B}^{\rm CMB}=6.21\pm0.12
,\end{equation}
both in units of $10^{-10}$.
A period of cosmic inflation, as suggested by the CMB, would have diluted any pre-inflationary asymmetry. Thus, the BAU must have been created dynamically afterwards. There are three conditions for the dynamical generation of a BAU (``baryogenesis''), known as Sakharov conditions \cite{Sakharov:1967dj}: Baryon number ($B$) violation, breaking of charge (C) as well as charge-parity (CP) symmetry and a deviation from thermal equilibrium. In principle, all of them are fulfilled in the SM: at the quantum level, baryon number is violated \cite{Adler:1969gk,Bell:1969ts,'tHooft:1976up} at $T>T_{EW}$ by nonperturbative sphaleron processes \cite{Kuzmin:1985mm}, C and CP are violated by the phase in the CKM matrix \cite{Kobayashi:1973fv} and the weak interaction \cite{Wu:1957my,Christenson:1964fg}, and a deviation from thermal equilibrium is caused by Hubble expansion. However, the CP violation and deviation from equilibrium are both too small in the SM to explain the observed $\eta_B$; see \cite{Canetti:2012zc} for a detailed account of the Sakharov conditions and baryogenesis in the SM.

RH neutrinos described by (\ref{L}) can fix both of these shortcomings. As gauge singlets, they can be out of equilibrium at temperatures when all other particles are tightly coupled by gauge interactions, and their Yukawa couplings contain several unconstrained CP-violating phases. 
This makes baryogenesis via leptogenesis possible \cite{Fukugita:1986hr}. 
In this scenario a matter-antimatter asymmetry is first generated in the leptonic sector by the CP-violating interactions of the $\nu_R$ and then transferred to the baryonic sector by electroweak sphaleron processes.\footnote{The term leptogenesis is often used in a wider sense, referring to all scenarios where the source of CP-violation lies in the leptonic sector. Here we only refer to those scenarios where the BAU is generated from the interactions in (\ref{L}).}
Sphalerons are only effective for $T>T_{EW}$. At these temperatures the Higgs field has a vanishing expectation value, hence leptogenesis operates in the symmetric phase of the SM. 
In this regime the fields $\nu_L$ behave like massless Weyl fields that can be described as the left chiral projection of a Dirac spinor [rather than the Majorana spinors (\ref{Nandupnudef})]. 
We can define the left handed lepton numbers $L_{L,\alpha}$ as the zero components of the currents $J_{L,\alpha}^\mu\equiv\bar{\nu}_{L,\alpha}\gamma^\mu\nu_{L,\alpha}+\bar{e}_{L,\alpha}\gamma^\mu e_{L,\alpha}$. 
Analogously, we define lepton numbers $L_{R,\alpha}$ from the current  $J_{R,\alpha}^\mu\equiv\bar{e}_{R,\alpha}\gamma^\mu e_{R,\alpha}$ for the RH fields. 
The total LH lepton number is $L_L\equiv\sum_\alpha L_{L,\alpha}$, 
and we furthermore define the  active lepton numbers $L_\alpha\equiv L_{L,\alpha}+L_{R,\alpha}$. 
For $F=0$ the $L_{\alpha}$ are exactly conserved.
The $\nu_R$ have a Majorana mass term $M_M$; in general they do not carry any conserved lepton charge, and interactions with them violate the charges $L_\alpha$.\footnote{For $T<T_{EW}$ the active neutrinos also receive a Majorana mass term (\ref{activeneutrinomasses}) that violates lepton number. This violation is, however, suppressed by $m_\nu/T$.
Similar arguments apply to thermal masses in the primordial plasma, which change the kinematics of quasiparticles, but respect the symmetries of the theory \cite{Weldon:1982bn,Weldon:1982aq}.
}
However, due to the smallness of $F$ these interactions are slow, and $L_\alpha$ are still approximately conserved charges for which one can define a chemical potential. 
If there is an (exactly or approximately) conserved charge in the $\nu_R$-sector, we may define ``lepton numbers'' for the sterile neutrinos. 
This is always the case at temperatures $T\gg M_I$, when helicity changing processes are suppressed by $M_I/T$; then one can interpret two helicity states of $N_I$ as ``particle'' and ``antiparticle'' and define sterile lepton charges $L_I$ as the difference of their occupation numbers, cf. (\ref{YN}).
They contribute to the RH lepton number $L_R\equiv\sum_I L_{I}+\sum_\alpha L_{R,\alpha}$. The total lepton number is $L\equiv L_L+L_R$.
If the flavour structure of $F$ and $M_M$ obeys a symmetry such that there is another conserved charge (exactly or approximately), it can be more convenient to use that charge to define ``lepton numbers'' in the sterile sector. 
A particularly interesting scenario of this kind is that in which total lepton number $L$ is almost conserved \cite{Wyler:1982dd,Mohapatra:1986bd,GonzalezGarcia:1988rw,Branco:1988ex,Dick:1999je,Kersten:2007vk,Abada:2007ux,Shaposhnikov:2006nn,Gavela:2009cd,Blanchet:2009kk}.

When one or more of the $N_I$ are out of equilibrium, they can generate lepton asymmetries via various processes, including decays and inverse decays, scatterings and flavour oscillations (the probability to decay/scatter/oscillate into leptons and antileptons is different at second order in $F$ due to quantum interferences between CP-violating processes).
Sphalerons violate $B$ and $L$ individually, but conserve $B-L$. At $T>T_{EW}$ they transform part of the lepton asymmetry into a baryon asymmetry; if all interactions are in equilibrium it is given by $B\sim \frac{28}{79}(B-L)$ \cite{Khlebnikov:1988sr,Laine:1999wv}.  $B$ is protected from washout after sphaleron freezeout around $T\sim T_{EW}$. Since sphalerons only couple to left chiral fields, they in fact only see the leptonic charge $L_L$ stored in these fields, which may differ from the total lepton asymmetry $L$. Thus, the BAU we observe is determined by $L_L$ at $T\sim T_{EW}\sim 140$ GeV.
The nonequilibrium condition can be fulfilled three times for each $N_I$: during their production, their freezeout, and their decay.

\subsection{Leptogenesis from $N_I$ freezeout and decay}\label{DecayLeptogenesis}
\begin{figure}[!h]
\centering
\includegraphics[width=15cm]{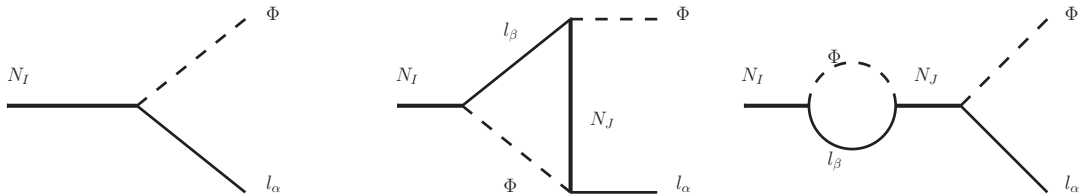}
\caption{A crucial contribution to the CP-violation in the simplest version of ``vanilla leptogenesis''  comes from the quantum interference between the tree level diagram and radiative corrections to the decay $N_I\rightarrow\Phi\l_\alpha$. Due to the CP-violation contained in the complex phases in $F$, the amplitudes for these processes with $l_\alpha$ in the final state differ from those with $\bar{l}_\alpha$. At tree level this difference amounts to an overall phase, which does not affect the physical absolute square of the amplitude, but in the interference terms between tree level and loop diagrams it makes a physical difference. 
The rightmost diagram gets resonantly enhanced for $M_I\sim M_J$ because then the $N_J$ propagator is nearly onshell. In more general scenarios, also scattering processes (such as $\Phi l_\alpha\rightarrow \Phi l_\beta$ with intermediate $N_I$) have to be considered. \label{leptogenesisfig}}
\end{figure}
It is common to split the lepton number violating processes into ``source'' and ``washout''. The source consists of all contributions that violate active lepton numbers in the absence of an existing asymmetry (e.g. $N_I$-decays); these can generate asymmetries in the presence of a deviation from thermal equilibrium. The ``washout'' consists of all other processes, which tend to eliminate existing asymmetries (e.g. inverse $N_I$-decays). In addition, there are ``spectator processes'' \cite{Buchmuller:2001sr,Nardi:2005hs} that do not violate $L$ themselves, but redistribute charges amongst different fields and thereby affect the time evolution of $L$.
If the $N_I$ have masses as suggested by GUT models, they are produced thermally, freeze out and decay at temperatures $T\gg T_{EW}$. 
Since $L$-violating processes are strongly suppressed at $T<M_I$,  $B-L\neq0$ is preserved from washout. 
This is the most studied version of leptogenesis, and there exist several detailed reviews, see e.g. \cite{Buchmuller:2005eh} or, more recently, \cite{Blanchet:2012bk,Fong:2013wr}. 
This setup is very appealing because it can easily be embedded into a GUT-framework, provides a ``natural'' explanation for the smallness of the neutrino masses $m_i$ via the seesaw mechanism and allows to probe at least some parameters of very high energy physics in low energy neutrino experiments via (\ref{SeesawNeutrinomass}). The downside is that such heavy  $N_I$-particles cannot be studied in the laboratory. 
Out of the $7n-3$ parameters in $F$ and $M_M$, one can under ideal conditions probe the $3$ neutrino masses and $6$ angles and phases in $U_\nu$ experimentally (being very optimistic in case of the Majorana phases). The generated asymmetry $\eta_B$ (the only observable number in leptogenesis) can in general depend on all 
$7n-3$ parameters. Thus, the perspectives to constrain $N_I$-properties from leptogenesis are very limited.
We therefore only recapitulate the basic ideas and refer the interested reader to the reviews named above. 
\paragraph{Minimal ``Vanilla'' leptogenesis} -
In the simplest scenario one assumes that the $N_I$ are very heavy, have hierarchical masses ($T_{EW}\ll M_1\ll M_{I>1}$) and the washout is strong.
The latter statement is usually parametrized in terms of the parameter $K_I=\Gamma_I|_{T=0}/H|_{T=M_I}\simeq (F^\dagger F)_{II}v^2/(M_I\times 10^{-3}{\rm eV})$, where $\Gamma_I$ is the thermal width of $N_I$ particles, $H\simeq1.66\sqrt{g_*}T^2/M_P$ the Hubble rate and $g_*$ the effective number of relativistic degrees of freedom in the plasma ($g_*= 106.75$ in the SM at $T\gg T_{EW}$). 
If $(F^\dagger F)_{11}\sim {\rm eV} \times M_1/v^2$, as suggested by the seesaw relation (\ref{activeneutrinomasses}), then $K_1\gg 1$ and one is in the {\it strong washout} regime.
In this case the BAU responsible for the observed $\eta_B$ may have been created by the freezeout and decay of the lightest sterile neutrino $N_1$ alone. This is possible if any pre-existing asymmetries,
the asymmetries generated during the production of the $N_I$ and the asymmetries from the $N_{I>1}$-decays are all washed out efficiently by $N_1$ \footnote{Though there  are no $N_{I>1}$ particles in the plasma at $T\ll M_{I>1}$, the existence of more than one RH neutrino is crucial to provide CP-violation in the lepton sector.}. 
This scenario has the advantage that $\eta_B$ is essentially independent of the initial conditions. A major contribution to the final asymmetry comes from the decay diagrams shown in figure \ref{leptogenesisfig}. In addition, there are lepton number violating scatterings. However, vanilla leptogenesis can be qualitatively understood by considering the decays and inverse decays only \cite{Buchmuller:2004nz}.
This mechanism can reproduce $\eta_B$ for $M_1\gtrsim4\times10^{8}$ GeV \cite{Davidson:2002qv}, which requires a rather large reheating temperature if $N_I$ are mainly produced thermally (``thermal leptogenesis'').
There is also a constraint on the mass of the lightest active neutrino as $m_1\lesssim 0.1$ eV \cite{Buchmuller:2002jk,Buchmuller:2003gz}; see \cite{Nardi:2011zz} for an review on the connection with neutrino masses. 
In the weak washout regime $K_1\lesssim 1$\footnote{This is possible in spite of (\ref{activeneutrinomasses}) because $M_M$ and $F$ are matrices.} the predictive power is much smaller because asymmetries from processes at earlier time (nonthermal production during reheating, thermal $N_I$ production, $N_{I>1}$ freezeout and decay...) are not washed out efficiently and contribute to $\eta_B$.  
\paragraph{Flavour effects} -
In the simplest scenario it is assumed that only one sterile neutrino $N_1$ dynamically participates in leptogenesis and all three active flavours can be treated equivalently (``unflavoured regime''). 
In the unflavoured regime, only the linear combination $\sim F_{\alpha 1} |l_{L \alpha}\rangle$ of active leptons that couples to $N_1$ is relevant for leptogenesis. In the corresponding flavour basis one can ignore the directions in flavour space perpendicular to that, and the problem is equivalent to the one flavour case.
Both of these assumptions do not hold in general.

Active flavours have to be treated separately when the Hubble rate drops below the rate at which interactions that distinguish active flavours, mediated by the charged Yukawa couplings, occur (``flavoured regime''). 
These interactions destroy the coherence of the flavour state that couples to $N_1$ because they have different strength for the different $l_{L,\alpha}$.
For the $\tau$-Yukawa, this happens below $T\sim 10^{12}$ GeV. 
The importance of flavour has been realized in \cite{Barbieri:1999ma} and \cite{Blanchet:2006be,Abada:2006ea,Abada:2006fw,Nardi:2006fx} and meanwhile been studied by various authors, see \cite{Blanchet:2012bk,Fong:2013wr} for details and references.
Once two flavour states are distinguishable, one has to treat the asymmetries stored in each individually. They can differ considerable and even have opposite signs \cite{Endoh:2003mz}.  
This can affect $\eta_B$ if e.g. the washout is very different for different flavours; it makes leptogenesis possible even if the source term does not violate $\sum_\alpha L_\alpha$. 
Flavour effects can reduce the lower bound on $M_1$ in generic seesaw scenarios by $1-2$ orders of magnitude \cite{Antusch:2009gn,Racker:2012vw}, which is still far out of experimental reach.

Also the assumption that the generation of today's BAU only involved $N_1$-dynamics does not hold in general. In the unflavoured regime, $N_1$ can only wash out the asymmetry that is stored in the combination $F_{\alpha 1} Y_\alpha$ in flavour space to which it couples. This singles out a particular direction in flavour space. If pre-existing asymmetries or those produced by $N_{I>1}$ have a component orthogonal to this, $N_1$ cannot wash them out efficiently \cite{DiBari:2005st,Engelhard:2006yg}. In fact, it is a rather special case that there is no such component. The effect of active flavours in scenarios where the heavier $N_I$ contribute has been discussed in \cite{Vives:2005ra,Blanchet:2008pw,Bertuzzo:2010et,Antusch:2010ms,Blanchet:2011xq}.

Flavour effects offer ways to circumvent the lower bound on $M_I$. One possibility is that the BAU originates from a purely flavoured asymmetry; i.e. the total asymmetry is vanishing or small ($L\simeq 0$), but the left handed asymmetry is non-vanishing ($|L_L|\gg |L|$) \cite{Blanchet:2009kk,Racker:2012vw}. Since sphalerons only couple to LH fields, they convert part of $L_L$ into a LH baryon asymmetry. This is possible for $M_I<v$ because in this case the seesaw relation (\ref{activeneutrinomasses}) enforces Yukawa couplings that are so small that $L_L$ and $L_R$ may not equilibrate before sphaleron freezeout. 
This makes leptogenesis possible even for $M_M=0$ \cite{Dick:1999je} (Dirac leptogenesis), and it is also the basis of scenarios discussed in the following section \ref{LeptogenfromOsc}.
The mass bound may also be circumvented if the generated asymmetry is resonantly enhanced by a degeneracy between sterile neutrino masses \cite{Pilaftsis:2003gt} (resonant leptogenesis). 
This enhancement is discussed from first principles in \cite{Garny:2011hg,Garbrecht:2011aw}; in \cite{Garny:2011hg} it is shown that the maximal enhancement (compared to the non-resonant case) of the total asymmetry $L$ can be expressed in terms of the parameter\footnote{See \cite{Liu:1993ds,Flanz:1996fb,Covi:1996fm,Pilaftsis:1997dr,Buchmuller:1997yu,Pilaftsis:2003gt,Garny:2011hg} for earlier discussions of this point.} 
\begin{equation}\label{resonantenhancement}
\frac{M_1 M_2}{2(M_1\Gamma_1+M_2\Gamma_2)}
.\end{equation}
The enhancement allows values of $M_M$ in the TeV range, which raises hope to probe $N_I$ in high energy experiments.
In this case it is of course not only the lightest sterile neutrino $N_1$ that generates the asymmetry, but the interplay between the two mass-degenerate states, and flavour effects in the sterile sector have to be taken into account.
This is generally the case when the spectrum of $M_I$ is not hierarchical ($|M_1-M_2|\lesssim M_1$), not only in the extreme case $M_1\simeq M_2$. 

This discussion and the quoted mass bounds apply to leptogenesis within the framework of a pure type-I seesaw. If one introduces new degrees of freedom beyond those in (\ref{L}), these bounds can weaken, see e.g. \cite{Giudice:2003jh,Fong:2010qh,Garbrecht:2013iga}, \cite{Garbrecht:2012qv} and \cite{Fong:2013gaa}.\\

\subsection{Leptogenesis during $N_I$ production}\label{LeptogenfromOsc} 
\begin{figure}[!h]
\centering
\begin{tabular}{c}
\includegraphics[width=12cm]{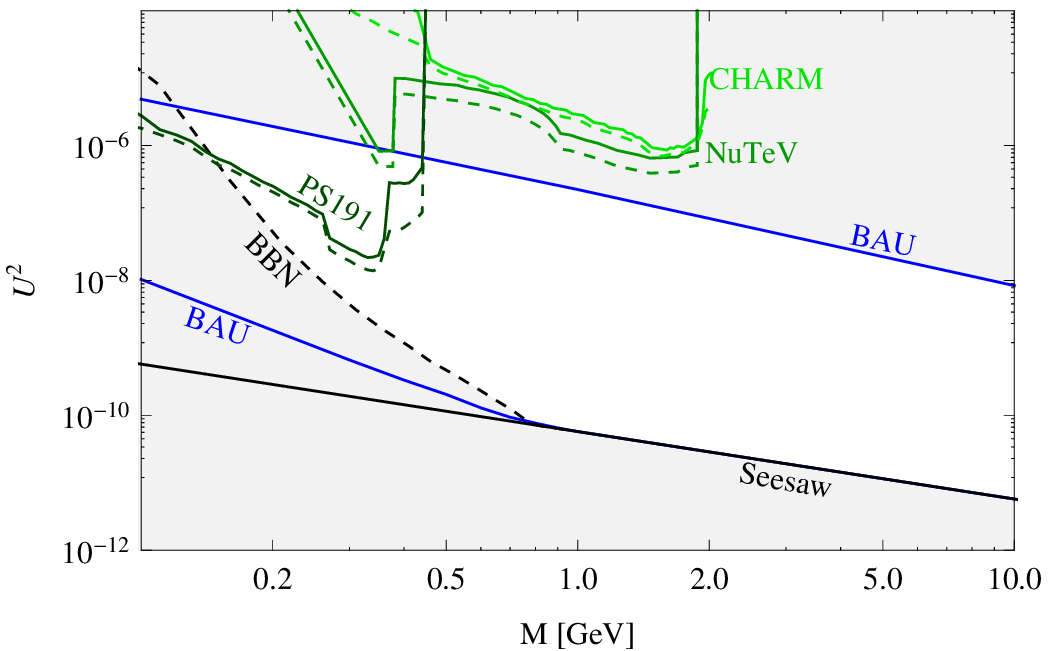} \\
\includegraphics[width=12cm]{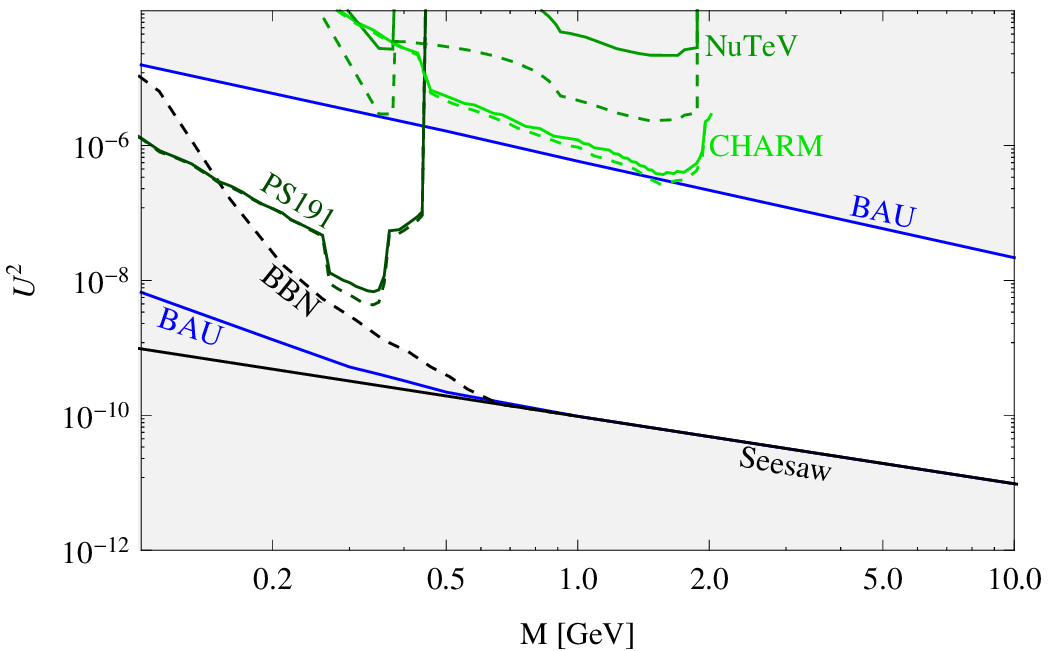}
\end{tabular}
\caption{Constraints on the sterile neutrino  masses and
mixing $U^2={\rm tr}(\theta^\dagger\theta)$ for $n=2$
from baryogenesis; upper panel - normal
hierarchy, lower panel - inverted hierarchy.
For the displayed case of $n=2$ RH neutrinos, baryogenesis can only be successful if their masses $M_1$ and $M_2$ are degenerate ($|M_1-M_2|\ll (M_1+M_2)/2$), the parameter $M$ refers to their mean value $(M_1+M_2)/2$. For $n\geq3$ RH neutrinos no mass degeneracy is required \cite{Drewes:2012ma}.
The observed BAU can be generated in the region between the
solid blue ``BAU'' lines. 
The regions below the solid black ``seesaw'' line and dashed black ``BBN''
line are excluded by neutrino oscillation experiments and BBN,
respectively. The areas above the green lines of different shade are
excluded by direct search experiments, as indicated in the plot. The
solid lines are {\it exclusion plots} for all choices of parameters, 
for the dashed lines the phases were chosen to maximize the asymmetry, 
consistent with the blue lines. Plots taken from \cite{Canetti:2012kh}.
 \label{BAUnuMSM}}
\end{figure}
If some masses $M_I$ are near the electroweak scale or below, the seesaw relation (\ref{activeneutrinomasses}) requires the entries $F_{\alpha I}$  to be very small to be consistent with bounds on the active neutrino masses.
That means that the rate $\Gamma_I\sim (F^\dagger F)_{II} T$ of thermal $N_I$ production is so small that these particles may not come into thermal equilibrium until $T\sim T_{EW}$. During the thermal production before equilibration, the nonequilibrium condition is fulfilled and asymmetries $L_\alpha$ are generated. The total asymmetry $L$ is suppressed by $M_I/T$ unless there is a resonant enhancement \`a la (\ref{resonantenhancement}), but the flavoured asymmetries $L_\alpha$ are unsuppressed \cite{Drewes:2012ma}, hence $|L_L|\gg |L|$ for $M_I\ll T_{EW}$. This is sufficient because sphalerons only see $L_L\simeq -L_R$.
In contrast to the scenarios with $M_I\gg T_{EW}$ discussed in section \ref{DecayLeptogenesis}, lepton number is not conserved near $T_{EW}$ because the sterile neutrinos are still present in the plasma; once the $N_I$ reach thermal equilibrium, the $L_\alpha$ are washed out. 
A baryon asymmetry can be generated from $L_L$ in spite of this (and the smallness of $|L|$) if the $N_I$ have not thermalized or the washout is not efficient enough to eliminate $L_L$ before sphaleron freezeout.
This scenario is often referred to as {\it baryogenesis via neutrino oscillations} \cite{Akhmedov:1998qx}, though individual oscillations are not always crucial \cite{Canetti:2010aw,Drewes:2012ma,Khoze:2013oga}.
One advantage of this mechanism is that the $N_I$ are light enough to be within reach of laboratory experiments.
This makes it one of the few known models of baryogenesis that are empirically testable (the other two much studied testable mechanisms are resonant leptogenesis and electroweak baryogenesis \cite{Morrissey:2012db}).
The perspectives to study the $N_I$ responsible for baryogenesis in the laboratory have been studied in detail in \cite{Canetti:2010aw,Canetti:2012vf,Canetti:2012kh,Drewes:2012ma}. 
If the $N_I$ interact only via the Yukawa couplings $F$, then a clear detection is realistically only possible for masses $M_I\lesssim$ a few GeV (unless $F$ has some special structure, leading to cancellations in (\ref{activeneutrinomasses}) that allow for larger individual elements $F_{\alpha I}$ \cite{Shaposhnikov:2006nn,Kersten:2007vk}).  
If they have additional interactions, masses up to a few TeV are in reach of high energy experiments, see section \ref{collidersec} and references therein for more details.
In the minimal scenario with $n=2$ sterile neutrinos a mass degeneracy $|M_1-M_2|/(M_1+M_2)\sim 10^{-3}$ is required to produce the observed BAU (or $\eta_B$). Bounds on mass and mixing from the baryogenesis requirement in this case are shown in figure \ref{BAUnuMSM} together with other experimental and astrophysical constraints.
For $n>2$ and $M_I$ in the GeV range no mass degeneracy is necessary \cite{Drewes:2012ma}.

\vspace{0.3cm}
The number of possible leptogenesis scenarios increases greatly when (\ref{L}) is extended by additional degrees of freedom or embedded into a bigger framework (e.g. supersymmetry); we do not discuss these here in detail and refer the interested reader to the reviews \cite{Buchmuller:2005eh,Blanchet:2012bk,Fong:2013wr}. 

Finally, the CP- and B-violation that make leptogenesis possible can also "work backwards" and rule out or constrain other baryogenesis scenarios because they wash out matter-antimatter asymmetries created by those mechanisms at higher temperatures.  
The mass spectrum of the $N_I$ can be constrained by the requirement that they do not wash out the baryon asymmetry in the early universe \cite{Blanchet:2008zg}, though these conclusions rely on assumptions about additional interactions of $N_I$. A similar argument was suggested in \cite{Hollenberg:2011kq} to constrain the properties of a fourth neutrino generation.
\subsection{Towards a quantitative treatment}\label{technical}
\paragraph{Transport equations} - Most quantitative studies of leptogenesis solve a set of momentum integrated Boltzmann equations (``rate equations'') to predict $\eta_B$ as a function of the parameters in (\ref{L}).
In the case of ``vanilla leptogenesis'', these can be written as  \cite{Buchmuller:2004nz}
\begin{eqnarray}
HX \frac{dY_1}{d X}&=&-\Gamma_1 (Y_1- Y_1^{eq}),\label{rate1}\\
HX \frac{dY_{B-L}}{d X}&=& \epsilon_1 \Gamma_1 (Y_1- Y_1^{eq}) - c_W\Gamma_1 Y_{B-L}\label{rate2}
.\end{eqnarray}
Here $Y_1$ is the abundance of $N_1$ particles, i.e. the momentum integral over the phase space distribution function divided by the entropy density $s$. $ Y_1^{eq}$ is its value in thermal equilibrium and $Y_{B-L}$ the difference between baryon and lepton abundances (particles minus antiparticles for each of them). 
We use the variable $X=M_1/T$ instead of time, which leads to a factor $dX/dt\simeq HX$ on the left hand side\footnote{The expression for $dX/dt$ can be complicated when $g_{*}$ changes during leptogenesis. This can e.g. affect the generation of lepton asymmetries at $T\ll T_{EW}$.}.
$\epsilon_1$ is a parameter that characterizes the amount of CP-violation,
\begin{equation}
\epsilon_1\simeq-\frac{3}{16\pi}\frac{1}{(F^\dagger F)_{11}}\sum_{I}{\rm Im}\left[\left(F^\dagger F\right)_{I1}^2\right]\frac{M_1}{M_I}\simeq \frac{3}{16\pi}\frac{M_1}{(F^\dagger F)_{11}v^2}{\rm Im}\left[(F^Tm_\nu F)_{11}\right]
\end{equation}
The constant $c_W$ in the simplest case (washout by inverse decays) is given by $1/2$ times the ratio between the number densities of $N_1$ and active leptons.
Calculation of the final $Y_{B-L}$ requires solving the rate equations (\ref{rate1}) and (\ref{rate2}), it can be estimated as  $Y_{B-L}\sim \kappa\epsilon_1/g_*$ \cite{Fukugita:1986hr},\footnote{Our notation is more close to that in \cite{Buchmuller:2005eh}.} where the number $\kappa<1$ is called efficiency factor and is related to the washout.

Equations (\ref{rate1}) and (\ref{rate2}) can be used when only one sterile neutrino dynamically contributes to the BAU and the SM leptons can be described in an effective one flavour treatment (unflavoured regime). In the fully flavoured regime flavour dependent interactions act sufficiently fast (compared to the time scale $\sim1/\Gamma_1$ related to the $N_1$ dynamics and $1/H$) to fully decohere the different contributions in the state $F_{\alpha 1} |l_{L \alpha}\rangle$ that couples to $N_1$. Then (\ref{rate2}) has to be replaced by three different equations for the asymmetries $Y_\alpha$ in the individual flavours. In the intermediate regime between these two cases, 
there is no full decoherence. In this case flavour oscillations have to be taken into account. This is usually done by {\it density matrix equations}  \cite{Sigl:1992fn}, see \cite{Barbieri:1999ma,Abada:2006fw} for early applications to leptogenesis. 
These are matrix valued generalizations of the rate equations (\ref{rate1}) and (\ref{rate2}) in which the lepton charges $Y_{\alpha\beta}$ carry two indices. The diagonal elements $Y_{\alpha\alpha}$ of these are simply describe the lepton abundance in flavour $\alpha$, the off-diagonal components describe correlations between different flavours.
A first principles approach that allows to study the flavoured, unflavoured and intermediate regimes consistently was presented in \cite{Beneke:2010dz}.

If more than one RH neutrino is relevant for leptogenesis, also the different sterile flavours have to be treated independently. For a hierarchical mass spectrum, this usually amounts to simply replacing (\ref{rate1}) by $n$ different equations for the abundances $Y_I$ of all relevant sterile flavours $N_I$. In scenarios of resonant leptogenesis or leptogenesis during $N_I$ production, oscillations between the different flavours may be relevant. In these scenarios, leptogenesis typically happens at temperatures $T\gg M_I$. In this situation transitions between the different helicity states are strongly suppressed and the different helicity states of $N_I$ evolve independently. Effectively, they act as ``particle'' and ``antiparticle'' for the $N_I$, though this notion can of course not be taken literally for a Majorana field\footnote{At $T\gg M_I$ the average momentum of particles is so large that the Majorana mass is kinematically negligible, hence $\nu_R$ effectively behaves like at massless Weyl field up to corrections $\mathcal{O}[M_M/T]$.}.
Correlations between different helicities are usually negligible, and it is justified to describe the $N_I$ by two matrices $Y_N$ and $Y_{\bar N}$ for the two helicity states. 
In the mass basis, the total abundances $Y_I$ can be identified with the elements $(Y_N)_{II}+(Y_{\bar{N}})_{II}$, while the ``lepton asymmetry'' $L_I$ stored in the sterile flavour $N_I$ is proportional to $(Y_N)_{II}-(Y_{\bar{N}})_{II}$. In the fully flavoured regime, where leptogenesis during $N_I$ production and resonant leptogenesis usually take place, the kinetic equations then read 
\begin{eqnarray}
\label{kinequ1}
i HX \frac{dY_{N}}{d X}&=&[H_N, Y_{N}]-\frac{i}{2}\{\Gamma_N, Y_{N} 
- Y_N^{eq}\} +\frac{i}{2} Y_\alpha{\tilde\Gamma^\alpha_N}~,\\
i HX \frac{dY_{\bar{N}}}{d X}&=& [H_N^*, Y_{\bar{N}}]-\frac{i}{2}
\{\Gamma^*_N, Y_{\bar{N}} - Y_{\bar{N}}^{eq}\} -
\frac{i}{2} Y_\alpha{\tilde\Gamma_N^{\alpha *}}~,\label{kinequ2}\\
i HX \frac{dY_\alpha}{d X}&=&-i\Gamma^\alpha_L Y_\alpha +
i {\rm tr}\left[{\tilde \Gamma^\alpha_L}(Y_{N} -Y_N^{eq})\right]
-i {\rm tr}\left[{\tilde \Gamma^{\alpha*}_L}(Y_{\bar{N}} 
 -Y_{\bar{N}}^{eq})\right]~.
\label{kinequ3}
\end{eqnarray}
The flavour-matrix $H_N$ is the dispersive part of the effective $N_I$-Hamiltonian, which leads to sterile neutrino oscillations; the rate-matrices $\Gamma_N$, $\tilde{\Gamma}_N$ and $\Gamma_L^\alpha$ form the dissipative part, which acts as collision term. 
These transport coefficients have to be computed from the real- and imaginary parts of the $N_I$ and $l_{L,\alpha}$ self energies in thermal field theory. More precise definitions are given in \cite{Canetti:2012kh}, along with a derivation of (\ref{kinequ1})-(\ref{kinequ3}).

The equations (\ref{rate1})-(\ref{kinequ3}) are {\it rate equations} for momentum integrated abundances. They provide a good approximation if the momentum distributions are proportional for equilibrium distributions (this is often called {\it kinetic equilibrium}). If this is not the case, each momentum mode has to be tracked independently by a Boltzmann equation, see e.g. \cite{HahnWoernle:2009qn,Asaka:2011wq}. In numerical simulations, one of course has to sample a finite number of representative momenta. In addition, one needs to calculate the collision term as a function of momentum and temperature. This makes the treatment technically much more challenging.

\paragraph{Conceptual issues} - A question that has received much attention in recent years is whether Boltzmann equations for particle abundances are in principle suitable to describe leptogenesis.
For instance, early calculations based on Boltzmann equations were plagued with a double-counting problem \cite{Kolb:1979qa}.
This problem arises if one ``naively'' plugs vacuum S-matrix elements as collision terms into the classical Boltzmann equations by hand. 
Doing so, one somewhat artificially distinguishes the external and internal lines of a Feynman diagram; only the external lines are associated with physical particles 
that appear in the phase space distribution functions.
If one, in addition to the $N_I$ decays and inverse decays shown in figure \ref{leptogenesisfig}, introduces collision terms for $N_I$-mediated scatterings in this way, then one counts the same processes twice: If the intermediate $N_I$ in a scattering is on-shell, then the scattering $l_\alpha\Phi\rightarrow \Phi l_\beta$ is identical to an inverse decay $l_\alpha\Phi\rightarrow N_I$ followed by a decay $N_I\rightarrow\Phi l_\beta$ and should not be counted independently.\footnote{This side of the problem is related to the bookkeeping of physically different processes in the plasma. In addition, the use of vacuum S-matrix elements neglects finite density corrections to the propagators.
The incomplete inclusion of quantum statistics can lead to an overestimate of the generated asymmetry, which has happened repeatedly in the literature. Most recently this was pointed out for the {\it soft leptogenesis} \cite{Grossman:2003jv,D'Ambrosio:2003wy,Grossman:2004dz} mechanism in \cite{Garbrecht:2013iga}. Using a more complete treatment, the authors there conclude that soft leptogenesis \cite{Fong:2009iu} cannot explain $\eta_B$ for generic parameter choices.}  
This particular problem can be fixed by hand within the Boltzmann approach by performing a \textit{real intermediate state subtraction} (RIS) and using finite temperature field theory to calculate the amplitudes\footnote{See \cite{Frossard:2012pc} and references therein for a detailed account of this issue.}, but it reflects a deep conceptual issue of Boltzmann equations in leptogenesis.

In leptogenesis, the leading order processes that contribute to the generation of a matter-antimatter asymmetry come from an interference with loop diagrams, as e.g. shown in figure \ref{leptogenesisfig}, hence leptogenesis is a pure quantum effect. This is in contrast to many other processes that are well-described by Boltzmann equations, such as CMB decoupling or BBN.
Boltzmann equations are semi-classical.
The dynamical quantities are phase space distribution functions for particles and antiparticles, i.e. classical quantities. 
The collision terms, on the other hand, are calculated from S-matrix elements.
This treatment is based on a number of assumptions that may be questionable in the dense primordial plasma. 
The definition of particle numbers and the S-matrix are both based on the notion of asymptotic states, the meaning of which is not clear in a dense plasma, where particles are never ``far away'' from their neighbours. 
Even if the plasma can be described as an ensemble of (quasi)particles, the dispersion relations of these differ considerably from those in vacuum. 
Finally, the collision terms are affected by thermodynamic effects (such as Bose enhancement, Pauli blocking, Landau-Pomeranchuk-Migdal effect, possible enhancements from multiple scatterings...) and  cannot be calculated from the (vacuum) S-matrix.
The range of validity of the Boltzmann equations and size of possible corrections cannot be estimated within this framework and requires a first principles treatment. This has lead to a great interest in nonequilibrium quantum field theory \cite{Prokopec:2003pj,Prokopec:2004ic,Anisimov:2008dz,Drewes:2010pf,Drewes:2012qw,Millington:2012pf} and applications to leptogenesis \cite{Buchmuller:2000nd,DeSimone:2007rw,Garny:2009rv,Cirigliano:2009yt,Garny:2009qn,Anisimov:2010aq,Gagnon:2010kt,Garbrecht:2010sz,Beneke:2010dz,Beneke:2010wd,Anisimov:2010dk,Garny:2010nz,Herranen:2010mh,Herranen:2011zg,Fidler:2011yq,Garbrecht:2011aw,Garny:2011hg,Drewes:2012qw,Garbrecht:2012pq,Frossard:2012pc,Frossard:2013bra,Garbrecht:2013iga} in recent years.

At a fundamental level, the state of any quantum system can be described by an infinite tower of n-point correlation functions. In practice, it is usually sufficient to consider two-point functions. The leptonic charge can conveniently be described in terms of a correlation function known as {\it statistical propagator}\footnote{Here we use the notation of \cite{Anisimov:2010dk}.}  
$S_{L}^+(x_1,x_2)_{\alpha\beta}=\frac{1}{2}\langle l_{L,\alpha}(x_1) \overline{l_{L, \beta}} (x_2) - \overline{l_{L, \beta}}(x_2)^T l_{L,\alpha}(x_1)^T\rangle$, where the transposition applies to spinor indices (which we have suppressed here). 
The average $\langle\ldots\rangle={\rm Tr}(\varrho \ldots)$ includes quantum mechanical and thermodynamic fluctuations ($\varrho$ is the density operator \cite{Neumann1,Neumann2}).
The definition of this two-point function does not depend on any notion of asymptotic states or (quasi)particles. 
From the Fourier transform of the statistical propagator in the relative coordinate $x_1-x_2$  one can define the momentum integrated lepton abundance matrices 
\begin{eqnarray}\label{Yl}
Y_l(t)\equiv-\int\frac{d^4p}{(2\pi)^4}
\frac{
{\rm tr}[\gamma^0 S_{L}^+(p;t)]
}{s}
=-\int d\textbf{p}\frac{\textbf{p}^2}{2\pi^2}
\int\frac{d p_0}{2\pi}
\frac{
{\rm tr}[\gamma^0 S_{L}^+(p;t)]
}{s}
.\end{eqnarray}
Here $x_i=(t_i,\textbf{x}_i)$, $t=(t_1+t_2)/2$
\footnote{In the homogeneous and isotropic early universe there is no dependence on $\textbf{x}_1+\textbf{x}_2$.}
and the trace runs over Dirac indices.
The diagonal elements $Y_l(t)_{\alpha\alpha}$ correspond the lepton numbers in flavour $\alpha$, the off-diagonals describe correlations between different flavours during flavour oscillations. 
For the sterile neutrinos we define the statistical propagator $G_{IJ}^+(x_1,x_2)=
\frac{1}{2}\langle 
N_{I}(x_1) \bar{N_{J}} (x_2) 
- \bar{N_J}(x_2)^T N_I(x_1)^T
\rangle$. 
The matrices $Y_N$ and $Y_{\bar N}$ can be extracted from it as 
\begin{eqnarray}\label{YN}
Y_N(t)\equiv-\int\frac{d^4p}{(2\pi)^4}\frac{
{\rm tr}[{\rm P}_+\gamma^0 G^+(p;t)]}{s} \
,
\
Y_{\bar{N}}(t)\equiv-\int\frac{d^4p}{(2\pi)^4}\frac{
{\rm tr}[{\rm P}_-\gamma^0 G^+(p;t)]}{s}
\end{eqnarray}
where ${\rm P}_\pm$ are helicity projectors. 
The time evolution of $G^+$ and $S_L^+$ is governed by the Kadanoff-Baym equations \cite{KBE}, which can be obtained from a Dyson-Schwinger equation on a complex time path \cite{Schwinger:1960qe,Bakshi:1962dv,Bakshi:1963bn,Keldysh:1964ud}. The Kadanoff-Baym equations are exact integro-differential equations; There are different ways to derive effective kinetic equations  for $Y_N$, $Y_{\bar N}$ and $Y_l$ from them \cite{Beneke:2010dz,Anisimov:2010dk,Garbrecht:2011aw,Fidler:2011yq,Frossard:2012pc,Drewes:2012qw}.
In the fully flavoured or unflavoured regime, one can neglect active flavour oscillations and use a flavour basis in which $Y_l$ is diagonal.
In this case we can, assuming that sphalerons act rapidly on the time scale related to the $N_I$-evolution with a rate that is in good approximation given by the equilibrium one\footnote{See \cite{Burnier:2005hp,D'Onofrio:2012ni} for some discussion of this point.}, relate $Y_l$ to the quantities in (\ref{kinequ1})-(\ref{kinequ3}) as $Y_\alpha=(Y_l)_{\alpha\alpha}-(n_B-n_{\bar{B}})/3$.

Though the above approach provides a controlled approximation scheme to formally obtain effective kinetic equations from first principles, it is not yet a complete theory of leptogenesis because the computation of the transport coefficients in practice can be complicated \cite{Asaka:2006rw,Laine:2008pg,Anisimov:2010gy,Laine:2011pq,Salvio:2011sf,Besak:2012qm,Garbrecht:2013gd,Garbrecht:2013bia,Laine:2013lka,Biondini:2013xua}. It requires knowledge of the quasiparticle spectrum in the plasma \cite{Giudice:2003jh,Kiessig:2010pr,Kiessig:2011fw,Kiessig:2011ga,Garbrecht:2011aw,Fidler:2011yq,Miura:2013fxa} and inclusion of all processes in the plasma \cite{Giudice:2003jh,Garbrecht:2010sz,Besak:2012qm,Garbrecht:2013bia}. 
In the regime $M_I\ll T$ it is still not clear which effect corrections from soft and collinear gauge boson exchange have \cite{Anisimov:2010gy,Besak:2012qm,Garbrecht:2013gd,Garbrecht:2013bia,Laine:2013lka}.

The conclusion on this point is that it is possible to describe leptogenesis 
in the unflavoured \cite{Anisimov:2010aq,Anisimov:2010dk}, flavoured \cite{Beneke:2010wd,Beneke:2010dz} and resonant \cite{Garbrecht:2011aw,Garny:2011hg} regimes
by effective kinetic equations. By that, we mean differential equations that are local and of first order in time \cite{Drewes:2012qw}, which is in contrast to the non-local second order fundamental Kadanoff-Baym equations. 
The situation is relatively simple if the lepton asymmetry is generated at $T\ll M_I$ and the $M_I$ spectrum is hierarchical.
Then first principles calculations suggest that
in the unflavoured \cite{Anisimov:2010aq,Anisimov:2010dk} and fully flavoured \cite{Beneke:2010dz} cases standard Boltzmann equations give results
which are correct up to factors $\mathcal{O}[1]$, provided that the RIS is performed consistently. 
When flavour oscillations are relevant, one has to use their matrix-valued generalisation, the density matrix equations, as e.g. derived in \cite{Beneke:2010dz,Fidler:2011yq,Canetti:2012kh}. For $T>M$ or when two masses are degenerate, one can still formulate effective kinetic equations, such as (\ref{kinequ1})-(\ref{kinequ3}). 
However, inserting S-matrix elements into the classical Boltzmann equation may not give a correct quantitative description. Instead, they should be derived using controlled approximations to the full nonequilibrium field theory, which take into account the modified (quasi)particle spectrum and dispersion relations in the plasma, the effect of (possibly multiple) scatterings with quanta from the thermal bath and other thermodynamic effects. Though great progress has been made in this regime \cite{Gagnon:2010kt,Anisimov:2010gy,Garny:2011hg,Garbrecht:2011aw,Drewes:2012ma,Besak:2012qm,Canetti:2012kh,Garbrecht:2013gd,Garbrecht:2013bia,Laine:2013vpa}, the accurate quantitative description is not complete at this stage.

\section{Sterile neutrinos as dark matter}\label{DMsection}
Over the past 80 years\footnote{The term ``dark matter'' was already used in \cite{Oort,Zwicky:1933gu}, but in that context simply referred to non-luminous matter.  
Zwicky concluded that the velocities in the Coma-system are an ``unsolved problem'', which, however, did not receive much attention for decades.}, overwhelming evidence has accumulated that most of the mass in the observable universe is not composed of {\it baryonic matter}\footnote{In this context all matter that is composed of SM fermions (including leptons) is usually referred to as ``baryonic''.} as we know it. The presence of large amounts of non-luminous matter can on one hand be inferred by comparing the observed gravitational potential\footnote{The gravitational potential can be studied by tracking the dynamics of astrophysical objects on various scales, such as stars in galaxies (in particular rotation curves) as well as galaxies and ionized gas clouds within galaxy clusters, and by gravitational lensing (weak and strong).} to the density of visible matter. 
Independently of that, it is also required in order to explain the clustering of matter and formation of structures in the universe, e.g. the power spectrum shown in figure \ref{matterpowerspectra}. This process can be studied by the analysis of density perturbations in the CMB\footnote{The fact that most of the matter started to cluster before the decoupling of photons implies that it had decoupled considerable earlier, hence is not composed of (electrically charged) SM particles.} and by observations of the structure in the universe, such as galaxy surveys, gravitational lensing or the absorption in the spectra of distant quasars (``Ly$\alpha$ forest''). 

Numerous attempts to simultaneously explain these phenomena by modifications of gravity or the presence of compact macroscopic objects (such as lonely planets, black holes or other non-luminous star remnants) have failed. In contrast to that, they can easily be understood if one assumes the existence of (one or several) new particles that are massive, electrically neutral, long lived (compared to the age of the universe), collisionless and have a free streaming length in the matter dominated era that is sufficiently small to be consistent with the observed structure in the universe.

One can qualitatively distinguish three types of DM candidates. {\it cold dark matter} (CDM) is composed of particles that were nonrelativistic at the time of decoupling. {\it hot dark matter} (HDM) particles were relativistic at the time of decoupling and remain so into the matter dominated epoch, when structures can grow nonlinearly due to gravitational collapse. {\it warm dark matter} (WDM) is relativistic at the time of freezeout, but becomes nonrelativistic during the radiation dominated epoch.  

CDM scenarios predict that smaller scale objects form first and then merge into bigger structures (``bottom up'') \cite{Blumenthal:1984bp,Davis:1985rj}. This is in good agreement with observations as well as numerical simulations of LSS. On smaller scales, it was noticed more than a decade ago \cite{Klypin:1999uc,Moore:1999nt} and remains an unsolved puzzle that CDM simulations tend to predict more objects (satellite galaxies and subhalos) than observed. 
Furthermore, they do not reproduce shape of galactic halos (``cusp/core issue''). 
However, at this stage this is far from being a reason to disregard CDM because the discrepancy may be related to the resolution or other technical limitations of the simulations (to date, most of them are pure DM simulations that do not include baryons consistently). Furthermore, small objects are hard to observe if they fail to confine gas and form stars, see e.g. \cite{Bullock:2000wn,Benson:2001at,Somerville:2001km,Maccio':2009dx}, though this may not explain the discrepancy for objects that are {\it too big to fail} to attract gas, see e.g. \cite{BoylanKolchin:2011de}. A recent summary on issues and possible solutions in the $\Lambda$CDM model can be found in \cite{Primack:2012id}.

In contrast to that, HDM predicts that large structures form first \cite{Zeldovich:1969sb} because primordial fluctuations in the gravitational potential have been erased due to the large free streaming length of DM particles (``top down''). This is in contradiction to the observed LSS, which rules out the only potential DM candidate within the SM, the neutrinos $\nu_L$.\footnote{The active neutrino background can be viewed as a small HDM contribution to the matter density in the universe.}
 In WDM scenarios, structure formation on scales larger than the particles' average free streaming length $\lambda_{DM}$ is similar to CDM, but differences on smaller scales are expected to be visible in various observables, such as the matter power spectrum, halo density profile, halo mass function, subhalo density profile and subhalo mass function. Qualitatively, structures on scales smaller than $\lambda_{DM}$ are suppressed due to the free streaming.

\begin{figure}[!h]
\centering
\includegraphics[width=8cm]{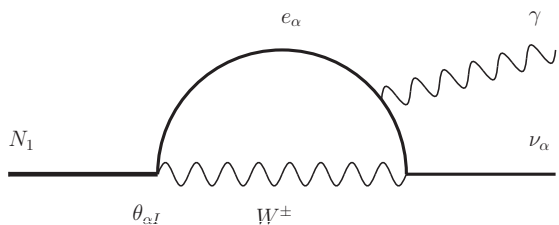}
\caption{Example for a contribution to the radiative decay of a DM sterile neutrino $N_1\rightarrow\gamma\upnu_\alpha$. The coupling of $N_1$ to the $W$-boson is suppressed by the mixing $\theta_{\alpha 1}$ as indicated.\label{DMdecay}}
\end{figure}

\begin{figure}
  \centering
    \includegraphics[width=12cm]{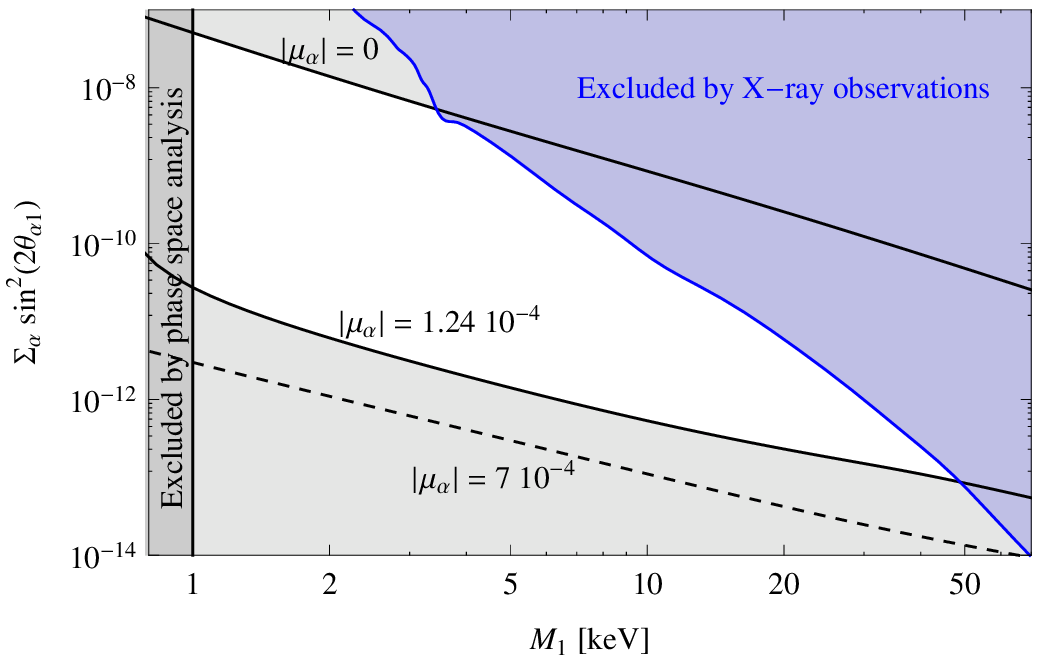}
    \caption{Different constraints on sterile neutrino DM mass and mixing, assuming $\nu_R$ have no interactions in addition to the Yukawa couplings in (\ref{L}). The blue region is excluded by X-ray observations, the dark grey region
$M< 1$ keV by the Tremaine-Gunn bound
\cite{Tremaine:1979we,Boyarsky:2008ju,Gorbunov:2008ka}.  
The solid black lines are ``production curves'' for thermal production.
For all points on the upper black line the observed $\Omega_{DM}$ 
is produced in the absence of lepton asymmetries (for $Y_{\alpha}=0$, non-resonant production) 
\cite{Laine:2008pg}.
Points on the lower solid black line yield the
correct $\Omega_{DM}$ for $|Y_{\alpha}|=1.24\cdot10^{-4}$ at $T=100$ MeV, the dashed line for $|Y_{\alpha}|=7\cdot10^{-4}$.
For these values the resonant production mechanism contributes.
The region between these lines is accessible for intermediate values of $Y_\alpha$.
We do not display bounds derived from structure formation because they depend on $Y_\alpha$ in a complicated way and there are considerable uncertainties. 
Plot taken from \cite{Canetti:2012vf}.\label{DMexclusion}}
\end{figure}

Sterile neutrinos are collisionless and can be very long lived, hence they are an obvious DM candidate. 
This scenario has been studied by a large number of authors, see e.g. \cite{Olive:1981ak,Dolgov:2000ew,Abazajian:2001nj,Abazajian:2001vt,Krauss:2002px,Viel:2005ha,Boyarsky:2005us,Boyarsky:2006fg,Boyarsky:2006ag,Watson:2006qb,Abazajian:2006yn,Abazajian:2006jc,RiemerSorensen:2006fh,Viel:2006kd,Hidaka:2006sg,Seljak:2006qw,Colin:2007bk,Belanger:2007dx,Boyarsky:2007ge,Boyarsky:2007ay,Boyarsky:2008xj,Boyarsky:2008mt,Boyarsky:2008ju,Loewenstein:2008yi,Gorbunov:2008ka,Kusenko:2009up,Herder:2009im,Abazajian:2009hx,Boyarsky:2009ix,Wu:2009yr,Boyanovsky:2010pw,Boyanovsky:2010sv,Kusenko:2010ik,Lovell:2011rd,Dunstan:2011bq,Merle:2012xq,Nemevsek:2012cd,Boyarsky:2006kc,Boyarsky:2006zi,Riemer-Sorensen:2006pi,Yuksel:2007xh,RiemerSorensen:2009jp,Boyarsky:2010ci,Watson:2011dw,Gorbunov:2008ka,deVega:2009ku,Destri:2012yn,Boyarsky:2009af,ESDark,Boyarsky:2012rt,Ishida:2013mva}
. 
Formally, sterile neutrinos are WDM candidates, though this classification is slightly ambiguous because momentum distribution may be very different from Fermi-Dirac.
Their properties are constrained  by the requirements outlined below. 
In what follows, we require that all DM is composed of sterile neutrinos. If they make up only a fraction of $\Omega_{DM}$, then the bounds weaken considerably. For instance, sterile neutrinos with eV masses could give a subdominant HDM contribution to $\Omega_{DM}$ if there are other CDM particles that ensure consistency with structure formation.
\paragraph{Stability} - $N_I$ particles are unstable and decay via the $\theta$-suppressed weak interaction. If they are DM, their lifetime must be longer than the age of the universe.
\begin{figure}[!h]
\centering
\includegraphics[width=10cm]{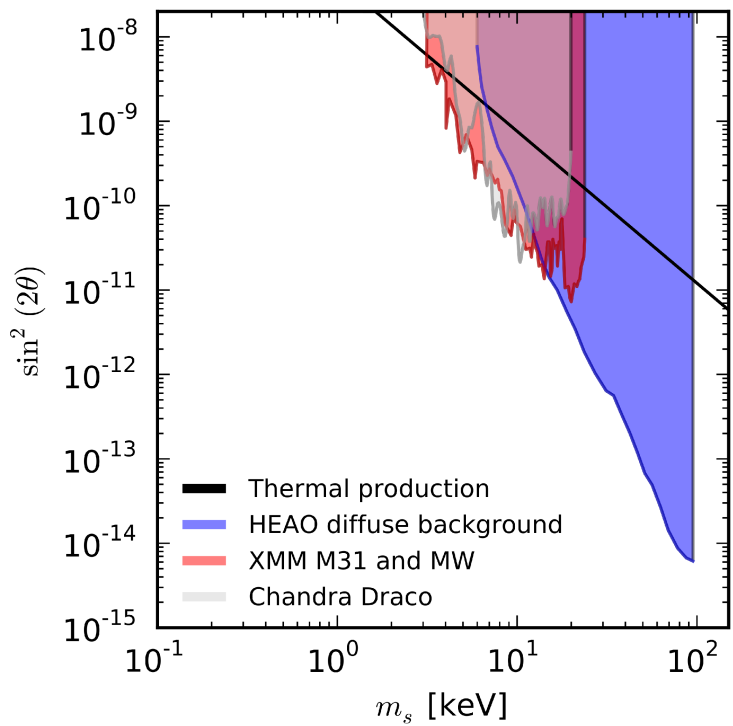}
\caption{Different constraints on sterile neutrino DM mass and mixing from X-ray observations. The constraints are take from  \cite{Boyarsky:2007ay} (red), rescaled
by a factor of two due to mass estimate uncertainties as recommended in \cite{Boyarsky:2009ix},  \cite{RiemerSorensen:2009jp} (grey) and  \cite{Boyarsky:2005us} (purple).
Some analyses have claimed stronger constraints,
but were later found to be too optimistic. In \cite{Boyarsky:2006fg,Abazajian:2006yn} it was found that \cite{Abazajian:2001vt} underestimated the flux by two orders of magnitude.
According to \cite{Boyarsky:2007ay} the mass was overestimated in \cite{Watson:2006qb} leading to too restrictive constraints. The constraints in \cite{Yuksel:2007xh} might be too restrictive due to the choice of source profile
\cite{Boyarsky:2007ge}. The spectral resolution seems to be overestimated in \cite{Watson:2011dw}, 
cf. Chandra Proposers Guide; this seems to be the main reason for the stronger bounds used in \cite{Merle:2013ibc}. 
Thanks to Signe Riemer-S\o
rensen for the plot and comments.
\label{signe}}
\end{figure}
\paragraph{X-ray bounds} - The radiative decay $N_I\rightarrow\gamma\upnu_\alpha$ via processes as shown in figure \ref{DMdecay}
predicts the emission of photons with energy $M_I/2$ from DM dense regions \cite{Boyarsky:2009af,Boyarsky:2006fg}\footnote{Here we assume that $\nu_R$ have (at least) the Yukawa interactions in (\ref{L}). If such interactions are suppressed, the RH neutrinos that constitute DM can be much heavier and may even be related to the observed excess \cite{Weniger:2012tx} of $\gamma$-ray emission near the galactic center  \cite{Bergstrom:2012bd}. }. 
The non-observation of such signal in the data of different X-ray observatories (such as XMM, Chandra and Suzaku) \cite{Abazajian:2001vt,Boyarsky:2005us,Boyarsky:2006kc,Abazajian:2006yn,Abazajian:2006jc,Boyarsky:2006zi,Boyarsky:2006ag,Watson:2006qb,Riemer-Sorensen:2006pi,Boyarsky:2006fg,Boyarsky:2007ay,Boyarsky:2007ge,Yuksel:2007xh,Loewenstein:2008yi,RiemerSorensen:2009jp,Boyarsky:2009ix,Boyarsky:2010ci,Watson:2011dw,Essig:2013goa} imposes upper bounds in the mass-mixing plane; these are displayed in figure \ref{signe}. 
In figure \ref{DMexclusion} X-ray bounds are combined with other constraints. 
For decaying DM, the signal strength scales only linear with the DM column density along the line of sight, hence one can expect a signal from wider range of astrophysical objects than for annihilating DM. 
This makes future searches promising despite the fact that there are various astrophysical sources in the keV range.
The perspectives for future searches have e.g. been outlined in \cite{Boyarsky:2009af,Herder:2009im,ESDark,Boyarsky:2012rt,Abazajian:2012ys}.
Finally, it is worth to emphasize that this scenario is falsifiable: if DM is made of RH neutrinos that are produced thermally in the early universe through their mixing with active neutrinos, their mass and mixing are constrained in all directions in figure \ref{DMexclusion} and this scenario can be found or falsified on human time scales.
\paragraph{Phase space analysis} - As a fermionic DM candidate, the mass of RH neutrinos is constrained by the Tremaine-Gunn bound \cite{Tremaine:1979we} on the phase space density in the Milky way's dwarf spheroidal galaxies~\cite{Boyarsky:2008ju,Gorbunov:2008ka}. This yields a lower bound of $M_N>1$ keV.
In \cite{deVega:2009ku} it has been shown that phase space arguments on the galactic scale indeed favour a keV DM particle. The same authors have argued in \cite{Destri:2012yn} that, due to the high phase space density, a quantum mechanical treatment is necessary on these scales, which may solve the ``cusp issue''.
\paragraph{Production in the early universe}
It is not known when and how DM was produced in the early universe. If it is composed of sterile neutrinos, then there are several possible production mechanisms.
\begin{itemize}
\item \textbf{Thermal production via mixing} (non-resonant) - If $\nu_R$ have nonzero Yukawa couplings $F$ (i.e. $m_D\neq0$), then they are produced thermally from the primordial plasma via their mixing with the SM neutrinos $\nu_L$ \cite{Dodelson:1993je}. If this production was efficient enough to bring them into thermal equilibrium, their density would be bigger than $\Omega_{DM}$ and ``overclose'' the universe unless they are diluted by entropy production at some later stage \cite{Asaka:2006ek,Bezrukov:2009th,Bezrukov:2012as}. This puts an upper bound on the mixing angle, given by the upper production curve in figure \ref{DMexclusion}. The production can yield the correct $\Omega_{DM}$ if $\theta$ is small enough that the DM sterile neutrino never reach thermal equilibrium. In that case they still have a momentum distribution that is proportional to a Fermi-Dirac spectrum \cite{Dodelson:1993je,DiBari:1999ha,Dolgov:2000ew,Asaka:2006nq}, which 
makes them a WDM candidate that is at least disfavoured  by structure formation arguments (see below).
\item \textbf{Resonant thermal production} - The properties of (quasi)particles in the primordial plasma are modified by the interactions with the medium \cite{Klimov:1982bv,Klimov:1981ka,Weldon:1982bn}. In the presence of a lepton asymmetry, the Mikheev-Smirnov-Wolfenstein effect \cite{Wolfenstein:1977ue,Mikheev:1986gs} can lead to a level crossing between active and sterile neutrino dispersion relations \cite{Shi:1998km,Laine:2008pg}. This results in a resonant enhancement of the sterile neutrino production rate. The resulting spectrum is nonthermal, with higher occupation numbers for low momentum modes \cite{Laine:2008pg}. Production curves for two different values of the asymmetry are also given in figure \ref{DMexclusion}. Note that these have to be taken with some care because in the calculation of the production efficiency \cite{Laine:2008pg} it was assumed that $Y_e=Y_\nu=Y_\tau$. In reality the asymmetries in different flavours may be different and can even have opposite sign (see e.g. \cite{Canetti:2012kh} for a particular model). 
\item \textbf{Thermal production beyond the SM} - The fields $\nu_R$ are singlet under SM the gauge group and only interact with SM particles via their Yukawa couplings $F$. However, if there exists and extended ``hidden sector'' \cite{Das:2010ts} or extended Higgs sector, they may have additional interactions with these. They may also be charged under some extended gauge group that is broken at high energies, as e.g. in the left-right-symmetric model \cite{Nemevsek:2012cd}. 
Such additional interactions would contribute to the thermal production \cite{Bezrukov:2009th} in the early universe. 
Since they usually increase the production rate, the universe may overclose unless the DM abundance is diluted by additional entropy production at later times \cite{Scherrer:1984fd}.
\item \textbf{Nonthermal production} - RH neutrinos can also be produced nonthermally, e.g. due to a coupling to an inflaton \cite{Shaposhnikov:2006xi,Anisimov:2008qs,Bezrukov:2009yw}, the SM Higgs \cite{Bezrukov:2008ut}, other scalars \cite{Petraki:2007gq,Shoemaker:2010fg} or modified gravity \cite{Gorbunov:2010bn}. 
\end{itemize} 
All these scenarios are constrained by the requirements to produce the correct $\Omega_{DM}$ and a momentum distribution that is consistent with the observed LSS.
\paragraph{Structure formation} - The masses dictated by the above bounds suggest that, if RH neutrinos mix with LH neutrinos, then they formally are WDM, but their momentum distribution can be non-thermal. Structure formation in WDM scenarios on scales above the free streaming length is similar to CDM. 
On smaller scales the formation of structures is affected in different ways, the study of which is in general complicated due to the nonlinear nature of the clustering process. Most importantly, one expects a cut in the matter power spectrum at scales below the free streaming length \cite{Bond:1980ha}. 
In addition, the suppression of small scale structures should be visible in the halo mass function (which counts the number of haloes per unit volume per unit mass at given redshift) \cite{Benson:2012su} and affect the gravitational collapse leading to the formation of the first stars \cite{SommerLarsen:2003bv,O'Shea:2006tp}.
Unfortunately, these arguments do not directly constrain $M_N$, but the free streaming length. 
Furthermore, structure formation is a highly nonlinear problem in the regime $\delta\rho_k/\rho>1$ that can only be studied quantitatively with expensive many body simulations.
With few exceptions \cite{Lovell:2011rd}, these simulations assume that the DM distribution is proportional to a Fermi-Dirac distribution \cite{Dolgov:2000ew}.
If this is the case,  then the comoving free streaming length $\lambda_{DM}$ is simply related to the mass by \cite{Bond:1980ha,Boyarsky:2012rt}
\begin{equation}\label{freestream}
\lambda_{DM}\sim 1 {\rm Mpc}\left(\frac{\rm keV}{M}\right) \frac{\langle p_N\rangle}{\langle p_\upnu\rangle}
,\end{equation}
where $p_N$ and $p_\upnu$ are the spatial momenta of sterile and active neutrinos and $M$ is the sterile neutrino mass. 
Such spectra are produced by the non-resonant thermal production from the primordial plasma.
In that case Ly$\alpha$ forest observations strongly constrain the viability of WDM \cite{Bode:2000gq,Hansen:2001zv,viel:063534,Viel:2006kd,Seljak:2006qw,Colin:2007bk,Viel:2007mv,Boyarsky:2008xj,deNaray:2009xj,Boyanovsky:2010pw,Boyanovsky:2010sv,VillaescusaNavarro:2010qy,Lovell:2011rd};
for sterile neutrinos they impose a lower bound of $M_N> 8$ keV \cite{Boyarsky:2008xj}. Then it may be very difficult to distinguish RH neutrino WDM from CDM observationally \cite{Boyarsky:2008xj,Schneider:2011yu}, see also \cite{Power:2013rpw}. 
If, on the other hand, RH neutrino DM is produced by some other mechanism and has a nonthermal spectrum, the relation (\ref{freestream}) between mass and free streaming length does not apply, and simulations of structure formation based on thermal WDM do not allow to draw any general conclusions. 
In case of the resonant thermal production mechanism, the resulting spectrum has been calculated for $Y_e=Y_\mu=Y_\tau$ \cite{Laine:2008pg,Boyarsky:2009ix} . It can be viewed as a superposition of a WDM component and a nonthermal cold component \cite{Boyarsky:2008mt,Boyarsky:2008xj}, where the Ly$\alpha$ forest bounds allow the warm component to be much ``warmer'' than in conventional WDM scenarios.  
First simulations \cite{Lovell:2011rd} indicate that this scenario may perform better than CDM in predicting the abundance of small scale structures (such as satellite/dwarf galaxies), but the question is not settled at this stage.
On one hand, only little is known about how much structure actually exists at small (sub-galactic) scales. While lensing flux and stellar streams suggest the existence of subhalos \cite{Primack:2012id}, there is no direct observation of such structures and their existence is disfavoured by some stability considerations \cite{Lora:2012hr}. 
On the other hand there are no systematic studies of structure formation in the nonlinear regime that include the nonthermal component. Arguments against sterile neutrino DM are essentially based on extrapolations of results from simulations that assume purely thermal spectra.

\section{A theory of almost everything}\label{nuMSMsec}
Right handed neutrinos $\nu_R$, described by the Lagrangian (\ref{L}), provide plausible explanations for the phenomena (I), (II), (III), (i) and (ii) named in the introduction. 
In this section we address the question how many of these phenomena can be explained {\it simultaneously} by RH neutrinos {\it alone}.

\paragraph{Minimal case $n=3$} - The Lagrangian (\ref{L}) with $n=3$ RH neutrinos and $M_M$ at or below the electroweak scale \cite{Asaka:2005an,Asaka:2005pn} is an extension of the SM motivated by the principle of minimality (or Ockham's razor); it makes only minimal modifications to the known principles and particles in nature. 
There is no modification to the SM gauge group, the number of fermion families is unchanged, there is no modification to the bosonic field content of the SM or the mechanism of electroweak symmetry breaking, and no new scale above the electroweak scale is introduced\footnote{This implies that the hierarchy problem of the SM can be avoided \cite{
Shaposhnikov:2007nj,Shaposhnikov:2008xb,Shaposhnikov:2008xi}.}. In \cite{Canetti:2012vf} it has been shown that this model can indeed simultaneously explain (I)-(III); that is, all known {\it evidence} for particle physics beyond the SM. This minimal scenario is known as Neutrino Minimal Standard Model ($\nu$MSM).
Various aspects of this model have been studied in the past by different authors \cite{Asaka:2005pn,Asaka:2005an,Boyarsky:2005us,
Asaka:2006rw,Bezrukov:2005mx,Sahu:2005fe,Shaposhnikov:2006nn,Asaka:2006ek,Boyarsky:2006zi,Boyarsky:2006fg,Boyarsky:2006jm,Shaposhnikov:2006xi,Asaka:2006nq,
Gorbunov:2007ak,Gorbunov:2007zz,Bezrukov:2007qz,Bezrukov:2007ep,Bezrukov:2008ut,GarciaBellido:2008ab,Shaposhnikov:2008xb,
Shaposhnikov:2008xi,Laine:2008pg,
Anisimov:2008qs,Shaposhnikov:2008pf,Boyarsky:2009ix,Gorkavenko:2009vd,Canetti:2010aw,Asaka:2010kk,
Asaka:2011pb,GarciaBellido:2011de,Bezrukov:2011sz,Ruchayskiy:2011aa,
Gorkavenko:2012mj,Ruchayskiy:2012si,Canetti:2012vf,Drewes:2012ma,Asaka:2012bb,Allison:2012qn}, see \cite{Boyarsky:2009ix,Canetti:2012kh,Canetti:2013qna} for a detailed introduction and review. 

In the $\nu$MSM, two sterile neutrinos ($N_{2,3}$) have degenerate masses between roughly a GeV and the electroweak scale; the third one ($N_1$) has a mass in the keV range and acts as DM candidate. $N_{2,3}$ generate active neutrino masses via the seesaw mechanism; at the same time, the CP-violating oscillations during their thermal production at temperatures $T>T_{EW}$ produce the baryon asymmetry of the universe via the mechanism outlined in section \ref{LeptogenfromOsc}. 
The lepton asymmetries generated during $N_{2,3}$ production are washed out after sphaleron freezeout at $T\sim T_{EW}$.
The freezeout ($T\sim M_{2,3}$ ) and decay ($T<M_{2,3}$) of the $N_{2,3}$ particles 
produce new, late time lepton asymmetries, which can be orders of magnitude bigger than the baryon asymmetry. The late time asymmetries are capable of enhancing the rate of $N_1$ production sufficiently to explain the observed $\Omega_{DM}$ for $|Y_\alpha|\gtrsim 10^{-5}$, while being in agreement with all known constraints on DM properties. 

The minimal extension of the particle content in the $\nu$MSM comes at the price of a "fine tuning" in the mass splitting of two RH neutrinos: $M_2$ and $M_3$ have to be equal to a level of precision $|M_2-M_3|/(M_2+M_3)\sim 10^{-13}$ \cite{Canetti:2012kh}. This degeneracy may be related to new symmetries \cite{Shaposhnikov:2006nn,Shaposhnikov:2008pf}, but cannot be explained in the framework of the $\nu$MSM itself. 
The tuning mainly arises from the requirement to produce a late time lepton asymmetry $|Y_\alpha|\gtrsim 10^{-5}$; this is necessary to sufficiently enhance the $N_1$ production to explain the observed $\Omega_{DM}$. If some of the asymmetries generated before sphaleron freezeout can be preserved from washout \cite{Boyarsky:2011uy} or $N_1$ is produced by some other mechanism (e.g. during reheating), then the required degree of degeneracy reduces to $\sim 10^{-3}$.
This weaker degeneracy is necessary for baryogenesis, i.e. to explain the observed $\Omega_B$ from CP-violating oscillations of two sterile neutrinos $N_{2,3}$. The third sterile neutrino $N_1$ cannot contribute significantly to leptogenesis if it is DM because in this case its couplings $F_{\alpha 1}$ are constrained to be tiny by X-ray bounds on its decay width, cf. section \ref{DMsection}.
For the same reason it also cannot make a significant contribution to neutrino masses via the seesaw mechanism (\ref{activeneutrinomasses}); this implies that the lightest active neutrino is effectively massless in the $\nu$MSM.
The smallness of the Yukawa couplings also implies that the lepton flavour violation they induce is too small to be observed in muon-to-electron conversion experiments in foreseeable time \cite{Canetti:2013qna}.
Constraints from to $0\nu\beta\beta$-decays currently also do not constrain the allowed region, but will be relevant when $m_{ee}$ is probed at the level of $10^{-2}$ eV \cite{Asaka:2013jfa}.\footnote{The contributions from $N_I$ exchange in the $\nu$MSM are negligible \cite{Bezrukov:2005mx} or negative \cite{Asaka:2011pb}.} 

\begin{figure}[!h]
\centering
\begin{tabular}{c}
\includegraphics[width=12cm]{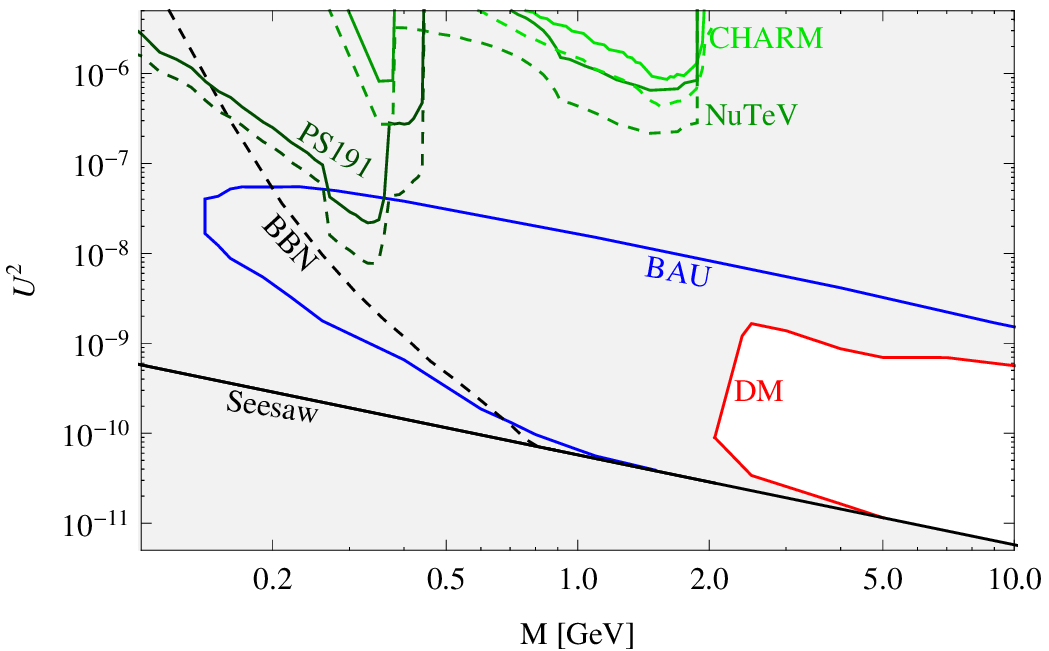} \\
\includegraphics[width=12cm]{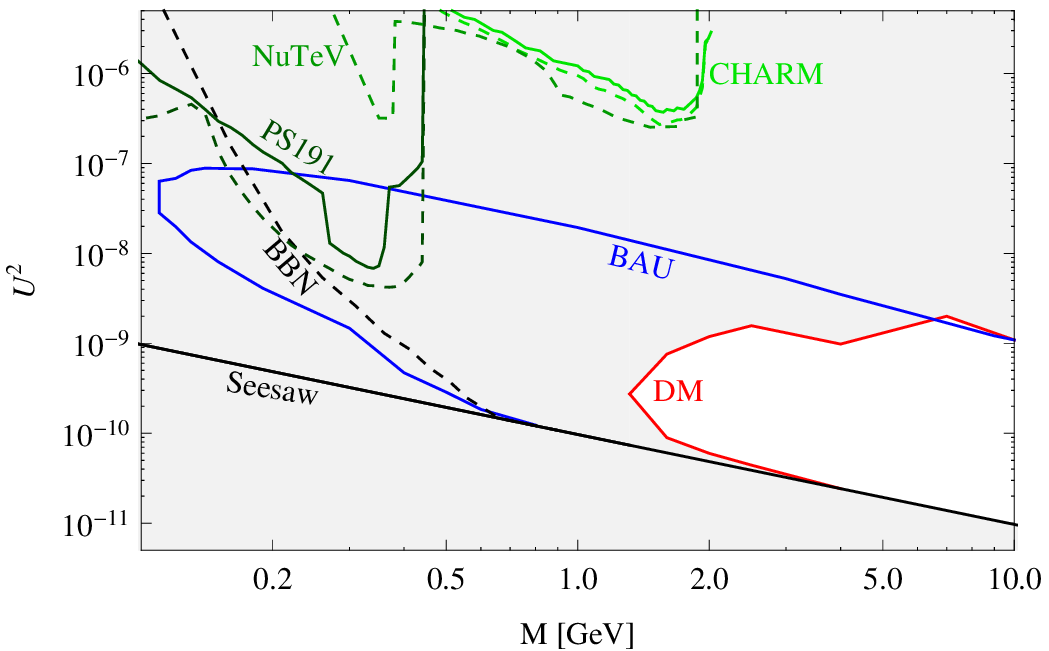}
\end{tabular}
\caption{Constraints on the mean value $M$ of the degenerate $N_{2,3}$ masses $M_{2,3}\simeq M$ and
mixing $U^2={\rm tr}(\theta^\dagger\theta)$ in the 
$\nu$MSM; upper panel - normal hierarchy, lower panel - inverted
hierarchy. In the region between the solid blue ``BAU'' lines, the
observed $\Omega_B$ can be generated. 
The observed $\Omega_{DM}$ can be explained in terms of sterile neutrinos inside the solid red
``DM'' line (there the lepton asymmetry at $T=100$ MeV can
be large enough to resonantly enhance the $N_1$ production).
The CP-violating phases were chosen to maximize the
asymmetry at $T=100$ MeV. The regions below the solid black ``seesaw''
line and dashed black ``BBN'' line are excluded by neutrino
oscillation experiments and BBN, respectively. The areas above the
green lines of different shade are excluded by direct search
experiments, as indicated in the plot. The solid lines are {\it
exclusion plots} for all choices of $\nu$MSM parameters, for the
dashed lines the phases were chosen to maximize the late time
asymmetry, consistent with the red line.
We do not show constraints from to $0\nu\beta\beta$-decays, which currently do not constrain the allowed region, but will be relevant when $m_{ee}$ is probed at the level of $10^{-2}$ eV \cite{Asaka:2013jfa}.
Plot taken from \cite{Canetti:2012vf}.
\label{BAUDMnuMSM}}
\end{figure}

The particles $N_2$ and $N_3$ can be searched for in collider experiments, cf. section \ref{collidersec}. They also lead to $L$-violation that may be seen in $0\nu\beta\beta$-decay searches \cite{Asaka:2011pb,Asaka:2013jfa}. Known bounds on their mass and mixing are summarized in figure \ref{BAUDMnuMSM}. The DM candidate $N_1$ is too weakly coupled to be studied at colliders, but can be found by indirect DM searches for the X-ray emission line from the decay $N_1\rightarrow \gamma\upnu$, cf. figure \ref{DMdecay}. Its properties are also constrained by phase space and structure formation considerations. Figure \ref{DMexclusion} summarizes different bounds on $N_1$ mass and mixing. 

\paragraph{More than three sterile neutrinos} - 
The required amount of ``fine tuning'' can be considerably reduced if there are more than three sterile neutrinos. For instance, if three sterile neutrinos participate in baryogenesis, then there is no need for degenerate masses even if they are as light as a few GeV \cite{Drewes:2012ma}. If one at the same time requires one sterile neutrino with keV mass to compose DM, then that means there must be at least $n=4$ of them altogether.
If one, in addition to the evidence (I)-(III), also wants to address the hints (i) and (ii) in this framework, then there must be least one more sterile neutrino with an eV mass. This light sterile neutrino, responsible for the oscillation anomalies, would leave the thermal history of the universe before BBN essentially unaffected due to the smallness of its couplings enforced by the seesaw relation (\ref{activeneutrinomasses}). 
At the same time it can be responsible for the $N_{\rm eff}>3$ inferred from BBN and the CMB, see section \ref{ThermalHistory}.
The late time lepton asymmetries generated by the freezeout and decay of its $M_I\gtrsim 1$ GeV siblings could help to suppress the efficiency of its thermal production and ease the tension between oscillation anomaly and dark radiation signals, see section \ref{RHasDR}.
Thus, $n\geq 4$ RH neutrinos described by the Lagrangian (\ref{L}) alone can be capable of explaining the points (I)-(III) as well as (i) and (ii) simultaneously without ``new physics'' above the electroweak scale; that is all confirmed fundamental physics phenomena that cannot be explained in the SM except those that are likely to involve accelerated cosmic expansion.
Hence, this framework at the current stage provides a phenomenologically complete description of particle physics.
\section{Conclusions}\label{conclusions}
Right handed neutrinos provide plausible explanations for neutrino oscillations,
the observed dark matter and the baryon asymmetry of the universe, which cannot be understood within the SM. 
They may also be responsible for the anomalies reported by some short baseline neutrino experiments and hints for dark radiation in cosmological data.
Right handed neutrinos can be embedded naturally into bigger frameworks, such as left-right symmetric theories, grand unified theories, supersymmetric extensions of the SM and various models of neutrino mass generation.
The different phenomena point towards different right handed neutrino mass ranges, which may of course coexist in nature.  
Several of these can be probed by direct and indirect laboratory searches, astrophysical observations and cosmological data in the near future.
We summarize well-motivated mass ranges and their phenomenological relevance in the overview table appendix \ref{overview}.

\section*{Acknowledgements}
I would like to thank Signe Riemer-S\o rensen and Yvonne Wong for valuable discussions on astrophysical and cosmological bounds as well as Fedor Bezrukov for his explanations on neutrinoless double $\beta$-decay. Thanks also to Oleg Ruchayskiy and Viviana Niro for their comments on the interplay between X-ray bounds and neutrinoless double $\beta$-decay.
I am very grateful to Bj\"orn Garbrecht, Pasquale Di Bari, Fedor Bezrukov, Alexander Merle, Signe Riemer-S\o rensen and Yvonne Wong for their comments on individual sections and to Mikhail Shaposhnikov for proof-reading of the manuscript. 
Finally, I would like to thank Nancy Lu for her patience on the numerous evenings I spent writing this review during the past few months.
This work was supported by the Gottfried Wilhelm Leibniz program of the Deutsche Forschungsgemeinschaft.

\appendix
\begin{landscape}
\section{Overview Table: Majorana mass scales and observables}\label{overview}
\setlength{\voffset}{20mm}
\scriptsize
\begin{center}
\begin{tabular}{| c || c | c | c | c | c | c | c |}
\hline
$M_M$ & Motivation  & $\upnu$-oscillations  & laboratory searches & CMB & BBN & DM & Leptogenesis \\
\hline\hline
$\lesssim$eV  & \begin{tabular}{c}$\upnu$-oscillations anomalies,\\ dark radiation\end{tabular} & 
\begin{tabular}{c}
\green{massses by seesaw,}$^a$\\
\green{explain anomalies}$^b$
\end{tabular}
& 
\begin{tabular}{c}
\green{oscillation anomalies,}\\
\green{$\beta$-decays} 
\end{tabular} & \green{explain $N_{\rm eff}>3$}$^b$ & \green{may explain $N_{\rm eff}>3$}$^b$ & \red{no} & \red{no} \\ 
\hline
keV  & DM & \red{no if DM}\green{$^c$} & 
\begin{tabular}{c}
\green{direct searches?}\red{$^d$},\\
\green{nuclear decays?}\red{$^d$} 
\end{tabular}
& \begin{tabular}{c}\green{act as DM,}\\ \red{no effect on $N_{\rm eff}$}\end{tabular} & \begin{tabular}{c} \red{effect on $N_{\rm eff}$}\\ \red{too small if DM}\end{tabular}  & \green{good candidate} & \red{no}  \\ 
\hline
MeV  & testability, why not? & \green{masses by seesaw}  & 
\begin{tabular}{c}
\green{intensity frontier},\\ 
\green{$0\nu\beta\beta$}
\end{tabular}
& \red{unaffected} & \begin{tabular}{c}\green{constrains}\\ \green{$M_I\gtrsim 200$ MeV}
\end{tabular} & \red{no}\green{$^e$}
& \begin{tabular}{c}\green{possible}\\ \green{(fine tuning)}\end{tabular} \\ 
\hline
GeV  & \begin{tabular}{c}testability,\\ minimality\end{tabular}   & \green{masses by seesaw} & \begin{tabular}{c}
\green{intensity frontier,}\\ \green{EW precision data,}\\ 
\green{$0\upnu\beta\beta$}\\
\end{tabular} 
& \red{unaffected} & \red{unaffected} & \red{no}\green{$^e$} & \green{possible} \\ 
\hline
TeV  & \begin{tabular}{c}minimality,\\testability\end{tabular}  & \green{masses by seesaw} & \green{LHC} & \red{unaffected} & \red{unaffected} & \red{no}\green{$^e$} & \green{possible} \\ 
\hline
$\gtrsim 10^9$GeV  & \begin{tabular}{c}grand unification,\\ ``naturally'' small $\upnu$-masses\end{tabular}  & \green{masses by seesaw} & \red{too heavy to be found} & \red{unaffected} & \red{unaffected} & \red{no}\green{$^e$} & \begin{tabular}{c}\green{works naturally}
\end{tabular}\\  
\hline
\end{tabular}
\end{center}
Colour code: \green{green = can affect}, 
\red{red = does not affect}\\
\black{$^a$ This assumes that the observed $\Delta m_{\rm atm}$ and $\Delta m_{\rm sol}$ are generated by $\nu_R$ other than those responsible for the anomalies, i.e. $n>2$.}\\
\black{$^b$ Sterile neutrinos with masses and mixings suggested by the oscillation anomalies would be in thermal equilibrium in the early universe, hence increase $N_{\rm eff}$ by one unit per species. In $\Lambda$CDM cosmology, this is in conflict with recent CMB data and HDM constraints. If some mechanism prevents the sterile neutrinos from thermalizing or there are deviations from the standard $\Lambda$CDM model, this conflict can be avoided.}\\
$^c$ $\nu_R$ with keV-masses can generate neutrino masses via the seesaw mechanism, but then they are too short lived to be DM. If they are DM, then their Yukawa coupling is too small to contribute to the seesaw - they can {\it either} be DM {\it or} contribute to the seesaw mechanism.\\
$^d$ It is disputed whether the signal can be distinguished from the active neutrino background; for the case that keV sterile neutrinos compose all DM the author considers it very unlikely that such searches are successful because of the astrophysical constraints on the mixing angle.\\
$^e$ This applies to sterile neutrinos thermally produced via their mixing. Sterile neutrinos with $M_I\gg$ keV can be DM if $F\simeq0$ ensures their stability and the production in the early universe is due to an unknown interaction.
\normalsize
\end{landscape}

\section{Dirac and Majorana masses}\label{appendixDiracMajorana}
This appendix summarizes some basic aspects of Majorana fermions, see \cite{Fukugita:2003en,Mohapatra:1998rq} for details. 
In the SM, matter is composed of fermions.
In relativistic quantum field theory these can be described by two component (Weyl) spinors $\psi_L$ and $\psi_R$, which transform under the (irreducible) left or right handed representations of the Poncair\'e group, respectively.
For historical 
\cite{Dirac:1928hu} 
and computational reasons fermions are, however, often represented by four component spinors \cite{Pal:2010ih}. 
In this review we adopt this common ``overnotation''; using the Weyl representation of $\gamma^\mu$ matrices, we define four component spinors $\Psi_L=(\psi_L,0)^T$ and $\Psi_R=(0,\psi_R)^T$. 
Consider sets of left and right handed spinors $\Psi_{L,i}$ and $\Psi_{R,j}$, where the indices $_i$ and $_j$ run from $1$ to $n$ and $m$, respectively, and label individual fields (``flavours'').
The most general free Lagrangian reads
\begin{equation}\label{freeL}
\frac{i}{2}\left(\overline{\Psi_L}\slashed{\partial}\Psi_L 
+ \overline{\Psi_R}\slashed{\partial}\Psi_R\right)
-\overline{\Psi_L}m_D\Psi_R
-\frac{1}{2}\left(
\overline{\Psi_L}m_M\Psi_L^c
+\overline{\Psi_R}M_M\Psi_R^c\right)
+h.c.  ,
\end{equation}
where we have suppressed flavour indices ($\Psi_R$, $\Psi_L$ are now flavour vectors with components $\Psi_{R,i}$ and $\Psi_{L,j}$). 
The matrices $m_D$, $m_M$ and $M_M$ can be combined into a $(n+m)\times(n+m)$ matrix $\mathfrak{M}$, the physical mass squares are the eigenvalues of $\mathfrak{M}\mathfrak{M}^\dagger$. The mass term reads
\begin{eqnarray}\label{massmatrixexplicit}
\frac{1}{2}
(\overline{\Psi_L} \  \overline{\Psi_R^c})
\mathfrak{M}
\left(
\begin{tabular}{c}
$\Psi_L^c$\\
$\Psi_R$
\end{tabular}
\right) + h.c.
\equiv
\frac{1}{2}
(\overline{\Psi_L} \ \overline{\Psi_R^c})
\left(
\begin{tabular}{c c}
$m_M$ & $m_D$\\
$m_D^T$ & $M_M^\dagger$
\end{tabular}
\right)
\left(
\begin{tabular}{c}
$\Psi_L^c$\\
$\Psi_R$
\end{tabular}
\right) + h.c. 
\end{eqnarray}
All fermions in the SM are charged under some gauge group, thus any intrinsic mass terms are forbidden by gauge invariance.
Terms as $m_D$ are generated by the Higgs mechanism. 
For $M_M=0$, $m_M=0$ one can combine pairs of left and right handed fields in (\ref{freeL}) into Dirac spinors $\Psi_i\equiv\Psi_{R,i} + \Psi_{L,i}$ and write $\mathcal{L}=\overline{\Psi_i}(i\slashed{\partial}-m_D)\Psi_i$.\footnote{For $m\neq n$ or if not all eigenvalues of $M_M$ are vanishing this is obviously not possible for all $\Psi_{L,i}$ and $\Psi_{R,j}$.} 
This is the reason why one can describe charged leptons and quarks by Dirac spinors though the basic building blocks of matter are Weyl fermions: the conserved charge related to the unbroken $U(1)$ subgroup in the electroweak theory
leads to  mass matrix that allows to combine all Weyl fields into Dirac spinors. 
If, on the other hand, $m_M, M_M\neq0$, then one can form $n+m$ Majorana spinors 
\cite{Majorana:1937vz}
. In the simplest case, when $m_D=0$ and $m_M$ and $M_M$ are diagonal, 
one can define these as 
$\chi_i=\Psi_{L,i}+\Psi_{L,i}^c$ and 
$\uppsi_j=\Psi_{R,j}+\Psi_{R,j}^c$. 
Formally $\chi_i$ and $\uppsi_i$ are four component objects, and the Lagrangian can be written in analogy to the Dirac equation, $\frac{1}{2}(\overline{\chi}(i\slashed{\partial}-m_M)\chi + \overline{\uppsi}(i\slashed{\partial}-M_M)\uppsi)$, but they obey the additional Majorana conditions $\chi_i=\chi_i^c$, $\uppsi_i=\uppsi_i^c$ and have only two independent components. Due to this property the neutral fermions can annihilate with themselves (are their own antiparticles).

Because of these considerations $m_D$ is usually referred to as {\it Dirac mass term}, while $m_M$ and $M_M$ are called {\it Majorana mass terms}. However, apart from the two special cases discussed above, they have not much to do with the appearance of Dirac or Majorana particles. Even for $m_D=0$ the particle spectrum may contain Dirac spinors. Consider, for instance, a set of three fields $\Psi_{L,1}$, $\Psi_{R,1}$ and $\Psi_{R,2}$ with $m_D,m_M=0$ and $M_M=M \mathbbm{1}_{2\times 2}$. Then one can combine $\Psi_{R,1}$ and $\Psi_{R,2}$ into a Dirac spinor $\Uppsi=(i\uppsi_1+\uppsi_2)/\sqrt{2}$.
On the other hand, one can split  any Dirac spinor into two Majorana spinors in this way. 

Thus, fermions can generically described by two component Weyl spinors. If one, for computational convenience, prefers to use four component spinors, one can use the chiral spinors $\Psi_{R,L}$. Only in special cases one can combine these into Dirac spinors. This is possible whenever the full mass matrix $\mathfrak{M}$ has degenerate eigenvalues (with opposite sign), and not necessarily related to the presence of ``Dirac'' or ``Majorana mass terms''.

\bibliographystyle{JHEP}
\bibliography{all}
\end{document}